\documentclass[journal]{IEEEtran}
\ifCLASSINFOpdf
\else
\fi

\usepackage{graphicx}

\usepackage{url}
\usepackage{subfigure}        
\usepackage[usenames]{color}            
\usepackage{amsfonts}%
\usepackage{amssymb}%
\usepackage{mathrsfs}
\usepackage{amsmath}%

\usepackage[latin1]{inputenc}

\newtheorem{definition}{Definition}
\newtheorem{theorem}{Theorem}
\newtheorem{proposition}{Proposition}

\newtheorem{pf}{Proof}

\begin{document}
%
\title{Tools for Analysis of Shannon-Kotel'nikov Mappings}
%
%
%

\author{Pål~Anders~Floor,
        Tor~A.~Ramstad,
\thanks{P. A. Floor is at NTNU Gjovik,
  Gjovik, Norway

  T. A.
Ramstad  is with the Department of Electronics and
Telecommunication, Norwegian University of Science and Technology
(NTNU), Trondheim,
Norway (e-mail: ramstad@iet.ntnu.no).}
}

\maketitle

\begin{abstract}
This document introduces tools from differential geometry needed for the analysis of a subset of Joint Source-Channel Codes (JSCC) named Shannon-Kotel'nikov (S-K) mappings. New results based on these concepts are further provided.
\end{abstract}


%
\IEEEpeerreviewmaketitle

\section{Introduction}\label{sec:Introduction}
This document extends the theory of S-K mappings~\cite{hekland_floor_ramstad_T_comm,HeklandThesis,FloorThesis} by introducing useful tools from differential geometry for analysis of  (hyper) surfaces  like \emph{curvature}, \emph{geodesics}, \emph{first fundamental form}, \emph{second fundamental form}, \emph{coordinate mappings}, \emph{ruled surfaces} and \emph{developable surfaces}. These results will complement the ones provided in~\cite{hekland_floor_ramstad_T_comm}. 

\section{Useful theorems and tools from differential geometry}\label{sec:Diff_geom}
This section explains some of the concepts and the terminology used in the extended analysis of S-K mappings.
Most results stated in this section are taken from E. Kreyszig's book \emph{Differential Geometry}~\cite{Kreyszig_DiffGeom91}.

\subsection{Fundamental concepts of Parametric Curves in $\mathbb{R}^3$}\label{ssec:DiffGeom_ParCurves}
Any continuous (\emph{class} $C^1$) curve $\mathcal{C}:\mathbf{S}\in\mathbb{R}^3$ can be represented perimetrically as the vector valued function
\begin{equation}\label{e:ParCurve_def}
\mathbf{S}(x)=[S_1(x), S_2(x), S_3(x)], \ x\in\mathbb{R}.
\end{equation}
In the following we denote the derivative w.r.t to a general parameter as ${\mathbf{S}}'$, ${\mathbf{S}}''$ and ${\mathbf{S}}'''$  and with respect to (w.r.t.) the special case of arch length as $\dot{\mathbf{S}}$, $\ddot{\mathbf{S}}$ and $\dddot{\mathbf{S}}$. That is, when we parameterize the curve using
\begin{equation}\label{e:CurveLength}
\ell(x)= \int_{x_0}^x \sqrt{\mathbf{S}'\cdot\mathbf{S}'} \mbox{d}x =\int_{x_0}^x \|\mathbf{S}'\| \mbox{d}x
\end{equation}
With this parametrization the norm of the derivative of the curve at any point, corresponding to the length of the tangent vector at that point, is unity. I.e.,  $\|\mathbf{t}\|=\|\dot{\mathbf{S}}\|=1$, $\forall x$ (See Appendix~\ref{sec:app_usp}).

There are three unit vectors connected to any curve $\mathcal{C}:\mathbf{S}\in\mathbb{R}^3$: The \emph{unit tangent vector} $\mathbf{t}=\dot{\mathbf{S}}=\mathbf{S}'/\|\mathbf{S}'\|$, the \emph{unit principal normal vector} $\mathbf{p}=\dot{\mathbf{t}}/\|\dot{\mathbf{t}}\|= \ddot{\mathbf{S}}/\|\ddot{\mathbf{S}}(x_0)\|$ , and the \emph{unit binormal vector} $\mathbf{b} =\mathbf{t} \times \mathbf{p}$.
The vectors $\mathbf{t}$, $\mathbf{p}$ and $\mathbf{b}$ make out a vector space of mutually orthogonal vectors named \emph{moving trihedron} which is so defined at each point along the curve $C$. This is nicely illustrated in Kreyszig's book~\cite[pp. 36-37]{Kreyszig_DiffGeom91}. These three vectors further define three mutually orthogonal planes: i) \emph{Osculating Plane} spanned by $\mathbf{t}$ and $\mathbf{p}$, ii) \emph{normal plane} spanned by $\mathbf{p}$ and $\mathbf{b}$, and iii) \emph{rectifying plane} spanned by $\mathbf{t}$ and $\mathbf{b}$.

For a parametric curve $\mathbf{S}(x)$ one can define the \emph{curvature}~\cite[p. 34]{Kreyszig_DiffGeom91} w.r.t. arch length as   $\kappa_0=\|\ddot{\mathbf{S}}(x_0)\|$. We then also have $\mathbf{p} = (1/\kappa)\ddot{\mathbf{S}} =\rho \ddot{\mathbf{S}}$.

The \emph{torsion}~\cite[p. 37-40]{Kreyszig_DiffGeom91} is defined as
\begin{equation}\label{e:torsion_def}
\tau(x) = -\mathbf{p}\cdot\mathbf{b} = \frac{\big|\dot{\mathbf{S}} \ \ddot{\mathbf{S}} \ \dddot{\mathbf{S}}\big|}{\|\ddot{\mathbf{S}}\|^2}.
\end{equation}
When $\tau = 0$, $\forall x$, we have a plane curve. Whenever $\tau\neq 0$, the curve is not a plane curve, but will "twist" up into space ($\mathbb{R}^3$). For a general parametrization the curvature and torsion are given by~\cite[pp.35,39]{Kreyszig_DiffGeom91}
\begin{equation}\label{e:curvature_curve_gen_coord}
\kappa = \frac{\sqrt{\|\mathbf{S}'\|^2\|\mathbf{S}''\|^2-(\mathbf{S}' \cdot \mathbf{S}'')^2}}{\|\mathbf{S}'\|^\frac{3}{2}},
\end{equation}
\begin{equation}\label{e:torsion_curve_gen_coord}
\tau = \frac{\big|\mathbf{S}' \ \mathbf{S}'' \ \mathbf{S}'''\big|}{\|\mathbf{S}'\|^2\|\mathbf{S}''\|^2-(\mathbf{S}' \cdot \mathbf{S}'')^2},
\end{equation}

The curvature can locally be interpreted as a circle of radius $\rho = 1/\kappa$, named \emph{radius of curvature}, lying in the osculating plane of $\mathbf{s}$. The corresponding circle is named \emph{Osculating Circle} and its center named \emph{centre of curvature}. I.e., the curvature in a $\epsilon$-neighborhood of a given point is equivalent to that of a circle with radius $\rho$. This concept of curvature is also valid for curves in $\mathbb{R}^M, M\geq3$.

The \emph{Formula of Frenet}~\cite[p. 41]{Kreyszig_DiffGeom91} relates the derivatives $\dot{\mathbf{t}}$, $\dot{\mathbf{p}}$ and $\dot{\mathbf{b}}$ to linear combinations of $\mathbf{t}$, $\mathbf{p}$, and $\mathbf{b}$ as follows:
\begin{equation}\label{e:FoF}
\begin{split}
\dot{\mathbf{t}} &= \kappa\mathbf{p}\\
\dot{\mathbf{p}} &= -\kappa \mathbf{t} + \tau \mathbf{b}\\
\dot{\mathbf{b}} &=  -\tau \mathbf{p}
\end{split}
\end{equation}

One can now obtain a \emph{canonical representation} of a curve. That is, its shape in the neighborhood of any of its points is related to the first three derivatives of the curve: With  a curve of class $\mathbf{S}\in C^3$, and $\ell$ the curve length, one has the Taylor expansion:
\begin{equation}\label{e:Taylor_Gen_curve_3order}
\mathbf{S}=\mathbf{S}(0)+\sum_{v=1}^3 \frac{\ell^v}{v!}\frac{\mbox{d}^v \mathbf{S}(0) }{\mbox{d}\ell^v}+O(\ell^3).
\end{equation}
From the formula of Frenet one can show that
\begin{equation}
\begin{split}
\ddot{\mathbf{S}}&=\dot{\mathbf{t}}=\kappa \mathbf{p}\\
\dddot{\mathbf{S}}&=\ddot{\mathbf{t}}=\dot{\kappa} \mathbf{p} +  \kappa \dot{\mathbf{p}}= \dot{\kappa} \mathbf{p} - \kappa^2\mathbf{t} + \kappa\tau\mathbf{b}.  \end{split}
\end{equation}
By aligning $\mathbf{t}(0)$, $\mathbf{p}(0)$ and $\mathbf{b}(0)$ with the positive rays of the $x_1$, $x_2$ and $x_3$ axes of $\mathbb{R}^3$ respectively, one obtains ($S_i$ is the i'th component of $\mathbf{S}$)~\cite[p.48]{Kreyszig_DiffGeom91}
\begin{equation}\label{e:Canon_repr}
\begin{split}
S_1(\ell) &= \ell - \frac{\kappa_0^2}{3!}\ell^3 + O(\ell^3)\\
S_2(\ell) &= \frac{\kappa_0}{2}\ell^2 + \frac{\dot{\kappa_0}}{3!}\ell^3 + O(\ell^3)\\
S_3(\ell) &=  \frac{\kappa_0 \tau_0}{3!}\ell^3 + O(\ell^3).
\end{split}
\end{equation}
By discarding all terms in each component except the dominating one we get the canonical representation
\begin{equation}\label{e:canon_curve_repr}
\mathbf{S}(\ell)\approx \bigg[\ell, \frac{\kappa_0}{2}\ell^2, \frac{\kappa_0\tau_0}{6}\ell^3\bigg]
\end{equation}

Another useful concept is that of the \emph{Osculating Sphere}~\cite[pp. 54-55]{Kreyszig_DiffGeom91}. A curve $\mathbf{S}\in \mathbb{R}^3$ of class $C^3$ (functions of continuous 3rd order derivatives) has 3rd order \emph{contact} (i.e. touches the curve up to 3rd order derivative) in the point $P$ with the osculating sphere, which is a sphere with center $\mathbf{a}=\mathbf{s}+\rho \mathbf{p} +\dot{\rho}/\tau \mathbf{b}$ and radius $R_s=\sqrt{\rho^2 + (\dot{\rho}/\tau)^2}$. For the special case where $\dot{\kappa}=0$ in $P$, the center of the osculating sphere will lie in the osculating plane of $\mathbf{s}$. That is, with constant curvature $\dot{\rho}=0$ and $R_s = \rho$ (like circles).

The above results imply that any curve of class $C^3$ (or above), $\mathbf{s}(x)\in\mathbb{R}^3$ has a \emph{spherical geometry} locally, where the sphere is uniquely determined by $\rho=1/\kappa$, $\dot{\rho}=1/\dot{\kappa}$ and $\tau$. Locally, we then have the approximation in~(\ref{e:canon_curve_repr}).

\subsection{Fundamental concepts of Parametric Surfaces in $\mathbb{R}^3$}\label{ssec:DiffGeom_ParSurfaces}
Many concepts from curves can be extended to surfaces, as any surface can be mathematically described as a set of (interrelated) \emph{coordinate curves}. However, many new concepts are necessary. We repeat key results from Chapter III in~\cite[pp.72-117 ]{Kreyszig_DiffGeom91} here.

\subsubsection{Fundamentals}\label{ssec:fundament_surf}
A surface (or manifold) $M\in\mathbb{R}^3$ can be described parametrically as
\begin{equation}\label{e:ParSurf_def}
\mathbf{S}(u^1,u^2)=[s_1(u^1,u^2),s_2(u^1,u^2),s_3(u^1,u^2)],
\end{equation}
where $u_1$ and $u_2$ are variables defined on a simply-connected bounded domain $B\subseteq\mathbb{R}^2$  (the $u^1 u^2$-plane). For~(\ref{e:ParSurf_def}) to represent a surface then:  1) $\mathbf{S}(u^1,u^2) \in C^r, \ r\geq1$ on $B$ and each point in the set $M$ represented by $\mathbf{S}(u^1,u^2)$ corresponds to just one ordered pair $(u^1,u^2)\in B$.  2) The Jacobian Matrix (see~(\ref{e:jacobian}) in Appendix~\ref{sec:app_one_mt}), $J$, is of rank 2 on $B$.

In the following, partial derivatives will be denoted
\begin{equation}\label{e:PartDer_Not}
\mathbf{S}_\alpha = \frac{\partial \mathbf{S}}{\partial u^\alpha}, \ \ \mathbf{S}_{\alpha\beta} = \frac{\partial^2 \mathbf{S}}{\partial u^\alpha \partial u^\beta}.
\end{equation}

By imposing a coordinate transformation
\begin{equation}\label{e:ACT}
u^\alpha = u^\alpha(\bar{u}^1,\bar{u}^2),
\end{equation}
a new parametric representation $\mathbf{S}(\bar{u}^1,\bar{u}^2)$ of $M$ is obtained.

\begin{definition}
An \emph{Allowable Coordinate Transformation} (ACT) is one where the functions in~(\ref{e:ACT}):

i) are defined in a domain $\bar{B}$ such that (s.t.) the corresponding range of values includes the domain $B$.

ii) are one-to-one of class $r\geq1$ ($\in C^r$) everywhere in $\bar{B}$.

iii) have Jacobian
\begin{equation}\label{e:Jacobian_ACT}
D=\frac{\partial(u^1,u^2)}{\partial (\bar{u}^1,\bar{u}^2)}=
\begin{vmatrix}
\frac{\partial{u^1}}{\partial{\bar{u}^1}}&
\frac{\partial{u^1}}{\partial{\bar{u}^2}}\\
\frac{\partial{u^2}}{\partial{\bar{u}^1}}&
\frac{\partial{u_2}}{\partial{\bar{u}^2}}
\end{vmatrix}
\neq 0,
\end{equation}
everywhere in $\bar{B}$.\hspace{6cm}$\square$
\end{definition}

A curve on a surface $\mathcal{S}: \mathbf{S}(u^1,u^2)$ can be determined by the parametric representation
\begin{equation}\label{e:SurfCurve_ParRepr}
u^1=u^1(t), \ \ u^2=u^2(t),
\end{equation}
of class $r\geq 1$, where the parameter $t\in \mathbb{R}$. Of special importance are the \emph{coordinate curves} on $\mathcal{S}$, $u^1 = $constant and  $u^2 = $constant, corresponding to parallels to the coordinate axes in $u^1 u^2$-plane. A set of curves on $\mathcal{S}$ that depend continuously on a parameter is said to be a \emph{one-parameter family of curves}. Two one-parameter family of curves are called a \emph{net of curves} on $\mathcal{S}$ if through every point $P$ of $\mathcal{S}$ there passes one and only one curve of each of these families, and if the two curves have a distinct direction at $P$.

The direction of the tangent to a curve $\mathcal{C}: u^1(t), u^2(t)$ on $\mathcal{S}: \mathbf{S}(u^1, u^2)$ is determined by:
\begin{equation}\label{e:tanget_dir_CurveSurf}
\mathbf{S}'=\frac{\mbox{d}\mathbf{S}}{\mbox{d} t}=\frac{\partial \mathbf{S}}{\partial u^1}\frac{\mbox{d} u^1}{\mbox{d} t}+\frac{\partial \mathbf{S}}{\partial u^2}\frac{\mbox{d} u^2}{\mbox{d} t}=\mathbf{S}_1 {(u^1)}' + \mathbf{S}_2 {(u^2)}',
\end{equation}
which depends on $t$. The tangent vector, $\mathbf{S}'$,  is a linear combination of the coordinate curves' tangent vectors $\mathbf{S}_\alpha$ at $P$. Whenever $\mathbf{S}_1$ and $\mathbf{S}_2$ are linearly independent, they span the tangent plane $E(P)$ of $\mathbf{S}$ at $P$. $E(P)$ therefore contains the tangent of any curve on $\mathbf{S}$ that passes through $P$. With $q^1,q^2$ coordinates of the points of $E(P)$, the tangent plane can be expressed as
\begin{equation}\label{e:tangent_plane_Eqn}
\mathbf{y}(q^1,q^2)=\mathbf{S}+q^1\mathbf{S}_1+q^2\mathbf{S}_2,\ \ \mathbf{S}_1\times \mathbf{S}_2 \neq 0.
\end{equation}
Since $\mathbf{y}-\mathbf{S}\in E(P)$, $\mathbf{y}-\mathbf{S}$, $\mathbf{S}_1$ and $\mathbf{S}_2$ are linearly dependent, giving the alternative expression for the tangent plane
\begin{equation}\label{e:TangentPlane_eq}
|\mathbf{y}-\mathbf{S}\ \mathbf{S}_1\  \mathbf{S}_2|=0.
\end{equation}

To cope with all the sum-operations that result when analyzing surfaces and coordinate transformations in general, it is convenient to express all sums through \emph{Einstein summation convention} to significantly shorten expressions~\cite[p.84]{Kreyszig_DiffGeom91}:

\begin{definition}\emph{Einstein summation convention:}
If in a product a letter figures twice, once a superscript and once a subscript, summation should be carried out from $1$ to $N$ w.r.t. this letter. \hspace{7cm}$\square$
\end{definition}

For example, for simple sums
\begin{equation}\label{e:summ_conv_vector}
\sum_{\alpha = 1}^N a^\alpha b_\alpha = a^\alpha b_\alpha,
\end{equation}
and for double sums
\begin{equation}\label{e:summ_conv_vector}
\sum_{\alpha = 1}^N \sum_{\beta = 1}^N a_{\alpha\beta} u^\alpha u^\beta = a_{\alpha\beta} u^\alpha u^\beta,
\end{equation}
and so on.

\textbf{Contravariant and Covariant vectors and tensors:}

\begin{definition}\emph{Contravariant and Covariant vectors:}

A contravariant vector (or 1st order contravariant tensor) at $P$ of $\mathbf{S}$ is a vector that under the coordinate transformation in~(\ref{e:ACT}), and its inverse, transform as
\begin{equation}\label{e:contravariant_vector}
\bar{a}^\beta = a^\alpha \frac{\partial \bar{u}^\beta}{\partial u^\alpha}\ \text{and}\ {a}^\gamma = \bar{a}^\beta \frac{\partial {u}^\gamma}{\partial \bar{u}^\beta}, \ \beta,\gamma = 1,\cdots,N,
\end{equation}
where $\bar{a}^\beta$ and $a^\alpha $ are vector components in the two coordinate systems. The components of a vector in one system is related to the components in the other system by \emph{parallel projection} onto the coordinate axis in the original system. See Fig.~\ref{fig:ContravarCovar} to the left.

A covariant vector (or 1st order covariant tensor) at $P$ of $\mathbf{S}$ is a vector that under the coordinate transformation in~(\ref{e:ACT}), and its inverse, transform as
\begin{equation}\label{e:covariant_vector}
\bar{b}^\beta = b^\alpha \frac{\partial {u}^\alpha}{\partial \bar{u}^\beta}\ \text{and}\ {b}^\gamma = \bar{b}^\beta \frac{\partial \bar{u}^\beta}{\partial {u}^\gamma}, \ \beta,\gamma = 1,\cdots,N,
\end{equation}
where $\bar{a}^\beta$ and $b^\alpha $ are vector components in the two coordinate systems. The components of a vector in one system is related to the components in the other system by \emph{orthogonal projection} onto the coordinate axis in the  original system. See Fig.~\ref{fig:ContravarCovar} to the left.\hspace{7cm}$\square$
\end{definition}

One example of a covariant vector field is the \emph{Gradient}, $\partial \phi /\partial u^\alpha$, of a scalar function, since
\begin{equation}\label{e:gradient}
\frac{\partial \phi}{\partial \bar{u}^\beta}=\frac{\partial \phi}{\partial {u}^\alpha}\frac{\partial u^\alpha}{\partial \bar{u}^\beta}.
\end{equation}

Note that in orthogonal coordinate systems, contravariant and covariant transformations are the same.

\begin{definition}\label{def:Tensor}\emph{2nd order Tensors:}

A \emph{2nd order contravariant tensor} at $P$ of $\mathbf{S}$ is an entity $a^{\alpha\beta}$ that under the coordinate transformation in~(\ref{e:ACT}), and its inverse, transform as
\begin{equation}\label{e:contravariant_2tensor}
\bar{a}^{\gamma\kappa} = a^{\alpha\beta} \frac{\partial \bar{u}^\gamma}{\partial u^\alpha}\frac{\partial \bar{u}^\kappa}{\partial u^\beta}, \ \gamma,\kappa = 1,\cdots,N,
\end{equation}
and
\begin{equation}\label{e:contravariant_2tensor_inv}
a^{\sigma\tau} = \bar{a}^{\gamma\kappa} \frac{\partial {u}^\sigma}{\partial \bar{u}^\gamma}\frac{\partial {u}^\tau}{\partial \bar{u}^\kappa}, \ \sigma,\tau = 1,\cdots,N,
\end{equation}
where $\bar{a}^{\gamma\kappa}$ and $ a^{\alpha\beta}$ are tensor components in the two coordinate systems.

A \emph{2nd order covariant tensor} at $P$ of $\mathbf{S}$ is an entity $a_{\alpha\beta}$ that under the coordinate transformation in~(\ref{e:ACT}), and its inverse, transform as
\begin{equation}\label{e:covariant_2tensor}
\bar{a}_{\gamma\kappa} = a_{\alpha\beta} \frac{\partial u^\alpha}{\partial \bar{u}^\gamma}\frac{\partial u^\beta}{\partial \bar{u}^\kappa}, \ \gamma,\kappa = 1,\cdots,N,
\end{equation}
and
\begin{equation}\label{e:covariant_2tensor_inv}
a_{\sigma\tau} = \bar{a}_{\gamma\kappa} \frac{\partial \bar{u}^\gamma}{\partial {u}^\sigma}\frac{\partial \bar{u}^\kappa}{\partial {u}^\tau}, \ \sigma,\tau = 1,\cdots,N,
\end{equation}
where $\bar{a}^{\gamma\kappa}$ and $ a^{\alpha\beta}$ are tensor components in the two coordinate systems.

A \emph{2nd order mixed tensor} at $P$ of $\mathbf{S}$ is an entity $a_\alpha^\beta$ that under the coordinate transformation in~(\ref{e:ACT}), and its inverse, transform as
\begin{equation}\label{e:mixed_2tensor}
\bar{a}_\gamma^\kappa = a_\alpha^\beta \frac{\partial u^\alpha}{\partial \bar{u}^\gamma}\frac{\partial \bar{u}^\kappa}{\partial u^\beta}, \ \gamma,\kappa = 1,\cdots,N,
\end{equation}
and
\begin{equation}\label{e:mixed_2tensor_inv}
a_\sigma^\tau = \bar{a}_\gamma^\kappa \frac{\partial \bar{u}^\gamma}{\partial {u}^\sigma}\frac{\partial {u}^\tau }{\partial \bar{u}^\kappa}, \ \sigma,\tau = 1,\cdots,N,
\end{equation}
where $\bar{a}_\gamma^\kappa$ and $ a_\alpha^\beta$ are tensor components in the two coordinate systems.\hspace{7.3cm}$\square$
\end{definition}

Note that the $N^2$ components of any 2nd order tensor can be arranged in a matrix.

The \emph{contravariant metric tensor} (which will be introduced in Section~\ref{ssec:Fundamental_Forms}) is an example of a contravariant tensor and the \emph{metric tensor} (or metric of the surface) as described in Appendix~\ref{sec:app_one_mt} is an example of a covariant tensor.

\subsubsection{Fundamental Forms and Curvature}\label{ssec:Fundamental_Forms}
The fundamental forms are crucial in order to analyze surfaces based on coordinate representations.

\textbf{First Fundamental Form (FFF):} In order to measure lengths and areas on a surface, introducing a metric on the surface is necessary. To determine an \emph{element of arc} (length differential) of a curve $C$~(\ref{e:SurfCurve_ParRepr}) on $\mathcal{S}$ one can use~(\ref{e:tanget_dir_CurveSurf}):
\begin{equation}\label{e:arc_length_element_surf}
\begin{split}
\mbox{d}\ell^2 &= (\mathbf{S}_1 \mbox{d}u^1 + \mathbf{S}_2 \mbox{d}u^2)\cdot(\mathbf{S}_1 \mbox{d}u^1 + \mathbf{S}_2 \mbox{d}u^2)\\
&= \mathbf{S}_1\cdot\mathbf{S}_1 (\mbox{d}u^1)^2 + 2\mathbf{S}_1\cdot\mathbf{S}_2 \mbox{d}u^1 \mbox{d}u^2 + \mathbf{S}_2\cdot\mathbf{S}_2 (\mbox{d}u^2)^2.
\end{split}
\end{equation}
We recognize the quantities $g_{\alpha\beta}=\mathbf{S}_\alpha\cdot\mathbf{S}_\beta$ as the components of the metric tensor in Appendix~\ref{sec:app_one_mt}. By the summation convention we have
\begin{equation}\label{e:FFF}
 \mbox{d}\ell^2  = g_{\alpha\beta} \mbox{d}u^\alpha \mbox{d}u^\beta.
\end{equation}
This \emph{quadratic form} is named First Fundamental Form (FFF), and defines a metric on $\mathcal{S}$ which allows us to measure arc lengths, angles and areas on $\mathcal{S}$. This is why the (covariant) tensor $g_{\alpha\beta}$ is named \emph{metric tensor}\footnote{Also named \emph{fundamental tensor} or \emph{Riemannian metric}.}. The metric tensor transforms covariantly under coordinate transformations (see Definition~\ref{def:Tensor})

\begin{figure}[h]
    \begin{center}
           \includegraphics[width=1.0\columnwidth]{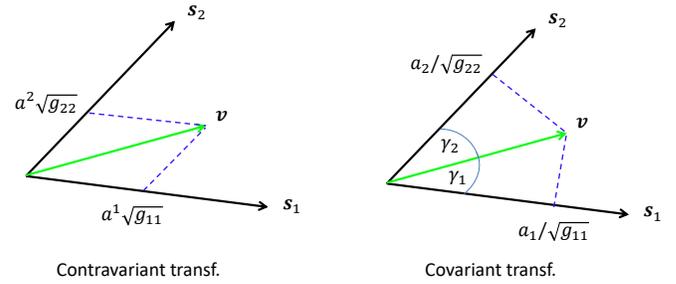}
    \end{center}
    \caption{Left: Decomposition into contravariant components $a^i$. Right: Decomposition into covariant components $a_i$.}\label{fig:ContravarCovar}
\end{figure}

Any vector $\mathbf{v}\in E(P)$ of $\mathcal{S}$ has a contravariant and covariant representation: A vector $\mathbf{v}$ on $\mathcal{S}$ at $P$ is in $E(P)$ and can therefore be decomposed as $\mathbf{v}=a^\alpha \mathbf{S}_\alpha$. The components $a^\alpha$ are the lengths of the respective parallel projections of $\mathbf{v}$ on the $\mathbf{S}_\alpha$ axis of the coordinate system in $E(P)$ measured in units of $\sqrt{g_{\alpha\alpha}}$ as shown to the left in Fig.~\ref{fig:ContravarCovar} (see section~\ref{ssec:fundament_surf}).
One may represent $\mathbf{v}$ by its orthogonal projection onto $\mathbf{S}_\alpha$. I.e., $\mathbf{v}=a_\alpha \mathbf{S}_\alpha$, where $a_\alpha = \mathbf{S}_\alpha \cdot\mathbf{v}$. With angle $\gamma$ between $\mathbf{v}$, and $\mathbf{S}_\alpha$, the length of the orthogonal projection of $\mathbf{v}$ upon $\mathbf{S}_\alpha$ will be $\|\mathbf{v}\|\cos \gamma = \mathbf{S}_\alpha\cdot \mathbf{v}/\sqrt{\mathbf{S}_\alpha\cdot \mathbf{S}_\alpha}=a_\alpha / \sqrt{g_{11}}$. I.e., $a_\alpha$ is the length of the orthogonal projection of $\mathbf{v}$ upon the $\mathbf{S}_\alpha$ axis of $E(P)$ measured in units of $1/\sqrt{g_{\alpha\alpha}}$ depicted to the right in Fig.~\ref{fig:ContravarCovar}.

Covariant and contravariant components of $\mathbf{v}$ are related:
\begin{equation}\label{e:covar_contravar_rel1}
a_\alpha = \mathbf{S}_\alpha \cdot \mathbf{v} = \mathbf{S}_\alpha \cdot a^\beta \mathbf{S}_\beta = g_{\alpha\beta} a^\beta.
\end{equation}
Since $g=\mbox{det}(g_{\alpha\beta})\neq 0$ one can solve~(\ref{e:covar_contravar_rel1}) w.r.t. $a^\beta$ to obtain
\begin{equation}\label{e:contravar_MetricTensor}
a_\beta = g^{\alpha\beta} a^\alpha,
\end{equation}
where
\begin{equation}\label{e:contravar_covar_rel}
g^{11}=\frac{g_{22}}{g}, \ g^{12}=g^{21}=- \frac{g_{12}}{g}, \ g^{22}=\frac{g_{11}}{g}
\end{equation}
and $g_{\alpha\beta}g^{\alpha\beta}=\delta_\alpha^\beta$.  $g^{\alpha\beta}$ are the components  of a 2nd order contravariant tensor named \emph{contravariant metric tensor}.

In a Cartesian coordinate system $g_{\alpha\beta}=\delta_{\alpha\beta}$ and $g^{\alpha\beta}=\delta^{\alpha\beta}$ and so $a^\alpha = a_\alpha$.

\textbf{Second Fundamental Form (SFF):} 
At any point of a curve $\mathcal{C}$ on a surface $\mathcal{S}$, the corresponding unit normal vector to $\mathcal{S}$
\begin{equation}\label{e:surface_unit_normal}
\mathbf{n}=\frac{\mathbf{S}_1\times \mathbf{S}_2}{\|\mathbf{S}_1\times \mathbf{S}_2\|}=\frac{\mathbf{S}_1\times \mathbf{S}_2}{-\sqrt{g}},
\end{equation}
lies in the normal plane to $\mathcal{C}$ which also contains the principal normal $\mathbf{p}$ to $\mathcal{C}$~\cite[p.118]{Kreyszig_DiffGeom91} (see Section~\ref{ssec:DiffGeom_ParCurves}). The angle $\gamma$ between these two vectors depend on the geometric shape of both $\mathcal{C}$ and $\mathcal{S}$ in a neighborhood of the point $P$ under consideration. We have two extremes:  i) $\gamma=\pi/2$ for all points along $\mathcal{C}$, implying that  $\mathbf{p}\perp \mathbf{n}$, and so $\mathcal{C}$ is a plane curve with $\mathcal{S}$ a plane (flat space with zero curvature). ii) $\gamma = 0$ for all points along $\mathcal{C}$, implying that  $\mathbf{p}|| \mathbf{n}$, and then $\mathcal{C}$  is a \emph{geodesic} on $\mathcal{S}$  (see discussion on geodesics in section~\ref{ssec:SK_ExpSurf}). That is the arc with the shortest possible length between its endpoints on $\mathcal{S}$. This is the equivalent of a straight line in a plane (like the great circles on a sphere), and is generally the path that minimizes \emph{energy}.

We now assume that $\mathcal{C}$ is represented by arc length parametrization $u^1(\ell), u^2(\ell)$. Since $\mathbf{n}$ and $\mathbf{p}$ are unit vectors, $\cos \gamma = \mathbf{p}\cdot\mathbf{n}$ ($\kappa > 0$). This scalar product will generally vary along  $\mathcal{C}$. From the formula of Frenet~(\ref{e:FoF}) we get
\begin{equation}\label{e:normals_CurveSurf_angle}
\kappa \cos \gamma = \ddot{\mathbf{S}}\cdot \mathbf{n}.
\end{equation}
From the product rule we have that
\begin{equation}\label{e:CurveSurf_arcderivative}
\dot{\mathbf{S}}= \frac{\partial \mathbf{S}}{\partial u^1} \frac{\mbox{d}u^1}{\mbox{d}\ell}+\frac{\partial \mathbf{S}}{\partial u^2} \frac{\mbox{d}u^2}{\mbox{d}\ell}=\mathbf{S}_\alpha \dot{u}^\alpha.
\end{equation}
Differentiating w.r.t. $\ell$ again, we get
\begin{equation}\label{e:CurveSurf_2nd arcderivative}
\ddot{\mathbf{S}}=\mathbf{S}_{\alpha\beta} \dot{u}^\alpha\dot{u}^\beta + \mathbf{S}_\alpha\ddot{u}^\alpha.
\end{equation}
Since $\mathbf{S}_\alpha \cdot \mathbf{n} = 0$, Eq.~(\ref{e:normals_CurveSurf_angle}) becomes
\begin{equation}\label{e:normals_CurveSurf_angle2}
\kappa \cos \gamma = (\mathbf{S}_{\alpha\beta} \cdot \mathbf{n})\dot{u}^\alpha\dot{u}^\beta.
\end{equation}
The term in the parentheses is an important quantity,
\begin{equation}\label{e:b_ab_coefs1}
b_{\alpha\beta}=\mathbf{S}_{\alpha\beta} \cdot  \mathbf{n}, \  \alpha,\beta=1,\cdots,N.
\end{equation}
The scalar products $b_{\alpha\beta}$ depend on $\mathcal{S}$ only (i.e.independent of choice of curve $\mathcal{C}$) and are symmetric due to the symmetry of $\mathbf{S}_{\alpha\beta}$. The quadratic form
\begin{equation}\label{e:SFF}
b_{\alpha\beta} \mbox{d}u^\alpha \mbox{d}u^\beta,
\end{equation}
is called the \emph{second fundamental form} (SFF). The SFF is invariant w.r.t allowable coordinate transformations which preserves the sign (or \emph{sense}) of $\mathbf{n}$. $b_{\alpha\beta}$ are therefore the components of a 2nd order covariant tensor (like $g_{\alpha\beta}$).

An alternative way to obtain $b_{\alpha\beta}$ is by differentiating $\mathbf{S}_\alpha \cdot \mathbf{n} = 0$. That is, $\mathbf{S}_{\alpha\beta}\cdot\mathbf{n}+\mathbf{S}_{\alpha}\cdot\mathbf{n}_\beta=0$, where $\mathbf{n}_\beta=\partial \mathbf{n}/\partial u^\beta$. Therefore
\begin{equation}\label{e:b_ab_coefs2}
b_{\alpha\beta}=-\mathbf{S}_{\alpha}\cdot\mathbf{n}_\beta, \  \alpha,\beta=1,\cdots,N.
\end{equation}

To compute $b_{\alpha\beta}$, the following relation is convenient to use:
\begin{equation}\label{e:compute_b_ab}
b_{\alpha\beta}= \mathbf{S}_{\alpha\beta} \cdot  \mathbf{n} = \mathbf{S}_{\alpha\beta}\cdot \bigg(\frac{\mathbf{S}_1\times \mathbf{S}_2}{\sqrt{g}}\bigg)=\frac{1}{\sqrt{g}}|\mathbf{S}_1\ \mathbf{S}_2\ \mathbf{S}_{\alpha\beta}|.
\end{equation}

Let now $t$ be any allowable parameter for the curve $\mathcal{C}$. Then
\begin{equation}\label{e:gen_par_surfcurve}
\dot{u}^\alpha = \frac{\mbox{d} u^\alpha}{\mbox{d} t}\frac{\mbox{d} t}{\mbox{d} \ell} = \frac{{u^\alpha}'}{\ell'}.
\end{equation}
Then, in the consequence of~(\ref{e:SFF}), Eq.~(\ref{e:normals_CurveSurf_angle}) becomes
\begin{equation}\label{e:normals_CurveSurf_angle3}
\kappa \cos \gamma = \frac{b_{\alpha\beta} {u^\alpha}' {u^\beta}'}{\ell'}=\frac{b_{\alpha\beta} {u^\alpha}' {u^\beta}'}{g_{\alpha\beta} {u^\alpha}' {u^\beta}'}=\frac{b_{\alpha\beta} \mbox{d}{u^\alpha} \mbox{d}{u^\beta}}{g_{\alpha\beta} \mbox{d}{u^\alpha} \mbox{d}{u^\beta}}.
\end{equation}

\textbf{Normal curvature, Principal curvature, Lines of curvature:}

We first seek a geometric interpretation of~(\ref{e:normals_CurveSurf_angle3}): For any fixed point $P\in\mathcal{S}$, $g_{\alpha\beta}$ and $b_{\alpha\beta}$ are fixed and independent of any curve $\mathcal{C}$ on $\mathcal{S}$ passing trough $P$. Therefore, the curvature $\kappa$ of  $\mathcal{C}$ at $P$ will only depend on the directions of the tangent $\mathbf{t}$ and the principal normal $\mathbf{p}$. Since the osculating plane is spanned by $\mathbf{t}$ and $\mathbf{p}$, then any curve that has the same osculating plane at $P$, will have the same curvature $\kappa$. This includes the curve of intersection between this common osculating plane and $\mathcal{S}$, and therefore one can restrict the investigation to plane curves on $\mathcal{S}$.

Consider now all curves that has the same tangent direction $\mathbf{t}_c$  at $P$. For these curves, the right hand side of~(\ref{e:normals_CurveSurf_angle3}) is constant implying that their curvature depend only on the angle $\gamma$. Therefore, for these curves, $\kappa\cos \gamma = \kappa_n$, with $\kappa_n$ a constant. If $\gamma=0$ then $\kappa = \kappa_n $, and if $\gamma=\pi$ then $\kappa = -\kappa_n $, implying that $|\kappa_n|$ is the curvature of the curve of intersection between $\mathcal{S}$ and a plane passing through both the tangent to $\mathcal{C}$ at $P$ and the normal to $\mathcal{S}$ at $P$. These curves of intersection are called \emph{normal sections} of $\mathcal{S}$.  $\kappa_n$ is called \emph{normal curvature} of $\mathcal{S}$ at $P$ corresponding to the direction of the tangent to the normal section at $P$. We then have the \emph{normal curvature vector} $\mathbf{k}_n=\kappa_n \mathbf{n}$. From this discussion and~(\ref{e:normals_CurveSurf_angle3}) we see that
\begin{equation}\label{e:normal_curvature}
\kappa_n =\frac{b_{\alpha\beta} \mbox{d}{u^\alpha} \mbox{d}{u^\beta}}{g_{\alpha\beta} \mbox{d}{u^\alpha} \mbox{d}{u^\beta}}.
\end{equation}
One can now set $\kappa_n=1/R$ where $|R|$ is the radius of curvature of the corresponding normal section at $P$. Since $\kappa = 1/\rho$ we can write
\begin{equation}\label{e:rel_radCurv_radNormsec}
\rho = R\cos \gamma.
\end{equation}
We have the Theorem of Meusnier~\cite[p.122]{Kreyszig_DiffGeom91}\\
\begin{theorem}\label{th:Meusnier}
The centre of curvature of all curves on $\mathcal{S}$ passing through an arbitrary point $P$ and whose tangent at $P$ have the same direction (different from an \emph{asymptotic direction}, where $b_{\alpha\beta} \mbox{d}{u^\alpha} \mbox{d}{u^\beta}=0$), lie on a circle $K$ of radius $|R|/2$ which lies in the normal plane and has (at least) first order contact with $\mathcal{S}$ at $P$. \hspace{4.6cm}$\square$\\
\end{theorem}
This implies that we can restrict the consideration to normal sections of $\mathcal{S}$ at $P$ without loss of generality.

A special point is when $b_{\alpha\beta}\sim g_{\alpha\beta}$ named \emph{navel point} or \emph{umbilic}. For points that are not umbilics, we seek directions where $\kappa_n$ has extreme values: The formula for normal curvature~(\ref{e:normal_curvature}) can be rewritten as
\begin{equation}\label{e:Norm_curvature_rewritten}
 (b_{\alpha\beta}- \kappa_n g_{\alpha\beta}) \mbox{d}{u^\alpha} \mbox{d}{u^\beta}=0.
\end{equation}
By differentiating w.r.t. $\mbox{d}{u^\gamma}$, treating $\kappa_n$ as a constant, one obtains~\cite[pp.128-129]{Kreyszig_DiffGeom91}
\begin{equation}\label{e:Principal_curvature_DiffEq1}
 (b_{\alpha\gamma}- \kappa_n g_{\alpha\gamma}) \mbox{d}{u^\alpha}=0, \ \gamma = 1,2.
\end{equation}
By eliminating $\kappa_n$, then
\begin{equation}\label{e:Principal_curvature_DiffEq2}
\begin{vmatrix}
g_{1\alpha}\mbox{d}{u^\alpha}&
b_{1\beta}\mbox{d}{u^\beta}\\
g_{2\alpha}\mbox{d}{u^\alpha}&
b_{2\beta}\mbox{d}{u^\beta}
\end{vmatrix}
=0
\end{equation}
or
\begin{equation}\label{e:Principal_curvature_DiffEq3}
\begin{vmatrix}
(\mbox{d}{u^2})^2& -\mbox{d}{u^1}\mbox{d}{u^2} & (\mbox{d}{u^1})^2\\
g_{11}& g_{12}& g_{22}\\
b_{11}& b_{12}& b_{22}
\end{vmatrix}
=0
\end{equation}
The roots of~(\ref{e:Principal_curvature_DiffEq2}) determine the directions for which $\kappa_n$ is extreme, named \emph{principal directions of normal curvature}  at $P$. The corresponding curvature values, $\kappa_1$ and $\kappa_2$, are the \emph{principal normal curvatures} of $\mathcal{S}$, corresponding the maximal and minimal curvature of $\mathcal{S}$ at $P$ respectively. We have the following theorem~\cite[p.129]{Kreyszig_DiffGeom91}:\\

\begin{theorem}\label{th:princ_dir_orthog}
The roots of~(\ref{e:Principal_curvature_DiffEq2}) are real, and at every point not an umbilic, the principal directions are orthogonal.
\end{theorem}

\emph{Proof:} See~\cite[p.129]{Kreyszig_DiffGeom91}.\hspace{5.0cm}$\square$\\

A curve on $\mathcal{S}$ whose direction at every point is a principal direction is known as a \emph{line of curvature} (LoC) of $\mathcal{S}$. The LoC is a solution to the differential equation in~(\ref{e:Principal_curvature_DiffEq3}). We have two LoC at each point $P$ (not umbilic), where the following theorem is satisfied~\cite[p.130]{Kreyszig_DiffGeom91}:\\

\begin{theorem}\label{th:LoC_Net}
The LoC on any surface $\mathcal{S}\in C^r$, $r\geq 3$, are real curves. If $\mathcal{S}$ has no umbilics, the LoC form an orthogonal net everywhere on $\mathcal{S}$.
\end{theorem}

\emph{Proof:} See~\cite[p.130]{Kreyszig_DiffGeom91}.\hspace{5.0cm}$\square$\\

One may always choose coordinates $u^1,u^2$ on $\mathcal{S}$ so that the LoC are allowable coordinates at any point of $\mathcal{S}$ not umbilic. We have~\cite[p.130]{Kreyszig_DiffGeom91}:\\

\begin{theorem}\label{th:LoC_Coordinates}
The coordinate curves of any allowable coordinate system on $\mathcal{S}$  coincide with the LoC $\Leftrightarrow$
\begin{equation}\label{e:LoC_Coord_cond}
g_{12}=0 \ \      \text{and} \ \ b_{12}=0,
\end{equation}
at any point where those coordinates are allowable.
\end{theorem}

\emph{Proof:} See~\cite[p.130]{Kreyszig_DiffGeom91}.\hspace{5.0cm}$\square$\\

To derive an analytical expression for the principal curvatures, multiply~(\ref{e:Principal_curvature_DiffEq1}) with $g^{\beta\gamma}$ and sum w.r.t. $\gamma$ to obtain
\begin{equation}\label{e:Princ_curvat_eq1}
b_\alpha^\beta \mbox{d}u^\alpha - \kappa_n\mbox{d}u^\beta = 0, \ \beta=1,2,
\end{equation}
which after some algebraic manipulation leads to~\cite[p.130]{Kreyszig_DiffGeom91}
\begin{equation}\label{e:Princ_curvat_eq2}
\kappa_n^2 - b_{\alpha\beta}g^{\alpha\beta}\kappa_n +\frac{b}{g}=0.
\end{equation}
The principal curvatures $\kappa_1$ and $\kappa_2$, are the roots of this equation.

We have two important definitions~\cite[p. 131]{Kreyszig_DiffGeom91}:\\

\begin{definition}\label{def:GaussianCurvature}
The product
\begin{equation}\label{e:GaussianCurvature}
K = \kappa_1\kappa_2 =\frac{b}{g},
\end{equation}
is the \emph{Gaussian Curvature} of $\mathcal{S}$ at $P$.\\
\end{definition}

\begin{definition}\label{def:MeanCurvature}
The arithmetic mean
\begin{equation}\label{e:MeanCurvature}
H = \frac{1}{2} (\kappa_1+\kappa_2)= \frac{1}{2}b_{\alpha\beta}g^{\alpha\beta} =\frac{1}{2}b_\alpha^\alpha,
\end{equation}
is the \emph{Mean Curvature} of $\mathcal{S}$ at $P$.\\
\end{definition}

Both $K$ and $|H|$ are invariant under allowable coordinate transformations.

When the coordinate curves are LoC,~(\ref{e:Princ_curvat_eq1}) holds with $\kappa=\kappa_1$, $\mbox{d}u^2=0$ and again with $\kappa=\kappa_2$, $\mbox{d}u^1=0$. Therefore
\begin{equation}\label{e:LoC_Coord_cond}
\kappa_1=b_1^1, \ b_1^2=0, \ \kappa_2=b_2^2, \ b_2^1=0.
\end{equation}
For this special case
\begin{equation}\label{e:LoC_Coord_Curvatures}
\kappa_1=\frac{b_{11}}{g_{11}}, \ \kappa_2= \frac{b_{22}}{g_{22}}, \ K=\frac{b_{11}}{g_{11}}\frac{b_{22}}{g_{22}}, \ H=\frac{1}{2}\bigg(\frac{b_{11}}{g_{11}}+\frac{b_{22}}{g_{22}}\bigg).
\end{equation}

The normal curvature $\kappa_n$ for any (tangent) direction can be represented in terms of the principal curvatures $\kappa_1$ and $\kappa_2$ according to the Theorem of Euler~\cite[p. 132]{Kreyszig_DiffGeom91}:\\

\begin{theorem}\label{th:Euler}
Let $\alpha$ be the angle between an arbitrary direction at $P$ and the direction at $P$ corresponding to $\kappa_1$. Then
\begin{equation}\label{e:LoC_Coord_cond}
\kappa_n=\kappa_1\cos^2\alpha+\kappa_2\sin^2\alpha.
\end{equation}
\end{theorem}

\emph{Proof:} See~\cite[p.132]{Kreyszig_DiffGeom91}.\hspace{5.0cm}$\square$\\

We have one last theorem  concerning LoC~\cite[p.191]{Kreyszig_DiffGeom91}:\\

\begin{theorem}\label{th:LoC_Ort_Conj}
With the exception of umbilics, the family of LoC for $\mathcal{S}$ of class $r \geq 3$ will be orthogonal and \emph{conjugate}.
\end{theorem}

\emph{Proof:} See~\cite[p.191]{Kreyszig_DiffGeom91}.\hspace{5.0cm}$\square$\\

Note: Two directions $\mbox{d}u^1:\mbox{d}u^2$ and $\delta^1:\delta^2$ are conjugate $\Leftrightarrow$  $b_{\alpha\beta}\mbox{d}u^\alpha\delta^\beta=0$ (see Theorem 60.2~\cite[p.191]{Kreyszig_DiffGeom91})

\subsubsection{Formulae of Weingarten and Gauss}\label{ssec:Fomula_Wein_Gauss_Geodes}
For curves in $\mathbb{R}^3$ one can at each point $P$ associate three orthogonal unit vectors, $\mathbf{t}$ $\mathbf{p}$  and $\mathbf{b}$. Any vector bound at $P$ will be a linear combination of these vectors, like the derivatives  $\dot{\mathbf{t}}$ $\dot{\mathbf{p}}$, $\dot{\mathbf{b}}$ related through the formulae of Frenet.  For surfaces in $\mathbb{R}^3$ one have similar relations for the vectors $\mathbf{S}_1$, $\mathbf{S}_2$ and $\mathbf{n}$. Relations for the derivatives of these vectors are provided by the \emph{formulae of Weingarten} (FoW) and \emph{formulae of Gauss} (FoG):

The FoW relates the derivative of the normal vector to $\mathbf{x}_\beta$ as shown in~\cite[p. 139]{Kreyszig_DiffGeom91}
\begin{equation}\label{e:FoW}
\mathbf{n}_\alpha \doteq \frac{\partial \mathbf{n}}{\partial u^\alpha} = -g^{\sigma\beta}b_{\alpha\sigma}\mathbf{S}_\beta =-b_\alpha^\beta\mathbf{S}_\beta, \ \alpha=1,2.
\end{equation}
If LoC coordinate curves are chosen, then FoW reduces to
\begin{equation}\label{e:FoW_LoC}
\mathbf{n}_\alpha = \frac{b_{\alpha\alpha}}{g_{\alpha\alpha}}\mathbf{S}_\alpha, \ \alpha=1,2.
\end{equation}

The FoG relates the 2nd derivatives $\mathbf{S}_{\alpha\beta}$ to $\mathbf{S}_\gamma$ and $\mathbf{n}$ as shown in~\cite[p.140-142]{Kreyszig_DiffGeom91},
\begin{equation}\label{e:FoG}
\mathbf{S}_{\alpha\beta} \doteq \frac{\partial^2 \mathbf{S}}{\partial u^\alpha\partial u^\beta} = \Gamma_{\alpha\beta}^\gamma \mathbf{S}_{\gamma}+b_{\alpha\sigma}\mathbf{n}, \ \alpha,\beta=1,2,
\end{equation}
where
\begin{equation}\label{e:Christoffel2}
\Gamma_{\alpha\beta}^\gamma = g^{\lambda\kappa} \Gamma_{\alpha\beta\lambda},
\end{equation}
are \emph{Christoffel symbols of the 2nd kind} (CS-2), and where
\begin{equation}\label{e:Christoffel1}
\Gamma_{\alpha\beta\lambda} = \frac{1}{2}\bigg[\frac{\partial g_{\beta\lambda}}{\partial u^\alpha} + \frac{\partial g_{\lambda\alpha}}{\partial u^\beta}-  \frac{\partial g_{\alpha\beta}}{\partial u^\lambda} \bigg],
\end{equation}
are \emph{Christoffel symbols of the 1st kind} (CS-1). Note that the Christoffel symbols are uniquely defined by  the  metric $g_{\alpha\beta}$ and combinations of its derivatives on $\mathcal{S}$, making them an intrinsic property of the surface. That is, a quantity that depends on the FFF only, independent of the SFF and the surrounding (embedding) space~\cite[p. 178]{Kreyszig_DiffGeom91}.

\section{New theorems and results concerning S-K mappings based on differential geometrical concepts}\label{sec:SK_map_New_Proofs}
We start off with results concerning curves, that is $1$:$N$- and $M$:$1$ mappings. Intuition obtained through the concept of curves can than be extended further to surfaces. The basics of S-K mappings are provided in~\cite{Floor_Ramstad09} where distortion analysis is formulated w.r.t. the metric tensor, or FFF. Here we also take curvature into account by considering the SFF (see Section~\ref{ssec:Fundamental_Forms}). We begin with the definition of S-K mappings:\\

\begin{definition}\label{def:sk_mapp}\emph{Shannon-Kotel'nikov mapping}\\
An S-K mapping $\mathcal{S}$ is a continuous or piecewise continuous nonlinear or linear mapping between $\mathbb{R}^M$ (source space) and $\mathbb{R}^N$ (channel space). There are three cases to consider:

1. Equal dimension $M = N$: $\mathcal{S}$ is a bijective\footnote{MMSE decoding is needed at low SNR in order to obtain optimality, effectively weakening this condition.}  mapping.

2. Dimension expansion $M<N$: $\mathcal{S}\subseteq \mathbb{R}^N$ is a mapping that can be realized by a hyper surface described by the parametrization\footnote{This is not a restriction, i.e. the mapping does not need to be described by a parametrization.}
\begin{equation}\label{e:par_surf_eq}
\mathbf{S}(\mathbf{x})=[S_1(\mathbf{x}),S_2(\mathbf{x}),\cdots,S_N(\mathbf{x})],
\end{equation}
where each source vector $\mathbf{x}$ should have a unique representation $\mathbf{S}(\mathbf{x})\in \mathcal{S}$. $\mathcal{S}$ is then an M dimensional (locally Euclidean) manifold embedded in $\mathbb{R}^N$.

3. Dimension reduction $M>N$:  $\mathcal{S}\subseteq \mathbb{R}^M$ is a mapping that can be realized by a hyper surface described by the parametrization
\begin{equation}\label{e:par_surf_eq2}
\mathbf{S}(\mathbf{z})=[S_1(\mathbf{z}),S_2(\mathbf{z}),\cdots,S_M(\mathbf{z})],
\end{equation}
where each channel vector $\mathbf{z}$ should have a unique representation $\mathbf{S}(\mathbf{z})\in \mathcal{S}$. $\mathcal{S}$ is then an $N$ dimensional (locally
Euclidean) manifold embedded in $\mathbb{R}^M$. \hspace{1.5cm}$\square$\\
\end{definition}

\subsection{Higher order distortion analysis of S-K mappings using parametric curves}\label{ssec:SK_curves}
We treat $M$:$1$ SK-mappings first. Then use the intuition gained to treat $1$:$N$ mappings.

\subsubsection{$M$:$1$ dimension reducing mappings}\label{ssec:SK_curves_M_1}
In order to account for curvature and higher other order behavior of S-K mappings one has to consider the Taylor expansion beyond 1st order. We consider 3rd order Taylor expansion of a curve in $\mathbb{R}^3$ here and draw the conclusion that 2nd order behaviour is adequate for the investigation of S-K mappings at high SNR in general.
%

In the following we denote the derivatives w.r.t. arch length as $\dot{\mathbf{S}}$, $\ddot{\mathbf{S}}$ and $\dddot{\mathbf{S}}$ and w.r.t to a general parameter as ${\mathbf{S}}'$, ${\mathbf{S}}''$ and ${\mathbf{S}}'''$. With $x_0$ transmitted and noise $n$ we have from the 3rd order Taylor expansion:
\begin{equation}\label{e:Taylor_M_1}
\mathbf{S}(x_0 + n) \approx \mathbf{S}(x_0) + n \mathbf{S}'(x_0) + \frac{n^2}{2} \mathbf{S}''(x_0)+\frac{n^3}{3!} \mathbf{S}'''(x_0).
\end{equation}
From this we can derive the \emph{channel distortion} as
\begin{equation}\label{e:Derive_ChDist_M1_3rdOrder}
{\varepsilon}_{ch}^2(x_0) = E\bigg\{\bigg\| n \mathbf{S}'(x_0) + \frac{n^2}{2} \mathbf{S}''(x_0)+\frac{n^3}{3!} \mathbf{S}'''(x_0) \bigg\|^2\bigg\}.
\end{equation}
To expand this expression and take the expectation, it is advantages to use arc length parametrization (see Appendix~\ref{sec:app_usp}). Then the first, second and third derivatives will at each point of $\mathbf{S}$ make out a vector space of mutually orthogonal vectors as explained in Section~\ref{ssec:DiffGeom_ParCurves} (and with more detail in~\cite[pp. 36-37]{Kreyszig_DiffGeom91}). Since $\dot{\mathbf{S}} \cdot \ddot{\mathbf{S}} = 0$, $\dot{\mathbf{S}} \cdot \dddot{\mathbf{S}} = 0$ and $\ddot{\mathbf{S}} \cdot \dddot{\mathbf{S}} = 0$, and using the fact that~\cite[p.148]{papoulis02}
\begin{equation}\label{e:Moments_Gaussian}
E\{n^a\}=
\begin{cases}
  1\cdot3\cdots (a-1)\sigma_n^a, & a \ \ \text{even}  \\0, &a \ \ \text{odd}.
\end{cases}
\end{equation}
for $n$ Gaussian, the expectation of the norm in~(\ref{e:Derive_ChDist_M1_3rdOrder}) can be found from straight forward calculations
\begin{equation}\label{e:ChDist_M1_3rdOrder}
{\varepsilon}_{ch}^2(x_0) = \sigma_n^2\|\dot{\mathbf{S}}(x_0)\|^2 + \frac{3\sigma_n^4}{4} \|\ddot{\mathbf{S}}(x_0)\|^2 + \frac{5\sigma_n^6}{12} \|\dddot{\mathbf{S}}(x_0)\|^2,
\end{equation}
where the first norm $\|\dot{\mathbf{S}}(x_0)\|=1$ according to Theorem~\ref{th:clpar} in Appendix~\ref{sec:app_usp}. Using the canonical representation defined in~(\ref{e:canon_curve_repr}) in Section~\ref{sec:Diff_geom} one can expresses the channel distortion up to 3rd order~\cite[p.48]{Kreyszig_DiffGeom91}, valid for any curve $\mathbf{s}\in \mathbb{C}^3$ (of class $C^r, r\geq3$):
\begin{equation}\label{e:ChDist_Canon_repr}
\begin{split}
{\varepsilon}_{ch}^2(x_0) &= E\bigg\{\bigg\| \bigg[n,\frac{\kappa_0}{2} n^2, \frac{\kappa_0\tau_0}{6} n^3\bigg] \bigg\|^2\bigg\}\\
 &= \sigma_n^2 + \frac{3}{4}\kappa_0^2 \sigma_n^4 + \frac{5}{12}\kappa_0^2 \tau_0^2\sigma_n^6.
\end{split}
\end{equation}
We can conclude from~(\ref{e:ChDist_M1_3rdOrder}) and~(\ref{e:ChDist_Canon_repr}) that the 1st order approximation dominates when $\sigma_n$ is small, that is at high SNR, as long as the curvature $\kappa$ and torsion $\tau$ are not large. That is, $\sigma_n^4$ and $\sigma_n^6$ goes to zero much faster than $\sigma_n^2$, making the 1st order channel distortion approximation accurate at high SNR. As we will see in Section~\ref{ssec:CanalSurf_curves}, one should not choose a curve with large curvature or torsion to obtain a good mapping (large relative to transmission power and the curvature of the "noise sphere"). With this choice one can safely discard the 3rd order term as the 2nd order term will dominate.

The ML-estimate error also increases with $\kappa$ and $\tau$ as seen from~(\ref{e:ChDist_Canon_repr}) as $\kappa_0=\tau_0=0$ minimizes the error. However, we cannot avoid nonzero $\kappa$ and $\tau$ if we are to fill the space in question in order to keep the approximation distortion low. One should therefore choose a curve that fills the space but at the same time has the smallest possible $\kappa$ and $\tau$.

In general, the curvature is given by~(\ref{e:curvature_curve_gen_coord}), leading to a more complicated relations than the ones above. Since we most often will consider arc length parametrization with an amplification $\alpha$, which we name \emph{scaled arc length parametrization}, we have $\|{\mathbf{S}}'(x_0)\|=\alpha\|\dot{\mathbf{S}}(x_0)\|=\alpha$, $\forall x_0$. Then~(\ref{e:curvature_curve_gen_coord}) reduces to $\kappa(x_0)=\|\mathbf{S}_0''(x_0)\|/\|\mathbf{S}_0'\|^2$ (since $\mathbf{S}_0' \perp \mathbf{S}_0''$, still). Therefore we can express the channel error up to second order as
\begin{equation}\label{e:ChDist_Final_2ndOrder}
\begin{split}
{\varepsilon}_{ch}^2(x_0) &= \sigma_n^2\|\mathbf{S}'_0\|^2 + \frac{3\sigma_n^4}{4}\frac{\|\mathbf{S}''_0\|^2}{\|\mathbf{S}'_0\|^4}= \alpha^2\sigma_n^2 + \frac{3\sigma_n^4 \kappa^2(x_0) }{4}.
\end{split}
\end{equation}

\subsubsection{$1$:$N$ dimension expanding mappings}\label{ssec:SK_curves_1_N}
It is more complicated to compute higher order terms for the error in the expansion case as the ML decoder is a projection onto $\mathbf{S}$. We illustrate this for 2nd order approximation:
\begin{figure}[h]
    \begin{center}
           \includegraphics[width=0.9\columnwidth]{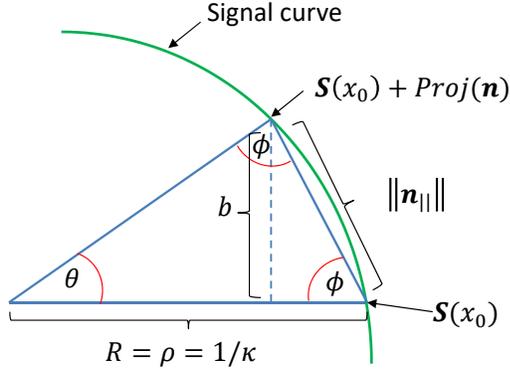}
    \end{center}
    \caption{Circle approximation for calculation of error up to 2nd order approximation.}\label{fig:PPM_curvature}
\end{figure}
From the Taylor series we have up to 2nd order
\begin{equation}\label{e:Taylor_1_N}
\mathbf{S}(x) \approx \mathbf{S}(x_0) + \mathbf{S}_0'\cdot(x - x_0) + \frac{\mathbf{S}_0''}{2}(x - x_0)^2,
\end{equation}
where
\begin{equation}\label{e:derivative_abbrv}
\mathbf{S}_0'=\frac{\mbox{d}\mathbf{S}}{\mbox{d}x}_{|x=x_0}.
\end{equation}
With $\mathbf{z}=\mathbf{S}(x_0)+\mathbf{n}$, the ML estimate seek to minimize $\|\mathbf{z}-\mathbf{S}(x)\|$, which up to 2nd order is the same as
\begin{equation}\label{e:ML_est_1_N_2nd_order}
\hat{x}_{ML}=\arg \min_x\bigg\|\mathbf{n} -  \mathbf{S}_0'\cdot(x - x_0) - \frac{\mathbf{S}_0''}{2}(x - x_0)^2 \bigg\|^2.
\end{equation}
It is straight forward to show that
\begin{equation}\label{e:derive_ML_2nd_order}
\begin{split}
&\frac{\mbox{d}}{\mbox{d} x}\bigg\|\mathbf{n} -  \mathbf{S}_0'\cdot(x - x_0) - \frac{\mathbf{S}_0''}{2}(x - x_0)^2 \bigg\|^2 \\
&=\frac{1}{2}\|\mathbf{S}_0''\|^2(x-x_0)^3 + \|\mathbf{S}_0'\|^2(x-x_0)-\cdots \\
&\mathbf{n}\cdot \mathbf{S}_0'- \mathbf{n}\cdot \mathbf{S}_0'' (x-x_0).
\end{split}
\end{equation}
Let now
\begin{equation}
a = \frac{2(\|\mathbf{S}_0'\|^2- \mathbf{n}\cdot \mathbf{S}_0'')}{\|\mathbf{S}_0''\|^2}, \ \ b=-\frac{2 \mathbf{n}\cdot \mathbf{S}_0'}{\|\mathbf{S}_0''\|^2},
\end{equation}
and $w = x-x_0$. Then the minimization problem in~(\ref{e:ML_est_1_N_2nd_order}) is the solution to $w^3+a w +b = 0$, which is a 3rd order polynomial equation that can be solved using \emph{Cardano's Method}. However, this does not result in a simple expression that one can  analyze further. It is therefore necessary to use another approach.

Using the concept of the osculating circle in Section~\ref{ssec:DiffGeom_ParCurves} we can apply the higher order analysis of distortion for \emph{pulse position modulation} (PPM) in~\cite[pp. 703-704]{wozandj65} to obtain a result valid for any $1$:$N$ mapping. PPM is geometrically described by a curve on a hyper sphere where the arc between any two coordinate axes is like a circle segment as depicted in Fig.~\ref{fig:PPM_curvature}.
The fact that any curve has a local spherical geometry implies that the analysis done for PPM is also valid locally for any other $1$:$N$ mapping under (scaled) arc length parametrization. In the following we name the signal curve segment in Fig.~\ref{fig:PPM_curvature} \emph{circle approximation}.

In polar coordinates the signal curve can be written
\begin{equation}\label{e:sign_curve_polar_coordinates}
\mathbf{S}(x)\approx [R(x),\theta(x)].
\end{equation}
Let $\mathbf{e}_r$ be a unit basis vector in the radial direction and $\mathbf{e}_\theta$ a unit basis vector in the angular direction in Fig.~\ref{fig:PPM_curvature}. We further divide the noise into two components $\text{Proj}(\mathbf{n})=\mathbf{n}_{||}$, that is the  projection onto the closest point on the circle in Fig.~\ref{fig:PPM_curvature}, and its normal, $\mathbf{n}_\perp$.

Further, for the circle approximation
\begin{equation}\label{e:RadiusOfCurvature_Expansion}
R(x)= \rho(x) = \frac{1}{\kappa(x)}=\frac{1}{\|\mathbf{S}_0''\|},
\end{equation}
iff $\mathbf{t}=\|\mathbf{S}_0'\|=1$. That is, under arc length parametrization. 
For PPM we have a constant radius $R(x)=R$ and the approximation in~(\ref{e:sign_curve_polar_coordinates}) becomes exact, that is, $\mathbf{S}=[R \cos(\theta(x)), R\sin(\theta(x))]$, and so
\begin{equation}\label{e:derivative_COC}
\frac{\mbox{d}\mathbf{S}}{\mbox{d} x} =\bigg[R\sin(\theta(x))\frac{\mbox{d}\theta(x)}{\mbox{d} x}, -R\cos(\theta(x))\frac{\mbox{d}\theta(x)}{\mbox{d} x} \bigg].
\end{equation}
By taking the norm we find that $\|{\mbox{d}\mathbf{S}}/{\mbox{d} x}\|^2=R^2 {\mbox{d}\theta(x)}/{\mbox{d} x}$. From this we get
\begin{equation}\label{e:PPM_theta_derivative}
\frac{\mbox{d}\theta(x)}{\mbox{d} x} = \frac{1}{R}\bigg\|\frac{\mbox{d}\mathbf{S}}{\mbox{d} x}\bigg\| = \frac{\|\mathbf{S}_0'\|}{\rho}=\alpha \kappa= \alpha \|\mathbf{S}_0''\|.
\end{equation}
That is, the rate of change of $\mathbf{S}$ is proportional to $\kappa$. Note that the two last equalities in~(\ref{e:PPM_theta_derivative}) are valid only for scaled arc length parametrization.


To determine the ML estimate for the circle approximation we rewrite~(\ref{e:PPM_theta_derivative}) as $\mbox{d}x = \mbox{d}\theta(x)/(\|\mathbf{s}_0'\|\|\mathbf{s}_0''\|)=\mbox{d}\theta(x)\rho(x_0)/\|\mathbf{s}_0'\|$. Therefore
\begin{equation}\label{e:ML_est_1_N_CircEst}
x_0 - \hat{x}_{ML} = \frac{\rho(x_0)}{\|\mathbf{s}_0'\|}\theta = \frac{\theta }{\kappa(x_0)\|\mathbf{s}_0'\|}
\end{equation}
We need to express $\theta$ in terms of the noise and $\rho$ (or $\kappa$). We use Fig.~\ref{fig:PPM_curvature} for this purpose. We have a right-legged triangle, i.e.  $\theta + 2\phi = \pi \Rightarrow \phi = (\pi-\theta)/2$. Further, $b=\rho \sin(\theta)$ is the right normal of the triangle. Therefore
\begin{equation}\label{e:PPM_angle_1a}
\sin(\phi) = \frac{b}{\|\mathbf{n}_{||}\|} \ \Rightarrow \ \|\mathbf{n}_{||}\| = \frac{b}{\sin(\phi)}=\frac{\rho \sin(\theta)}{\sin(\phi)}. 
\end{equation}
Further,
\begin{equation}\label{e:PPM_angle_1b}
\sin(\phi) = \sin\bigg(\frac{\pi}{2}-\frac{\theta}{2}\bigg)=\cos\bigg(\frac{\theta}{2}\bigg).
\end{equation}
Since $\sin 2y = 2\sin y \cos y$, with $y=\theta/2$, then $\sin(\theta) = 2\sin({\theta}/{2})\cos({\theta}/{2})$, implying that
\begin{equation}\label{e:PPM_Noise_1}
\|\mathbf{n}_{||}\| = \frac{2\rho\sin(\theta/2) \cos(\theta/2)}{\cos(\theta/2)}=2\rho\sin\bigg(\frac{\theta}{2}\bigg).
\end{equation}
Therefore
\begin{equation}\label{e:Theta_afo_noise}
\theta = 2\sin^{-1}\bigg(\frac{\|\mathbf{n}_{||}\|}{2\rho(x_0)}\bigg).
\end{equation}
By the Taylor expansion~\cite[p. 117]{Rottmann91}
\begin{equation}\label{e:InvSine_Expansion_1}
\sin^{-1}(x)= x + \frac{1}{2\cdot 3} x^3+\frac{1\cdot 3}{2\cdot 4 \cdot 5} x^5+\cdots
\end{equation}
we get $\theta$ up to 2nd order as
\begin{equation}\label{e:Theta_afo_noise_approx}
\theta \approx \frac{\|\mathbf{n}_{||}\|}{\rho(x_0)}\bigg(1+\frac{\|\mathbf{n}_{||}\|^2}{24\rho^2(x_0)}\bigg).
\end{equation}
One can then compute the error (in the absence of anomalies) up to second order from~(\ref{e:ML_est_1_N_CircEst}) and~(\ref{e:Theta_afo_noise_approx})
\begin{equation}\label{e:WeakNoiseDist_2ndOrder_pre}
\begin{split}
{\varepsilon}_{wn}^2 &= E\big\{(x-\hat{x}_{ML})^2 | x=x_0\big\} =E\bigg\{\bigg(\frac{\rho(x_0)}{\|\mathbf{s}_0'\|}\theta \bigg)\bigg\}\\
 &= \frac{\rho^2(x_0)}{\|\mathbf{s}_0'\|^2} E\bigg\{\bigg(\frac{\|\mathbf{n}_{||}\|}{\rho(x_0)}\bigg(1+\frac{\|\mathbf{n}_{||}\|^2}{24\rho^2(x_0)}\bigg)\bigg)^2\bigg\}\\
 &=\frac{1}{\|\mathbf{s}_0'\|^2} E\bigg\{\|\mathbf{n}_{||}\|^2  + \frac{2\|\mathbf{n}_{||}\|^4}{24\rho^2(x_0)}+\frac{\|\mathbf{n}_{||}\|^6}{24^2\rho^4(x_0)}\bigg\}.
\end{split}
\end{equation}
From~(\ref{e:Moments_Gaussian}), we then get
\begin{equation}\label{e:WeakNoiseDist_2ndOrder_1}
\begin{split}
{\varepsilon}_{wn}^2 &= \frac{\sigma_n^2}{\|\mathbf{s}_0'\|^2}\bigg(1+\frac{\sigma_n^2}{4\rho^2(x_0)}+\frac{5\sigma_n^4}{48\rho^4(x_0)}\bigg)\\
 &=\frac{\sigma_n^2}{\|\mathbf{s}_0'\|^2}\bigg(1+\frac{1}{4}\sigma_n^2 \kappa^2(x_0)+\frac{5}{48}\sigma_n^4 \kappa^4(x_0)\bigg)
\end{split}
\end{equation}
Note that the relation $\kappa(x_0)=\|\ddot{\mathbf{s}}(x_0)\|$ is valid under arc length parametrization, that is whenever $\|\mathbf{s}'(x_0)\|=\|\dot{\mathbf{s}}(x_0)\|=1$, $\forall x_0$. In general, the curvature is given by~(\ref{e:curvature_curve_gen_coord}), leading to a more complicated expression. Since we most often consider arc length parametrization with an amplification $\alpha$, that is $\|{\mathbf{S}}'(x_0)\|=\alpha\|\dot{\mathbf{S}}(x_0)\|=\alpha$, $\forall x_0$, then~(\ref{e:curvature_curve_gen_coord}) reduces to $\kappa(x_0)=\|\mathbf{S}_0''(x_0)\|/\|\mathbf{S}_0'\|^2$ (since $\mathbf{S}_0' \perp \mathbf{S}_0''$ still). In that case we can express the error in terms of the signal curve's derivatives as
\begin{equation}\label{e:WeakNoiseDist_2ndOrder_2}
{\varepsilon}_{wn}^2 = \frac{\sigma_n^2}{\|\mathbf{S}_0'\|^2}\bigg(1+\frac{1}{4}\sigma_n^2 \frac{\|\mathbf{S}_0''\|^2}{\|\mathbf{S}_0'\|^4}+\frac{5}{48}\sigma_n^4 \frac{\|\mathbf{S}_0''\|^4}{\|\mathbf{S}_0'\|^8}\bigg)
\end{equation}
We will argue in Section~\ref{ssec:CanalSurf_curves} that it is convenient to let $\sigma_n^2 << \rho^2(x_0)$ at high SNR.
Then at high SNR as $\sigma_n^2 << 1$, we can make the approximation
\begin{equation}\label{e:WeakNoiseDist_2ndOrder_Final}
\begin{split}
{\varepsilon}_{wn}^2 &\approx \frac{\sigma_n^2}{\|\mathbf{S}_0'\|^2}\bigg(1+\frac{1}{4}\sigma_n^2 \kappa^2(x_0)\bigg)\\ &= \frac{\sigma_n^2}{\|\mathbf{S}_0'\|^2}\bigg(1+\frac{1}{4}\sigma_n^2 \frac{\|\mathbf{S}_0''\|^2}{\|\mathbf{S}_0'\|^4}\bigg)
\end{split}
\end{equation}
Note that at high SNR we can stretch the curve significantly, that is  $\|\mathbf{S}_0'\|=\alpha >> 1$, and we see from~(\ref{e:WeakNoiseDist_2ndOrder_Final}) that the first term clearly dominates, making the accuracy of the 1st order term increasingly dominant with SNR. As was seen in Section~\ref{ssec:SK_curves_M_1}, higher order terms will contribute even less,  and therefore the circle approximation above will do.

Further note the similarity to the $M$:$1$ case given in~(\ref{e:ChDist_Final_2ndOrder}), the distortion is scaled by the curvature in nearly the same manner. Also, distortion in the expansion case is proportional to $1/\|\mathbf{S}_0'\|^2$  whereas distortion in the reduction case is proportional to $\|\mathbf{S}_0'\|^2$,  as expected for the weak noise regime.

\begin{figure}[h]
    \begin{center}
           \includegraphics[width=1.0\columnwidth]{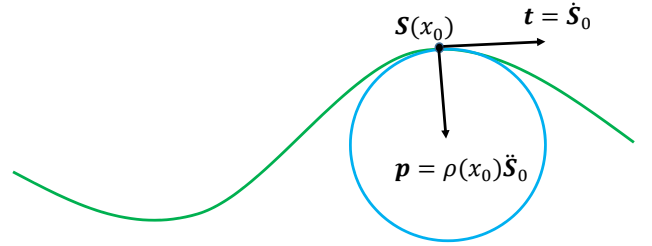}
    \end{center}
    \caption{How a general $1$:$2$ mapping locally can be represented by the circle approximation (osculating circle).}\label{fig:CircApprox_Curve}
\end{figure}

It is straight forward to see that the circle approximation above will be valid locally for any signal curve $\mathbf{S}$ using the concept of osculating circle. We illustrate with Fig.~\ref{fig:CircApprox_Curve}.
For a general curve in polar coordinates we have by the product rule
\begin{equation}\label{e:diff_polar_coord_gen_curve}
\begin{split}
&\frac{\mbox{d}\mathbf{S}}{\mbox{d}x}=\bigg[R'(x)\cos(\theta(x))+\cdots\\
& +R(x)\sin(\theta(x)),R'(x)\sin(\theta(x))-R(x)\cos(\theta(x))\bigg].
\end{split}
\end{equation}
When $\rho(x)=\rho=$constant we get the result derived above for all $x$. Looking at the curve locally, $R'(x)$ is small if the curvature of $\mathbf{S}$ changes slowly with $x$. The more slowly $\kappa(x)$ changes with $x$, the more accurate the circle approximation becomes. As we shall see in Section~\ref{ssec:CanalSurf_curves}, it is not convenient to choose a curve with a rapidly changing curvature $\kappa$. This implies that Shannon's suggestion for a $1$:$2$ mapping in~\cite{shann49}
 may not be the most convenient due to the "half circles" connecting each line segment.

To be a bit more rigorous: For PPM we had $\mathbf{S}(x_0)=S_r \mathbf{e}_r$. In general we have $\mathbf{S}(x_0)=S_r \mathbf{e}_r + S_\theta \mathbf{e}_\theta $. For the case $\|\mathbf{S}_0'\|=\alpha$, $\forall x$, we have seen that $\mathbf{S}_0'\cdot \mathbf{S}_0'' = 0$. Then locally around $\mathbf{S}(x_0)$ we again have $\mathbf{S}_0'=\alpha \mathbf{e}_\theta$,
and since $\mathbf{S}_0'\perp \mathbf{S}_0''$ we must have that $\mathbf{S}_0''=\kappa\mathbf{e}_r$. From the formula of Frenet (FoF) we then have
\begin{equation}
\ddot{\mathbf{S}}_0 = \dot{\mathbf{t}}=\kappa(x_0) \mathbf{p}.
\end{equation}
That is, the circle approximation is locally describing a general curve $\mathbf{S}(x)$.

\subsubsection{Canal Surfaces and maximal curvature}\label{ssec:CanalSurf_curves}
This section will justify the claim that large curvature or rapidly changing curvature is not convenient when constructing $1$:$N$ mappings, as this will lead to large errors. For this we will consider a special surface named \emph{canal surface} described in~\cite[pp. 266-267]{Kreyszig_DiffGeom91}.

A Canal Surface is the \emph{envelope}, $E$, to the family, $F$, of congruent spheres. The envelope is the set of all \emph{characteristics} to $F$, which is the point set represented by~\cite[p. 263]{Kreyszig_DiffGeom91}
\begin{equation}\label{e:CaSu_charachteristic}
S_c(z_i,\ell)=0, \ \ \frac{\partial S_c(z_i,\ell)}{\partial \ell}=0,
\end{equation}
where $S_c=0$  defines a surface in $\mathbb{R}^3$ (or hypersurface in $\mathbb{R}^N$). The characteristic is therefore a curve in $\mathbb{R}^3$ (or a hypersurface of dimension $N-2$ in $\mathbb{R}^N$). The \emph{characteristic points} of  the canal surface is the intersection of the characteristic (which here will be points of intersection of circles) given by~\cite[p. 266]{Kreyszig_DiffGeom91}
\begin{equation}\label{e:CaSu_CharPoints}
S_c(z_i,\ell)=0, \ \ \frac{\partial S_c(z_i,\ell)}{\partial \ell}=0, \ \ \frac{\partial S_c^2(z_i,\ell)}{\partial \ell^2}=0.
\end{equation}

The family $F$ of spheres with constant radius $r$ with center on a curve $C: \mathbf{y}(\ell)$, can be represented as
\begin{equation}\label{e:CaSu_Def}
S_c(z_i,\ell)=(\mathbf{z}-\mathbf{y}(\ell))\cdot(\mathbf{z}-\mathbf{y}(\ell))-r^2=0.
\end{equation}

To recapitulate in the terminology of S-K mappings one can set $\mathbf{y}(\ell)=\mathbf{S}(\ell)$ and $S_c(z_i;\ell)$, where $\mathbf{z}$ is the \emph{channel coordinates} and $\ell$ is the arc length of $\mathbf{S}$. Further, let the curvature of $\mathbf{S}$ be $\kappa_s$ with corresponding radius of curvature $\rho_s = 1/\kappa_s$. For $1$:$N$ S-K mappings we seek a canal surface that does not intersect itself in order to avoid anomalous distortion. This is the case if and only if the following conditions are satisfied

i) The minimum distance between two folds of $\mathbf{S}$ is $\geq 2 r$

ii) $\rho_s > r$, as shown in  Proposition~\ref{prop:canal_surf_cond} below.\\

\begin{proposition}\label{prop:canal_surf_cond}
For $\rho_s=1/\kappa_s$, radius of curvature for $\mathbf{S}(x)$, and $r$, the radius of the hypersphere $\mathbb{S}^{N-2}$. Then the corresponding canal surface will not intersect itself, that is the canal surface will have no characteristic points  $\Longleftrightarrow$ $\rho_s > r$. 
\end{proposition}

\begin{pf} (See~\cite[pp.266-267]{Kreyszig_DiffGeom91})
The spheres $\mathbb{S}^{N-2}$ in $F$ can be represented as
\begin{equation}\label{e:congruent_sphere_def}
S_c(z_i,\ell)=(\mathbf{z}-\mathbf{S}(\ell))\cdot(\mathbf{z}-\mathbf{S}(\ell))-r^2 = 0, \ \ i=1,\cdots,N.
\end{equation}
Further,
\begin{equation}\label{e:Diff_congruent_sphere}
\frac{\partial S_c}{\partial \ell} = - 2 \dot{\mathbf{S}}\cdot(\mathbf{z}-\mathbf{S})=0, \ \dot{\mathbf{S}}=\frac{\partial  \mathbf{S}}{\partial \ell}.
\end{equation}
this implies that $(\mathbf{z}-\mathbf{S})\perp \dot{\mathbf{s}}$ and so,
\begin{equation}\label{e:Diff2_congruent_sphere}
\frac{\partial^2 S_c}{\partial \ell^2} = - 2 \ddot{\mathbf{S}}\cdot(\mathbf{z}-\mathbf{S})+2 = -2\kappa_s \mathbf{p}\cdot(\mathbf{z}-\mathbf{S})+2 = 0,
\end{equation}
where the last equality is due to the formula of Frenet~(\ref{e:FoF}) with $\rho_s=1/\kappa_s$ we get
\begin{equation}\label{e:NoCharPointCond}
\mathbf{p}\cdot(\mathbf{S}-\mathbf{z})-\rho_s = 0.
\end{equation}
$\Rightarrow:$ The condition follows directly from~(\ref{e:Diff2_congruent_sphere}). $\Leftarrow:$ Since $\|\mathbf{p}\cdot (\mathbf{S}-\mathbf{z})\| = \|\mathbf{S}-\mathbf{z}\|=r$, one can see that if $\rho_s > r\ \forall x$, then the last equation in~(\ref{e:Diff2_congruent_sphere}) does not have any real solution.  I.e., no real characteristic  points exists. \hspace{2.8cm}$\square$\\
\end{pf}

\begin{figure}[h]
    \begin{center}
    \includegraphics[width=0.9\columnwidth]{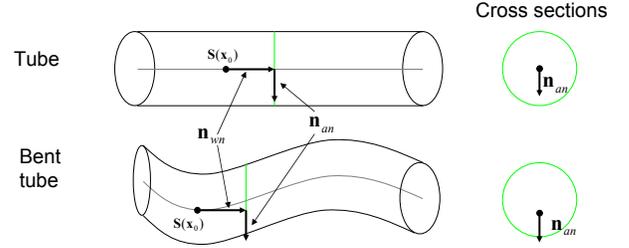}
    \end{center}
\caption{The tube $\mathbb{L}\times \mathbb{S}^{1}$ together with a tube that bends, that is a Canal Surface. The signal curve is at the tubes center.} \label{fig:tube}
\end{figure}

Note that the above concept is more or less the same as a $1$:$N$ signal curve in Gaussian noise, where the noise vectors will approximately lie within congruent spheres along $\mathbf{S}(x)$. We will elaborate a bit further for a $1$:$3$ mapping: Fig.~\ref{fig:tube} shows a canal surface surrounding a signal curve in $\mathbb{R}^3$.
The radius of the tube is linked to Gaussian noise. A correspondence to the the noise in the above discussion results by setting $\mathbf{n}_{wn}=\mathbf{n}_{||}$ and $\mathbf{n}_{wn}=\mathbf{n}_{\perp}$. We can see from this than any bending of the tube increases the probability for anomalous errors, implying that a straight line has the lowest possible probability. However, it is well known that at high SNR this will produce a sub-optimal mapping due to increased weak noise error.  By linking to Proposition~\ref{prop:canal_surf_cond} and the above discussion, it is easy to see that bending poses no problem if the radius of curvature of $\mathbf{S}$ is small enough: By setting
\begin{equation}\label{e:canal_surface_noise_cond_1_N}
 \rho_s > r = \rho_n \geq \sqrt{\frac{N-1}{N} b_n\sigma_n^2},
 \end{equation}
then no characteristic points will exist, and the tube will not intersect itself (if condition i) above is also satisfied).

Of course with noise, $\rho_n$ usually has no limitation, and the radius of the noise sphere is always linked to a certain probability. However, by setting $b_n>4$, then the probability that $\rho_n>\rho_s$ will be very small. However, when proving optimality in the limit of infinite dimensionality, then $\rho_n \rightarrow \sigma_n$ with unit probability, meaning that the canal surface concept above is crucial in proving optimality in the limiting case. The concept of canal surfaces extended to general signal hyper surfaces will be crucial when setting down criteria for the construction of well performing mappings  later.

To summarize the discussion on curves we see that as the  SNR gets high, the 1st order approximation (what we later name \emph{weak noise regime}) takes off with $\sigma_n^2$, whereas the largest higher order term takes off with $\sigma_n^4$, and therefore  becomes  negligibly small as $\sigma_n\rightarrow 0$ (at high SNR) if the curvature is not large,  and the 1st order approximation dominates. From the discussion of maximum likelihood estimation and canal surfaces above, one should keep the curvature as small as possible, never larger than $1/\rho_n$  (in fact $\rho_s >> \rho_n$ in a practical situation). This will keep the curvature $\kappa_s$  in eqns~(\ref{e:ChDist_Final_2ndOrder}) and~(\ref{e:WeakNoiseDist_2ndOrder_Final}) at a small enough level for the 2nd order term to vanish much faster than the 1st order term.

\subsection{S-K mappings using parametric surfaces}\label{ssec:SK_Surf}
Many new concepts concerning differential geometry of surfaces will be applied here. All relevant concepts are described in Section~\ref{ssec:DiffGeom_ParSurfaces}. A more thorough treatment is provided in~\cite{Kreyszig_DiffGeom91}. 

\subsubsection{$M$:$N$ dimension expanding mappings}\label{ssec:SK_ExpSurf}
We start off with surfaces in $\mathbb{R}^3$, that is, $2$:$3$ mappings.
According to \emph{theorem of Meusnier} and \emph{theorem of Euler} the curvature can in all cases be described by a combination of the \emph{principal curvatures} $\kappa_1$ and $\kappa_2$ (see Theorems~\ref{th:Meusnier} and~\ref{th:Euler} in Section~\ref{ssec:Fundamental_Forms}). These correspond to the maximal and minimal curvatures at any point $P\in \mathcal{S}$ respectively. Since the principal curvatures are extremal curvatures, they can be used to analyze higher order behavior of S-K mappings beyond curves: As it is the maximal curvature, $\kappa_1$, that bounds the probability of anomalous errors, it has to be limited to a certain maximum value. To analyze curvature of S-K mappings one has to consider both FFF and SFF introduced in Section~\ref{ssec:Fundamental_Forms}.

To make the math simpler one can choose \emph{lines of curvature} (LoC) as coordinates on $\mathcal{S}$. That is, the coordinates are aligned with the direction of $\kappa_1$ and $\kappa_2$ at any point (i.e., normal sections along principal directions). We first develop a expression for ${\varepsilon}_{wn}^2$ when the coordinate curves are chosen as LoC, since this is the most direct and simple generalization of the $1$:$N$ case.  We will then discuss the impact of  performance in choosing LoC as coordinates at the end of this section.

Assume that $x_1$ and $x_2$ are along the LoC. Then, according to Theorem~\ref{th:LoC_Coordinates} in Section~\ref{ssec:DiffGeom_ParSurfaces}, $g_{12}=b_{12}=0$ has to be satisfied.  Then the curvature is given solely in terms of $g_{\alpha\alpha}$ and $b_{\alpha\alpha}$ as
\begin{equation}\label{e:PrincCurvat_SK_Exp}
\kappa_i(\mathbf{x}_0)=\frac{b_{ii}(\mathbf{x}_0)}{g_{ii}(\mathbf{x}_0)}, \ \text{where} \ g_{ii}=\bigg\|\frac{\partial \mathbf{S}}{\partial x_i}\bigg\|^2,
\end{equation}
and where (see Eq.~(\ref{e:compute_b_ab}))
\begin{equation}\label{e:SFF_SK_Exp}
b_{ii}=\frac{1}{\sqrt{g}}\bigg|\frac{\partial\mathbf{S}}{\partial x_1}\ \frac{\partial\mathbf{S}}{\partial x_2}\ \frac{\partial^2\mathbf{S}}{\partial x_i^2}\bigg|,
\end{equation}
with $g=\det(G)=g_{11}g_{22}-g_{12}^2$.  These expressions are a direct consequence of~(\ref{e:LoC_Coord_Curvatures}) and~(\ref{e:compute_b_ab}).

For $2$:$N$ mappings, as $x_1$, $x_2$, $n_1$ and $n_2$ are independent and i.i.d., and we have two orthogonal coordinate curves, then \begin{equation}\label{e:WeakNoiseDist_2ndOrder_2_N}
\begin{split}
{\varepsilon}_{wn}^2(\mathbf{x}_0) &\approx \frac{\sigma_n^2}{2} \sum_{i=1}^2 \bigg\{\frac{1}{g_{ii}(\mathbf{x}_0)}\bigg(1+\frac{\sigma_n^2}{4} \kappa_i^2(\mathbf{x}_0)\bigg)\bigg\}\\ &= \frac{\sigma_n^2}{2}\sum_{i=1}^2 \bigg\{\frac{1}{g_{ii}(\mathbf{x}_0)}\bigg(1+\frac{\sigma_n^2}{4} \frac{b_{ii}^2(\mathbf{x}_0)}{g_{ii}^2(\mathbf{x}_0)}\bigg)\bigg\},
\end{split}
\end{equation}
which is a straight forward generalization of~(\ref{e:WeakNoiseDist_2ndOrder_Final}).

For $M$:$N$ mappings, the generalization follows directly as we have $M$ (orthogonal) LoC as coordinate curves:
\begin{equation}\label{e:WeakNoiseDist_2ndOrder_M_N}
\begin{split}
{\varepsilon}_{wn}^2(\mathbf{x}_0) &\approx \frac{\sigma_n^2}{M} \sum_{i=1}^M \bigg\{\frac{1}{g_{ii}(\mathbf{x}_0)}\bigg(1+\frac{\sigma_n^2}{4} \kappa_i^2(\mathbf{x}_0)\bigg)\bigg\}\\ &= \frac{\sigma_n^2}{M}\sum_{i=1}^M \bigg\{\frac{1}{g_{ii}(\mathbf{x}_0)}\bigg(1+\frac{\sigma_n^2}{4} \frac{b_{ii}^2(\mathbf{x}_0)}{g_{ii}^2(\mathbf{x}_0)}\bigg)\bigg\}.
\end{split}
\end{equation}
with $g_{ii}$  and $b_{ii}$ the components of FFF and SFF respectively (as previously defined).

\subsubsection{$M$:$N$ dimension reducing mappings}\label{ssec:SK_RedSurf}
A direct generalization of Section~\ref{ssec:SK_curves_M_1} is to use the 2nd order Taylor expansion for vector-valued functions $\mathbf{S}$:
\begin{equation}\label{e:Taylor_VecFunc_SKred}
\mathbf{S}(\mathbf{z}_0 + \mathbf{n})\approx \mathbf{S}(\mathbf{z}_0)+ J_\mathbf{S}(\mathbf{z}_0)\mathbf{n} + \frac{1}{2} \mathcal{D}_\mathbf{n}^2\mathbf{S}(\mathbf{z}_0)
\end{equation}
where
\begin{equation}\label{e:2ndOrder_VecTaylor_Def}
\mathcal{D}_\mathbf{n}^2\mathbf{S}(\mathbf{z}_0) = \big[\mathbf{n}^T D^2\mathbf{S}_1(\mathbf{z}_0)\mathbf{n}\ \cdots \  \mathbf{n}^T D^2\mathbf{S}_M(\mathbf{z}_0)\mathbf{n} \big],
\end{equation}
with $D^2\mathbf{S}_i$ the \emph{Hessian matrix}~\cite[p. 582]{nocedal/wright} for the function $\mathbf{S}_i$, that is
\begin{equation}\label{e:Hessian_Def}
D^2\mathbf{S}_i=\frac{\partial^2 \mathbf{S}_i(\mathbf{z}_0)}{\partial z_j \partial z_k}, \ j,k=1,\cdots,N.
\end{equation}
Ideally, one can then calculate the channel error up to 2nd order as
\begin{equation}\label{e:Channel_Dist_2ndOrder_Exact}
\begin{split}
\varepsilon^2_{ch}(\mathbf{z}_0)&=\frac{1}{M} E\bigg\{\big(J_\mathbf{S}(\mathbf{z}_0)\mathbf{n} + \frac{1}{2} \mathcal{D}_\mathbf{n}^2\mathbf{S}(\mathbf{z}_0)\big)^T\\
&\big(J_\mathbf{S}(\mathbf{z}_0)\mathbf{n} + \frac{1}{2} \mathcal{D}_\mathbf{n}^2\mathbf{S}(\mathbf{z}_0)\big)  \bigg\},
\end{split}
\end{equation}
directly. However, this leads to a very complicated and lengthy expression making it hard, if at all possible, to draw clear conclusions. Therefore it is more convenient to consider LoC. By choosing LoC as coordinates on a $M$:$2$ mapping $\mathcal{S}$, then, as for the $M<N$ case, we get the the most direct and simple generalization of the $M$:$1$ case.

Assume that $z_1$ and $z_2$ are along the LoC. Then, according to Theorem~\ref{th:LoC_Coordinates}, $g_{12}=b_{12}=0$, has to be satisfied.  Therefore the curvature is given solely in terms of $g_{\alpha\alpha}$ and $b_{\alpha\alpha}$
\begin{equation}\label{e:PrincCurvat_SK_Red}
\kappa_i(\mathbf{z}_0)=\frac{b_{ii}(\mathbf{z}_0)}{g_{ii}(\mathbf{z}_0)}, \ \text{where} \ g_{ii}=\bigg\|\frac{\partial \mathbf{S}}{\partial z_i}\bigg\|^2,
\end{equation}
and where
\begin{equation}\label{e:SFF_SK_Red}
b_{ii}=\frac{1}{\sqrt{g}}\bigg|\frac{\partial\mathbf{S}}{\partial z_1}\ \frac{\partial\mathbf{S}}{\partial z_2}\ \frac{\partial^2\mathbf{S}}{\partial z_i^2}\bigg|,
\end{equation}
again with $g=\det(G)=g_{11}g_{22}-g_{12}^2$.
For $M$:$2$ mappings, since $z_1$, $z_2$, $n_1$ and $n_2$ are independent and we have two orthogonal coordinate curves we get
\begin{equation}\label{e:ChannelDist_2ndOrder_M_2}
{\varepsilon}_{ch}^2(\mathbf{z}_0) \approx \frac{\sigma_n^2}{M} \sum_{i=1}^2 \bigg({g_{ii}(\mathbf{z}_0)}+\frac{3\sigma_n^2}{4} \kappa_i^2(\mathbf{z}_0)\bigg).
\end{equation}
For $M$:$N$ mappings, the generalization follows directly
\begin{equation}\label{e:ChannelDist_2ndOrder_M_N}
{\varepsilon}_{ch}^2(\mathbf{z}_0) \approx \frac{\sigma_n^2}{M} \sum_{i=1}^N \bigg({g_{ii}(\mathbf{z}_0)}+\frac{3\sigma_n^2}{4} \kappa_i^2(\mathbf{z}_0)\bigg),
\end{equation}
with $g_{ii}$  and $b_{ii}$ the components of FFF and SFF respectively (as previously defined).
%

\subsubsection{Extended Canal Surfaces and maximal curvature}\label{ssec:CanalSurf_surfaces}
With the definition of principal curvatures it is straight forward to extend the concept of a non-intersecting canal surfaces to hyper surfaces, that is \emph{canal hyper surfaces}. We provide conditions for $N > M$. The $M < N$ case is similar.\\

\begin{proposition}\label{prop:canal_surf_cond_surfaces}
Consider $M$:$N$, $N > M$ S-K mappings. For $\rho_{\text{min}}=1/\kappa_{\text{max}}$, with $\kappa_{\text{max}}$ the maximal principal curvature of $\mathcal{S}$, and $r$, the radius of the hyper-sphere $\mathbb{S}^{N-M-1}$, the corresponding canal hypersurface (that is, the envelope of $\mathbb{S}^{N-M-1}$), will not intersect itself at any point. That is, the canal surface will have no characteristic points  $\Longleftrightarrow$ $\rho_{\text{min}} > r$ for all points of $\mathcal{S}$. 
\end{proposition}

\begin{pf}
This result follows directly from the proof of Proposition~\ref{prop:canal_surf_cond} by letting a curve $\mathcal{C}$ be the LoC with maximal principal curvature for all points of $\mathcal{S}$. That is $\mathcal{C}$ is always in the direction of the maximal curvature on $\mathcal{C}$.
\hspace{0.5cm}$\square$\\
\end{pf}


\subsubsection{Choice of coordinates}\label{ssec:coordinates}
The problem with LoC coordinates is that one source is always getting a less exact ML-estimate than the other(s) due to the maximal curvature it has at every point. This is poses no significant problem (especially at high SNR) as long as a surface with relatively low curvature is chosen. The alternative is to look at general normal sections with approximately the same curvature (on average) for each source over $\mathcal{S}$. Generally, we have from~(\ref{e:normal_curvature})
\begin{equation}\label{e:normal_curvature_SK}
\kappa_n =\frac{b_{\alpha\beta} \mbox{d}{u^\alpha} \mbox{d}{u^\beta}}{g_{\alpha\beta} \mbox{d}{u^\alpha} \mbox{d}{u^\beta}},
\end{equation}
with $b_{\alpha\beta}$ and $g_{\alpha\beta}$ as defined before in Section~\ref{ssec:DiffGeom_ParSurfaces}. $\kappa_1$ and $\kappa_2$ are then found as the solution to the differential equation
\begin{equation}\label{e:PrincipalCurvat_SK_2}
\kappa_n^2 - b_{\alpha\beta} g^{\alpha\beta}\kappa_n + b/g = 0,
\end{equation}
or by the differential equations~(\ref{e:Principal_curvature_DiffEq2}) or~(\ref{e:Principal_curvature_DiffEq3}) in Section~\ref{ssec:Fundamental_Forms}. Then, from the Theorem of Euler (Theorem~\ref{th:Euler} in Section~\ref{ssec:DiffGeom_ParSurfaces}),
\begin{equation}\label{e:NormSectGen_SK_exp}
\kappa_n=\kappa_1\cos^2\alpha+\kappa_2\sin^2\alpha.
\end{equation}

An important special case of coordinates is \emph{Geodesic Coordinates} (GeCo). A geodesic is the curve with shortest possible length between any two points on a surface $\mathcal{S}$. They are therefore convenient as coordinate curves for dimension reducing S-K mappings as they lead to the smallest possible $g_{ii}$ among all curves on $\mathcal{S}$ and therefore minimize \emph{weak channel distortion}, $\bar{\varepsilon}_{ch}^2$,  as defined in~(\ref{e:mseort_mean_DimRed}) below. 

The geodesic is found by the minimization of an energy functional, which is a solution to the \emph{Euler-Lagrange Equations} (ELE)~\cite[p. 161]{Kreyszig_DiffGeom91}. With coordinate curves denoted $u^\tau$, from~\cite[p. 162]{Kreyszig_DiffGeom91}, assuming arc length parametrization
\begin{equation}\label{e:ParCurve_Geodesic_CL_Par}
\mathbf{S}(u^1(\ell), u^2(\ell)), \ \ \ell=\ell(x_i),
\end{equation}
a Geodesic curve on $\mathcal{S}$ is given by the differential equation
\begin{equation}\label{e:Geodecic_Eq_ELE}
\ddot{u}^\tau + \Gamma_{\nu \sigma}^\tau \dot{u}^\nu \dot{u}^\sigma = 0,
\end{equation}
where $\Gamma_{\nu \sigma}^\tau$ is \emph{Christoffel symbol of second kind} (CS-2) as defined in~(\ref{e:Christoffel2}) in Section~\ref{ssec:Fomula_Wein_Gauss_Geodes}. Further, one can show that~(\ref{e:Geodecic_Eq_ELE}) is equivalent to~\cite[pp. 160-162]{Kreyszig_DiffGeom91}
\begin{equation}\label{e:Geodesic_ELE_cmpkt_GeCu}
\kappa_g = \big|\dot{\mathbf{S}}\ \ddot{\mathbf{S}}\ {\mathbf{n}}\big| = 0,
\end{equation}
where $\mathbf{n}=\mathbf{S}_1\times \mathbf{S}_2$ is the normal vector to $\mathcal{S}$ at $P$. $\kappa_g$  is the \emph{geodesic curvature} of the relevant curve, which corresponds to the curvature of the projection of the relevant curve on $\mathcal{S}$ onto the tangent plane of $\mathcal{S}$ at $P$ (see~\cite[pp. 154-155]{Kreyszig_DiffGeom91} for further explanation).  That is, another equivalent definition of a geodesic is a curve with zero geodesic curvature. It is possible to show that this equation can be formulated solely w.r.t. the metric tensor of the surface as the following theorem states~\cite[p. 156]{Kreyszig_DiffGeom91}:\\

\begin{theorem}\label{th:GeCo_FFF}
The geodesic curvature, $\kappa_g$, of a curve $\mathcal{C}$ on a surface $\mathcal{S}$ depends on the FFF of $\mathcal{S}$ only (as well as the curve itself).
\end{theorem}

\emph{Proof:} See~\cite[p.156]{Kreyszig_DiffGeom91}.\hspace{5.0cm}$\square$\\

Considering LoC coordinates, we have the following two important Theorems:\\

\begin{theorem}\label{th:GeCo_1}
The coordinate curves $u^1=$ constant and $u^2=$ constant on a portion of a surface $\mathcal{S}: \mathbf{S}(u^1,u^2)\in C^r$, $r \geq 2$ are geodesics $\Leftrightarrow$
\begin{equation}\label{e:Christoffel_GeCo}
\Gamma_{22}^1=0 \ \text{and} \ \Gamma_{11}^2=0,
\end{equation}
respectively.
\end{theorem}

\emph{Proof:} See~\cite[p.156-158]{Kreyszig_DiffGeom91}.\hspace{4.5cm}$\square$\\

\begin{theorem}\label{th:GeCo_2}
If a geodesic, $C_G$, on $\mathcal{S}$ is a LoC on $\mathcal{S}$, then $C_G$ is a plane curve. Further, any plane geodesic on $\mathcal{S}$ is a LoC on $\mathcal{S}$.
\end{theorem}

\emph{Proof:} See~\cite[p.158-159]{Kreyszig_DiffGeom91}.\hspace{4.5cm}$\square$\\


Ideally, with geodesic coordinates and arc length parametrization, one could obtain a diagonal metric tensor with independent (and constant) $g_{ii}$, that is
\begin{equation}\label{e:diagonal_indep_MT}
G=
\begin{bmatrix}
g_{11}&g_{12}\\
g_{21}&g_{22}
\end{bmatrix}
=
\begin{bmatrix}
c_1& 0\\
0& c_2
\end{bmatrix},
\end{equation}
with $c_1, c_2$ positive constants. However, this is not necessarily possible to obtain over the whole of $\mathcal{S}$, only over portions of it: Two common ways of obtaining GeCo is through \emph{geodesic parallel coordinates} (GPC) or \emph{geodesic polar coordinates} (GPoC)~\cite[pp. 162-168]{Kreyszig_DiffGeom91}.

As we shall see in Section~\ref{sec:Ex_SK_surf_32_and_23}, there are subsets of surfaces named \emph{developable surfaces} that can be mapped \emph{isometrically} to the Euclidean plane, as well as  subsets of surfaces that can be mapped \emph{conformally} to the Euclidean plane, where~(\ref{e:diagonal_indep_MT}) is obtained all over $\mathcal{S}$.

\subsection{Definition of weak noise regime for S-K mappings}\label{ssec:SK_Precise}
We are now able to make a more rigourous definition of the \emph{weak noise regime} connected to the distortion analysis of S-K mapping:\\

\subsubsection{Weak noise distortion}\label{ssec:WeakNoiseDist_new}
We begin with the definition of \emph{weak noise distortion} for dimension expanding S-K mappings:  Let $\mathbf{S}_{lin}(\mathbf{x})$ denote the linear approximation of $\mathbf{S}(\mathbf{x})$ at $\mathbf{x}_0$
\begin{equation}\label{e:lin_apr_exp}
\mathbf{S}_{lin}(\mathbf{x})= \mathbf{S}(\mathbf{x}_0)+J(\mathbf{x}_0)(\mathbf{x}-\mathbf{x}_0),
\end{equation}
where $J(\mathbf{x}_0)$ is the Jacobian matrix (see Appendix~\ref{sec:app_one_mt}) of $\mathbf{S}$ evaluated at $\mathbf{x}_0$. We have the following proposition providing the exact distortion under linear approximation:\\

\begin{proposition}\label{th:weak_noise}\emph{Minimum weak noise distortion}\\
For any continuous i.i.d. source $\mathbf{x}\in \mathbb{R}^M$ with unimodal pdf $f_{\mathbf{x}}(\mathbf{x})$ communicated on an i.i.d. Gaussian channel of dimension $N$ using a continuous dimension expanding S-K mapping $\mathbf{S}$ where $S_i\in
C^r(\mathbb{R}^M), \ r\geq 1, \  i=1,..,N$,  the minimum distortion under the linear approximation in~(\ref{e:lin_apr_exp}) is given by
\begin{equation}\label{e:mseort_mean_exp}
\bar{\varepsilon}_{wn}^2=\frac{\sigma_n^2}{M}\iint
\cdots
\int_{\mathcal{D}}\sum_{i=1}^M\frac{1}{g_{ii}(\mathbf{x})}f_{\mathbf{x}}(\mathbf{x})\mbox{d}\mathbf{x},
\end{equation}
obtained when the \emph{metric tensor} $G$ of $\mathbf{S}$ (Appendix~\ref{sec:app_one_mt}) is diagonal.
$g_{ii}=\|\partial\mathbf{S}(\mathbf{x})/\partial x_i\|^2$ denote the diagonal components of $G$, i.e. the squared norm of the tangent vector along  $\mathbf{S}(x_i)$.
\end{proposition}

\emph{Proof:} See~\cite{Floor_Ramstad09}.\hspace{6.0cm}$\square$\\

The name \emph{weak noise distortion} is due to Definition~\ref{def:weak_noise_exp} given below.

Consider now the weak noise error given by the 2nd order approximation to $\mathbf{S}$ in~(\ref{e:WeakNoiseDist_2ndOrder_M_N}).\\

\begin{definition}\label{def:weak_noise_exp}\emph{Weak noise regime (dimension expansion)}\\
Let $\mathbf{x}_0$ denote the transmitted vector and $\mathbf{S}(\mathbf{x}_0)$ its representation in the channel space. We say that we are in the \emph{weak noise regime} whenever the second order term (as well as higher order terms) in~(\ref{e:WeakNoiseDist_2ndOrder_M_N}), i.e., the terms containing $\kappa_i$, are negligible compared to the 1st order term. That is,~(\ref{e:lin_apr_exp}) is a close approximation to  $\mathbf{S}$ and the weak noise distortion in~(\ref{e:mseort_mean_exp}) provides an accurate approximation to the actual distortion in the absence of anomalies.\hspace{7cm}$\square$\\
\end{definition}

From~(\ref{e:WeakNoiseDist_2ndOrder_M_N}) it is then clear why a linear system is convenient at low SNR, as  $\kappa_i=0$ and the higher order terms vanishes. For higher SNR one would seek nonlinear mappings with low curvature $\forall \mathbf{x}$.  And therefore the weak noise regime is a good approximation for any reasonably chosen mappings at high SNR.


\subsubsection{Weak channel distortion}\label{ssec:ChannelDist_new}
The received vector $\hat{\mathbf{z}}=\mathbf{z}+\mathbf{n}$ must be
passed through $\mathcal{S}$ to reconstruct $\mathbf{x}$. When the noise is weak enough, distortion analysis can be done by considering the tangent space of $\mathcal{S}$. That is, one can consider the linear approximation $\mathbf{S}_{lin}(\mathbf{z}_0)$ of $\mathbf{S}(\mathbf{z})$ at $\mathbf{z}_0$
\begin{equation}\label{e:lin_appr_red}
\mathbf{S}_{lin}(\mathbf{z}_0+\mathbf{n})=\mathbf{S}(\mathbf{z_0})+J(\mathbf{z_0})\mathbf{n},
\end{equation}
We have the following proposition providing the exact distortion under linear approximation:\\

\begin{proposition}\label{th:channel_dist}\emph{Minimum Weak Channel Distortion}\\
For any continuous i.i.d. Gaussian channel of dimension $N$ and any dimension reducing S-K mapping $\mathbf{S}$ where $S_i\in
C^r(\mathbb{R}^M), \ r\geq 1, \ i=1,...,M$,  the distortion under the linear approximation in~(\ref{e:lin_appr_red}) is given by
\begin{equation}\label{e:mseort_mean_DimRed}
\bar{\varepsilon}_{chw}^2=\frac{\sigma_n^2}{M}\iint
\cdots \int_{\mathcal{D}_c}\sum_{i=1}^N
g_{ii}({\mathbf{z}})f_{{\mathbf{z}}}({\mathbf{z}})\mbox{d}{\mathbf{z}},
\end{equation}
where $f_{{\mathbf{z}}}({\mathbf{z}})$ is the channel signal pdf, and $g_{ii}$ are the diagonal components of the metric tensor $G$ of $\mathbf{S}$.
\end{proposition}

\emph{Proof:} See~\cite{Floor_Ramstad09}.\hspace{6.2cm}$\square$\\

The name \emph{weak channel distortion} is due to Def.~\ref{def:weak_noise_red} given below.

Consider now the channel error given by the 2nd order approximation to $\mathbf{S}$ in~(\ref{e:ChannelDist_2ndOrder_M_N}).\\

\begin{definition}\label{def:weak_noise_red}\emph{Weak noise regime (dimension reduction)}\\
Let $\mathbf{z}_0$ denote the transmitted vector and $\mathbf{S}(\mathbf{z}_0)$ its representation in the source space. We say that we are in the \emph{weak noise regime} whenever the second order term (and higher order terms) in~(\ref{e:ChannelDist_2ndOrder_M_N}), i.e., the terms containing $\kappa_i$, are negligible compared to the 1st order term. That is,~(\ref{e:lin_appr_red}) is a close approximation to  $\mathbf{S}$ and the weak channel distortion in~(\ref{e:mseort_mean_DimRed}) provides an accurate approximation to the actual distortion due to channel noise.\hspace{3cm}$\square$\\
\end{definition}

%



\section{New results on mapping construction using surfaces}\label{sec:Ex_SK_surf_32_and_23}
There are a myriad of possible surfaces, as exemplified in the \emph{Encyclopedia of Analytical Surfaces}~\cite{Surf_Encyclopedia}. Many of these will correspond with Definition ~\ref{def:sk_mapp}. Trying out many of these as potential S-K mappings is not a constructive approach. However, the criteria laid out above as well as in this section will rule out most surfaces as good candidates for well-performing S-K mappings. Having many specific criteria that has to be satisfied, constraining the set of relevant surfaces, can bring an advantage to numerical solutions to \emph{variational calculus} approaches~\cite{Akyol2014_TIT,Mehmetoglu_Akyol_Rose}, which tend to be stuck in local minima if not already well performing mappings are provided as input. In fact, with these criteria "baked" into the solution process of the variational calculus problem, one can better constrain the solution, and this will likely produce mappings closer to the global optimum.

We provide results here for surfaces in $\mathbb{R}^3$, some of which can easily be extended to higher dimensional surfaces and spaces. Our earlier investigations~\cite{Floor_Ramstad09,FloorThesis} showed that a constant metric tensor with independent components like in~(\ref{e:diagonal_indep_MT}) would be advantageous, as it would lead to a \emph{shape preserving mapping}. However, to incorporate a general pdf, it may be better to let
\begin{equation}\label{e:diagonal_indep_MT_pdfopt}
G(x_1,x_2)=
\begin{bmatrix}
g_{11}(x_1)&0\\
0&g_{22}(x_2)
\end{bmatrix},
\end{equation}
where $g_{ii}(x_i)$ could be "adjusted" to the sources pdf. Both of these should ideally lead to the correct \emph{scaling} with channel SNR. That is, the performance of the relevant mapping has the same slope as the corresponding OPTA curve as the SNR increases when its free variables are chosen optimally. However, things are not that straight forward.


For more general surfaces we need further results from~\cite{Kreyszig_DiffGeom91}. We first consider \emph{isometric surfaces}, that is, surfaces that can be mapped onto each other in a length preserving manner. \\

\begin{theorem}\label{Th:Isometric_Surf1}
An allowable mapping of a PoS $\mathcal{S}$ onto a PoS $\mathcal{S}^\ast$ is isometric  $\Longleftrightarrow$ $g_{\alpha\beta}=g^\ast_{\alpha\beta}$ at corresponding points when referred to the same coordinate system on $\mathcal{S}$ and $\mathcal{S}^\ast$.
\end{theorem}

\emph{Proof:} See~\cite[pp.176-177]{Kreyszig_DiffGeom91}.\hspace{4cm}$\square$\\

\begin{theorem}\label{Th:Isometric_Surf2}
Isometric surfaces have the same Gaussian curvature at corresponding points. Corresponding points on those surfaces have the same geodesic curvature, $\kappa_g$, at corresponding points.
\end{theorem}

\emph{Proof:} See~\cite[p.177]{Kreyszig_DiffGeom91}.\hspace{5.0cm}$\square$\\

A subset of surfaces named \emph{developable surfaces} (DS) can be mapped isometrically to the Euclidean plane~\cite[p.189]{Kreyszig_DiffGeom91}. This implies a metric like~(\ref{e:diagonal_indep_MT}) for any such surface. The DS is a special case of a \emph{ruled surface} (RS). A RS is obtained by a set of straight lines, named \emph{generators} $\mathbf{z}(\ell)$, which are related through a space curve $\mathbf{y}(\ell)$, named \emph{indicatrix}, as follows
\begin{equation}\label{e:RuSu}
\mathbf{S}(\ell,t)=\mathbf{y}(\ell)+ t \mathbf{z}(\ell),
\end{equation}
where $\mathbf{z}$ is a unit vector linearly independent of the tangent $\dot{\mathbf{y}}$, that is $\dot{\mathbf{y}}\times \mathbf{z}\neq 0$. The indicatrix acts like the trajectory for a straight line through space. $\mathbf{z}$ and $\mathbf{y}$ also act as coordinate curves. An example on a RS is the \emph{Helicoid} (or \emph{Archimedes Screw}) which is depicted in Fig.~\ref{fig:s3_2} below. Here one can see a straight line $\mathbf{z}$ moved along a helix $\mathbf{y}$.

If an  RS should also be a DS, the tangent planes need to be the same at all points along the same generator $t \mathbf{z}$~\cite[p. 181]{Kreyszig_DiffGeom91}. Only then will $\mathbf{S}$ be isometric to the Euclidean plane. The following theorem formally state the condition needed for a RS to be a DS~\cite[p.182]{Kreyszig_DiffGeom91}:\\

\begin{theorem}\label{Th:RS_Cond_DS}
A ruled surface $\mathbf{S}(\ell,t)=\mathbf{y}(\ell)+ t \mathbf{z}(\ell)$ is a developable surface $\Leftrightarrow$
\begin{equation}
\big|\dot{\mathbf{y}} \ {\mathbf{z}} \ \dot{\mathbf{z}} \big| = 0.
\end{equation}
\end{theorem}

\emph{Proof:} See~\cite[p.182]{Kreyszig_DiffGeom91}.\hspace{5.0cm}$\square$\\

An example of a DS is shown in Fig.~\ref{fig:RCASD_Surface} below. Here a straight line is moved along a Archimedes spiral, which is a plane curve.

We have two important theorems for surfaces with constant metric tensor $G$~\cite[pp. 188-189]{Kreyszig_DiffGeom91}:\\

\begin{theorem}\label{Th:DS_GaussCurv}
A portion of a surface (PoS) $\mathbf{S}\in C^r$, $r\geq2$ is a PoS of a DS $\Leftrightarrow$ The Gaussian curvature $K=0$ everywhere on $\mathbf{S}$
\end{theorem}

\emph{Proof:} See~\cite[p.188]{Kreyszig_DiffGeom91}.\hspace{5.0cm}$\square$\\

The definition of Gaussian curvature is given in Section~\ref{ssec:DiffGeom_ParSurfaces}.\\

\begin{theorem}\label{Th:IsometryPlane_DS}
A sufficiently small PoS of $\mathbf{S}\in C^r$, $r\geq2$,  can be mapped isometrically to a plane $\Leftrightarrow$ $\mathbf{S}$ is a PoS of a DS.
\end{theorem}

\emph{Proof:} See~\cite[p.189]{Kreyszig_DiffGeom91}.\hspace{5.0cm}$\square$\\

A DS always has Gaussian curvature $K=0$, implying that either $\kappa_1$ or $\kappa_2$ is zero. Also, when $K=0$ we always have a DS. This implies that only DS can be mapped onto a plane without distorting/changing  distances. A DS will therefore guarantee a constant metric tensor $G$.

We have three examples of RS: Let $\mathbf{t}=\dot{\mathbf{y}}$ (tangent vector), $\mathbf{p}=\ddot{\mathbf{y}}$ (principal normal vector) and $\mathbf{b}=\mathbf{t}\times\mathbf{p}$ (binormal vector):

i) Tangent Surface (TS):
\begin{equation}\label{e:TaSu}
\mathbf{S}(\ell,t)=\mathbf{y}(\ell)+ t \mathbf{t}(\ell),
\end{equation}

ii) Principal Normal Surface (PNS):
\begin{equation}\label{e:PnSu}
\mathbf{S}(\ell,t)=\mathbf{y}(\ell)+ t \mathbf{p}(\ell),
\end{equation}

iii) Binormal Surface (BNS):
\begin{equation}\label{e:BnSu}
\mathbf{S}(\ell,t)=\mathbf{y}(\ell)+ t \mathbf{b}(\ell),
\end{equation}

Under certain conditions these surfaces are also DS~\cite[p.182]{Kreyszig_DiffGeom91}\footnote{A PNS in ii) was applied in~\cite{Floor_Kim_2013_Entropy} for a bivariate Gaussian to utilize correlation among two variables. This provided a significant gain over linear and piecewise linear approaches.}:\\

\begin{theorem}\label{Th:TS_PNS_BNS_Char}
The TS is always a DS. PNS and BNS are DS $\Leftrightarrow$  $\mathbf{y}(\ell)$ is a plane curve, i.e., the torsion $\tau=0$.
\end{theorem}

\emph{Proof:} See~\cite[p.182]{Kreyszig_DiffGeom91}.\hspace{5.0cm}$\square$\\


%

\emph{Proof:} See~\cite[p.185]{Kreyszig_DiffGeom91}.\hspace{5.0cm}$\square$\\

From the above results it seems like DS is the correct choice of surface for $3$:$2$ and $2$:$3$  mappings. However, as we show in the following, this is not true in general.

\begin{figure}[h!]
    \begin{center}
            \includegraphics[width=1\columnwidth]{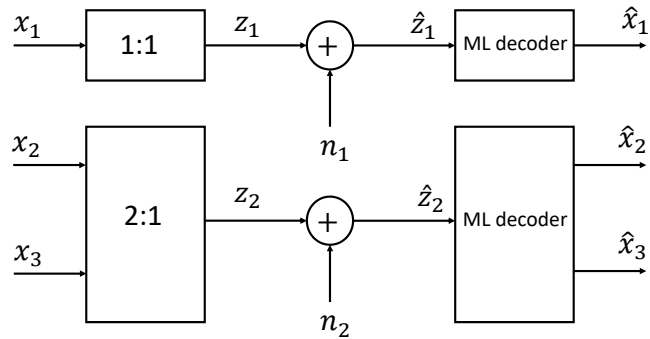}
    \end{center}
    \caption{Splitting of a $3$:$2$ mapping into $1$:$1$ and $2$:$1$ sub-mappings.}\label{fig:MapSplit3_2}
\end{figure}

A DS is one of many mappings that can be \emph{decomposed} into sub-mappings. Consider the $3$:$2$ system in Fig.~\ref{fig:MapSplit3_2},
consisting of a $1$:$1$ and $2$:$1$ system. For a DS, in particular, this sub-system consists of a linear $1$:$1$ mapping ($\mathbf{z}$) and a nonlinear $2$:$1$ mapping ($\mathbf{y}$). However, many decomposable systems may also have a nonlinear $1$:$1$ mapping (or general nonlinear subsystems of $M$:$N$ mappings). In the following we will show that any mapping that can be split into sub-mappings will not obtain the optimal slope in SDR as SNR increases. That is, they \emph{diverge} away from OPTA at high SNR.

Consider the general form of OPTA for $M$:$N$ mapping:
\begin{equation}\label{e:OPTA_gen}
\text{SDR}_{M:N}=\frac{\sigma_x^2}{D_{M:N}}=\bigg(1+\frac{P_{M:N}}{\sigma_n^2}\bigg)^\frac{N}{M}
\end{equation}
We have that the minimal distortion for the $1$:$1$ system is (solve~(\ref{e:OPTA_gen}) w.r.t. $D_t$ for $N/M=1$)
\begin{equation}\label{e:OptDist_1_1}
D_{1:1} = \frac{\sigma_x^2}{1+\frac{P_{1:1}}{\sigma_n^2}}
\end{equation}
and for the $2$:$1$ system is (solve~(\ref{e:OPTA_gen}) w.r.t. $D_t$ for $N/M=1/2$)
\begin{equation}\label{e:OptDist_2_1}
D_{2:1} = \bigg(\frac{\sigma_x^2}{1+\frac{P_{2:1}}{\sigma_n^2}}\bigg)^\frac{1}{2}
\end{equation}
With $P_t$ the total power of the $3$:$2$ system one can allocate power to the two sub-systems through a factor $\kappa\in[0,1]$. Let SNR$=P_t/\sigma_n^2$ and use the fact that $1+x\approx x$ as $x$ becomes large, then the total distortion in the limit of large SNR becomes,
\begin{equation}\label{e:OptDist_SplitComp3_2_tot}
\begin{split}
&\lim_{\text{SNR}\rightarrow \infty} D_{t(3:2)} =
\\ &\lim_{\text{SNR}\rightarrow \infty}\frac{\sigma_x^2}{3}\bigg[\bigg(1+\kappa \text{SNR}\bigg)^{-1} + \bigg(1+(1-\kappa)\text{SNR}\bigg)^\frac{1}{2}\bigg]\\
 &\frac{\sigma_x^2}{3}\lim_{\text{SNR}\rightarrow \infty} \bigg[\frac{1}{\kappa\text{SNR}} + \frac{1}{\sqrt{(1-\kappa)\text{SNR}}}\bigg].
\end{split}
\end{equation}
According to the laws of limits $\lim_{x\rightarrow \infty} \kappa x = \kappa \lim_{x\rightarrow \infty} x$, and so the power allocation factor can be moved outside the limit. Then, since $\sqrt{\text{SNR}}$ grows more slowly than SNR, $D_{t(3:2)}$ will be dominated by $1/\sqrt{\text{SNR}}$ as SNR$\rightarrow \infty$. This implies that a $3$:$2$ system realized by a DS will have the slope of  $2$:$1$  system at high SNR. The generalization to a general dimension reducing mapping ($M>N$) follows along the same lines.

As DS can be seen as a straight line ($1$:$1$ system) moved along a curve $\mathbf{y}$ (a $M$:$1$ or $1$:$N$ system) a DS has the wrong slope and will therefore diverge from the OPTA bound as SNR grows large (this will include the suggestion for higher dimensional mappings in~\cite{Hu_garcia_lamarca_tcom}).

\begin{figure}[h!]
    \begin{center}
            \includegraphics[width=0.98\columnwidth]{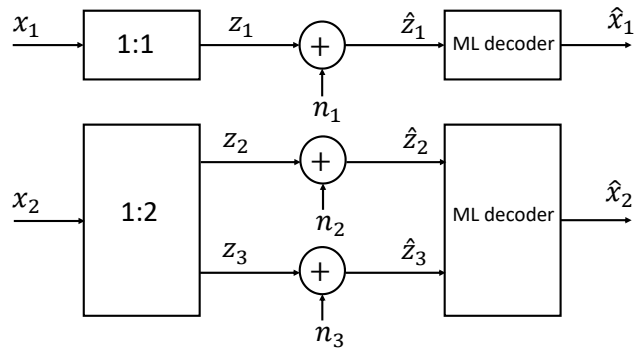}
    \end{center}
    \caption{Splitting of a $2$:$3$ mapping into $1$:$1$ and $1$:$2$ sub-mappings.}\label{fig:MapSplit2_3}
\end{figure}

We show that this is also the case for expanding mappings: Consider the $2$:$3$ system in
Fig.~\ref{fig:MapSplit2_3},
consisting of a $1$:$1$ and $1$:$2$ system. The minimal distortion for the $1$:$1$ system is as in~(\ref{e:OptDist_1_1}). For the $1$:$2$ system we have
\begin{equation}\label{e:OptDist_1_2}
D_{1:2} = \bigg(\frac{\sigma_x^2}{1+\frac{P_{2:1}}{\sigma_n^2}}\bigg)^2
\end{equation}
With $P_t$ the total power of the $2$:$3$ system one can allocate to the two sub-systems according to a allocation factor $\kappa\in[0,1]\subseteq \mathbb{R}$. The total distortion in the limit of large SNR becomes
\begin{equation}\label{e:OptDist_SplitComp2_3_tot}
\begin{split}
&\lim_{\text{SNR}\rightarrow \infty} D_{t(2:3)} =
\\ &\lim_{\text{SNR}\rightarrow \infty}\frac{\sigma_x^2}{2}\bigg[\bigg(1+\kappa \text{SNR}\bigg)^{-1} + \bigg(1+(1-\kappa)\text{SNR}\bigg)^{2}\bigg]\\
 &\frac{\sigma_x^2}{2}\lim_{\text{SNR}\rightarrow \infty} \bigg[\frac{1}{\kappa\text{SNR}} + \frac{1}{((1-\kappa)\text{SNR})^2}\bigg].
\end{split}
\end{equation}
By moving the power allocation factor outside the limit, then, since ${\text{SNR}}^2$ grows faster than SNR, $D_{t(2:3)}$ will be dominated by $1/{\text{SNR}}$ as SNR$\rightarrow \infty$. This implies that a $2$:$3$ system realized by a DS will have the slope of  $1$:$1$  system at high SNR. The generalization to a general dimension expanding mapping ($M > N$) follows along the same lines.

We have the more general Proposition:\\

\begin{proposition}\label{prop:Map_Split}\emph{Splitting of Mappings}\\
Any $m+n$:$2$ or $2$:$m+n$ mapping composed of lower dimensional (curve-based) sub-mappings will always have a smaller slope in SDR as a function of SNR than that of the OPTA curve, a slope always corresponding to the system with the smallest exponent.
\end{proposition}

\emph{Proof:} Take the dimension reduction case: The minimal distortion for a $n$:$1$ system is found by solving~(\ref{e:OPTA_gen}) w.r.t. $D_t$ for $N=1,M=n$,
\begin{equation}\label{e:OptDist_1_1}
D_{n:1} = \bigg(\frac{\sigma_x^2}{1+\frac{P_{n:1}}{\sigma_n^2}}\bigg)^\frac{1}{n}.
\end{equation}
Similarly, for an $m$:$1$ system, we obtain the same formula by exchanging $n$ with $m$.
With $P_t$, the total power of the $(m+n)$:$2$ system, one can allocate power to the two sub-systems through a factor $\kappa\in[0,1]$ so that $P_{n:1}=\kappa P_t$. Let SNR$=P_t/\sigma_n^2$ and use the fact that $1+x\approx x$ as $x$ becomes large, then the total distortion in the limit of large SNR becomes,
\begin{equation}\label{e:OptDist_SplitCompMN_2_tot}
\begin{split}
&\lim_{\text{SNR}\rightarrow \infty} D_{t(m+n:2)} =\\ &\lim_{\text{SNR}\rightarrow \infty}\frac{\sigma_x^2}{m+n}\bigg[\bigg(1+\kappa \text{SNR}\bigg)^\frac{1}{n} + \bigg(1+(1-\kappa)\text{SNR}\bigg)^\frac{1}{m}\bigg]\\
  &=\frac{\sigma_x^2}{m+n}\lim_{\text{SNR}\rightarrow \infty} \bigg[\frac{1}{\kappa^{1/n}\text{SNR}^{1/n}} + \frac{1}{(1-\kappa)^{1/m}\text{SNR}^{1/m}}\bigg],
\end{split}
\end{equation}
According to the laws of limits $\lim_{x\rightarrow \infty} \kappa x = \kappa \lim_{x\rightarrow \infty} x$, and so the power allocation factor(s) can be moved outside the limit. Then, for $m>n$ since ${\text{SNR}}^{1/m}$ grows more slowly than ${\text{SNR}}^{1/n}$, $D_{t(m+n:2)}$ will be dominated by the second term in~(\ref{e:OptDist_SplitCompMN_2_tot}) as SNR$\rightarrow \infty$.

In the expansion case, a similar derivation leads to
\begin{equation}\label{e:OptDist_SplitExpM_N_tot}
\begin{split}
\lim_{\text{SNR}\rightarrow \infty} D_{t(2:m+n)} =\frac{\sigma_x^2}{m+n}\lim_{\text{SNR}\rightarrow \infty} \bigg[\frac{1}{\kappa\text{SNR}^{n}} + \frac{1}{(1-\kappa)\text{SNR}^{m}}\bigg].
\end{split}
\end{equation}
If $m>n$, the first term in~(\ref{e:OptDist_SplitExpM_N_tot}) dominates as SNR$\rightarrow \infty$.
\hspace{0cm}$\square$\\

Despite these results, DS provide a simple mapping which is based on combinations of lower dimensional mappings which may provide quite good performance in the \emph{small to medium} SNR range as illustrated in Sections~\ref{ssec:Ex_SK_surf_32} as well as in~\cite{Hu_garcia_lamarca_tcom}.

To avoid the problem of wrong slope, one has to widen the set of relevant mappings, but still keep a similar type metric tensor. A more general and larger set of surfaces having a similar metric tensor is a subset of those that can me mapped \emph{conformally} to the plane~\cite[pp. 193]{Kreyszig_DiffGeom91}:\\

\begin{theorem}\label{Th:ConformalMap}
An allowable mapping of a PoS of $\mathcal{S}$ onto a PoS of $\mathcal{S}^\ast$ is  conformal $\Leftrightarrow$ when on $\mathcal{S}$ and $\mathcal{S}^\ast$ the same coordinate systems have been introduced, the coefficients of the FFF $g_{\alpha\beta}$ and $g^\ast_{\alpha\beta}$ of $\mathcal{S}$ and $\mathcal{S}^\ast$, respectively, are proportional:
\begin{equation}\label{e:ConformalMap_Cond}
g^\ast_{\alpha\beta}=\eta(u^1,u^2)g_{\alpha\beta}, \ \ \eta>0, \ \ \alpha\beta=1,2.
\end{equation}
\end{theorem}

\emph{Proof:} See~\cite[p.193-194]{Kreyszig_DiffGeom91}.\hspace{4.5cm}$\square$\\

\begin{theorem}\label{Th:ConformalMap_to_Plane}
Any simply-connected portion $\mathcal{S}$ of a surface which has a representation of class $r\geq3$ can be conformally mapped into a plane.
\end{theorem}

\emph{Proof:} See~\cite[p.196-198]{Kreyszig_DiffGeom91}.\hspace{4.5cm}$\square$\\


Another subset of  surfaces that is of importance is  \emph{minimal surfaces.} A minimal surface is a surface $\mathcal{S}$ of class $r\geq 2$ whose mean curvature
\begin{equation}\label{e:MeanCu_MiSu}
H =  \frac{1}{2} b_\alpha^\alpha = \frac{1}{2} b_{\alpha\beta} g^{\alpha\beta} = 0
\end{equation}
at every point of $\mathcal{S}$. We have the following theorem~\cite[p. 244]{Kreyszig_DiffGeom91}\\

\begin{theorem}\label{Th:MiSu_meaning}
Let $\mathcal{C}$ be a simple closed curve. If among all portions of surfaces of class $r\geq2$ bounded by $\mathcal{C}$ there exists a portion $\mathbf{x}(u^1,u^2)$ of minimum area then $\mathbf{x}(u^1,u^2)$ is necessarily a portion of a minimal surface.
\end{theorem}

\emph{Proof:} See~\cite[p.244-245]{Kreyszig_DiffGeom91}.\hspace{4.5cm}$\square$\\

That is, minimal surfaces are natural solution candidates for the Euler-Lagrange equations for surfaces. However, One cannot conclude that a surface needs to be minimal in order for S-K mappings to be optimal as we will see in the following sections.

Finally, we state a condition that has to be satisfied if a \emph{uniform} $M$:$N$ dimension reducing mapping\footnote{Uniform S-K mapping is defined in~\cite{Floor_Ramstad09}, Def. 6} ($M>N$) is to follow the slope of OPTA: From the proof of Proposition 6, i.e., Eq. (58) in~\cite{Floor_Ramstad09}, we have that
\begin{equation}\label{e:D_tot_Uniform_M_N_DimRed}
\begin{split}
 D_{t} &=\bar{\varepsilon}_a^2 + \bar{\varepsilon}_{ch}^2\\
 &=\frac{M-N}{4M(M-N+2)}\Delta^2 + M^{\frac{M}{N}-1}\tilde{B}^\frac{2}{N}\\
&\bigg(\frac{\Delta}{2}\bigg)^{-2\frac{M-N}{N}}
\sigma_x^{2\frac{M}{N}}
\bigg(1+\frac{P_N}{\sigma_n^2}\bigg)^{-1}.
 \end{split}
\end{equation}
Then, for a $3$:$2$ mapping, we have that
\begin{equation}\label{e:Uniform_M_N_red_OptExp}
\bar{\varepsilon}_{chw}^2\sim \frac{1}{\Delta} \  \text{when} \  \bar{\varepsilon}_a^2 \sim \Delta^2,
\end{equation}
in order to obtain  SDR$\sim\text{SNR}^{2/3}$.

\subsection{Examples on $3$:$2$ mappings}\label{ssec:Ex_SK_surf_32}
We analyze and simulate several $3$:$2$ mappings selected based on the characteristics of surfaces presented in the previous sections. For comparison we consider OPTA, \emph{Block Pulse Amplitude Modulation} (BPAM)~\cite{LeePetersen76} and \emph{Power Constrained Channel Optimized Vector Quantizer} (PCCOVQ)~\cite{fulds97a,fuldsethThesis}. BPAM is the optimal linear mapping (i.e., a plane in the source- or channel space) for any $r=N/M \in \mathbb{Q}$. The PCCOVQ is a numerically optimized discrete mapping (for any $r=N/M \in \mathbb{Q}$) that often replicate continuous or piece-wise continuous curves or surfaces when the number of points in its constellations is large. Obviously, one would like any nonlinear mapping to rise well above BPAM as the SNR increases for any $r\neq 1$. The PCCOVQ is often hard to beat with many points in the constellation since one is more free to move points in space than what is the case when one is stuck to a given curve or surface. Approaching or beating a PCCOVQ system with many points is therefore a good indication of a well performing mapping.

\subsubsection{Helicoid}\label{ssec_Helicoid}
\begin{figure}[h!]
    \begin{center}
            \includegraphics[width=1\columnwidth]{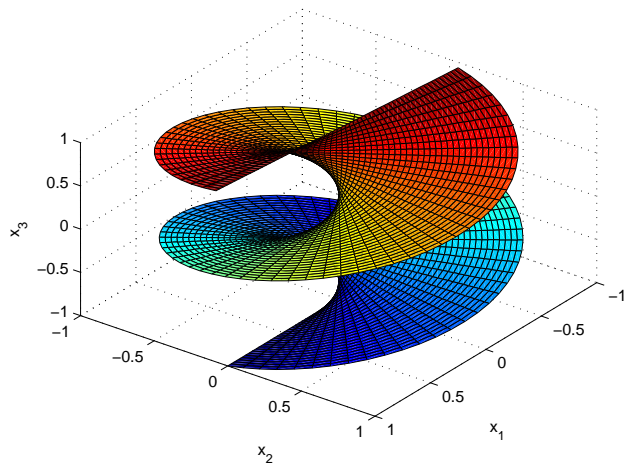}
    \end{center}
    \caption{$3$:$2$ Helicoid mapping shown in the source space. The structure corresponds to
    equation~(\ref{e:Helicoid}) where $R=\Delta=1$, $\alpha_1=\alpha_2=1$, and $z_1,z_2\in [-\pi,\pi]$.}\label{fig:s3_2}
\end{figure}

The first structure chosen for $\mathbf{S}$ is the \emph{Helicoid} (also known
as \emph{Archimedes' screw}) depicted in Fig.~\ref{fig:s3_2}.
 The Helicoid has a parametric equation $\mathbf{S}(z_1,z_2)$ with components 
\begin{equation}\label{e:Helicoid}
\begin{split}
S_1(z_1,z_2) &= \frac{R\alpha_1}{\pi} z_1 \cos(\alpha_2 z_2)\\
S_2(z_1,z_2) &= \frac{R\alpha_1}{\pi} z_1 \sin(\alpha_2 z_2), \ S_3(z_2) = \frac{\Delta \alpha_2}{\pi} z_2,
\end{split}
\end{equation}
where $\Delta$ corresponds to the distance between two
\emph{helical discs} of the surface (at some distance from the $x_3$ axis). $R$ is
determining the size of the ``radius'' of the surface in the $x_1
x_2$-plane (see Fig.~\ref{fig:s3_2} from the "top"). To obtain a
symmetric channel signal both positive and negative parameter values
are used and $(z_1,z_2)=(0,0)$ corresponds to the origin in
$\mathbb{R}^3$. The source vectors $\mathbf{x}$ are projected onto
the Helicoid. The point $(y_1,y_2)$, corresponding to the point on
the Helicoid, represents the two channel signals. The channel signals
are further scaled with a factor $\alpha_i$ to satisfy a given
power constraint, that is $z_i=y_i/\alpha_i$, $i=1,2$.

\textbf{Channel power}: The channel power can be found by
considering equation~(\ref{e:Helicoid}) and Fig.~\ref{fig:s3_2}. The
derivation is done leaving out the scaling factors $\alpha_1$ and
$\alpha_2$, so the statistics of the
channel signals $y_1$ and $y_2$ are considered.

Looking at the Helicoid from above ($z_3$ direction), one can see a
disc in the $x_1 x_2$ plane. The radius of this disc is given by
$\rho=(R/\pi) y_1$ (i.e., directrix lines, following a helical indicatrix), while
$y_2$ traces out all angles (the ``parallel helices'', i.e., the indicatrix). Thus $y_1$
is given by the function
\begin{equation}\label{e:h_radius}
y_1=\pm\frac{\pi}{R}\sqrt{x_1^2+x_2^2}.
\end{equation}
According to~\cite[p.190]{papoulis02}, the mapping of two
Gaussian random variables by this function gives a Rayleigh distribution. But due to
the $\pm$ sign a zero mean ``double Rayleigh'' distribution is obtained
\begin{equation}
f_{y_1}(y_1)=\frac{1}{2}\frac{R^2|y_1|}{\pi^2\sigma_x^2}e^{-\frac{R^2
y_1^2}{2\pi^2\sigma_x^2}}.
\end{equation}
To get the constants right, one has to use the fact that the random
variable $h(y)=cy$ has distribution $f_h(h)=(1/|c|) f_y(h/c)$
~\cite[p.131]{papoulis02}. Further in~\cite[p 148]{papoulis02} it is
shown that the second moment of the Rayleigh distribution
$(y/\upsilon^2)e^{-y^2/(2\upsilon^2)}$ is $E\{y^2\}=2\upsilon^2$,
and so the signal variance on channel one is given by
\begin{equation}
\sigma_{y_1}^2=2\bigg(\frac{\sigma_x \pi}{R}\bigg)^2
\end{equation}

The other parameter, $y_2$,  will trace out parallel helices with
"radius" given by~(\ref{e:h_radius}). Consider first  $y_1=0$.
Then $y_2$ will lie on the $x_3$ axis and have a Gaussian
distribution since then $x_3=(\Delta/\pi)y_2$, and so the variance of $y_2$ will be
\begin{equation}\label{e:sigma2_pre}
{\sigma_{y_2}^2}_{|y_1=0}=\bigg(\frac{\pi \sigma_x}{\Delta}
\bigg)^2.
\end{equation}
Considering other cases ($y_1\neq 0$) it seems like the variance
(and the pdf) stays more or less the same, independent of the value of
$y_1$. Although $y_2$ will be on a helix, it seems to trace out all
of $z_3$ in a very similar way. The little discrepancy
from~(\ref{e:sigma2_pre}) seem to be more or less constant, equal to $1/2$, except when $\Delta$ is very close to zero, so that
\begin{equation}\label{e:po_2}
\sigma_{y_2}^2=\bigg(\frac{\pi \sigma_x}{\Delta}
\bigg)^2+\frac{1}{2}.
\end{equation}
Taking the scaling, $\alpha_1$ and $\alpha_2$, into account,  the total channel power (per channel dimension) becomes
\begin{equation}\label{e:ch_pov_helicoid}
P=
\frac{1}{2}\bigg(\frac{\sigma_{y_1}^2}{\alpha_1^2}+\frac{\sigma_{y_2}^2}{\alpha_2^2}\bigg)
=\frac{1}{2}(\sigma_{z_1}^2+\sigma_{z_2}^2).
\end{equation}

\textbf{Channel distortion}: To determine the channel distortion, the diagonal
components of the metric tensor of $\mathbf{S}$ is needed
\begin{equation}\label{e:g11n}
\begin{split}
g_{11}&=\bigg\|\frac{\partial{\mathbf{S}}}{\partial{z_1}}\bigg\|^2=\bigg\|\frac{R\alpha_1}{\pi}\cos{(\alpha_2
z_2)},\frac{R\alpha_1}{\pi}\sin{(\alpha_2
z_2)},0\bigg\|^2\\
&=\bigg(\frac{R\alpha_1}{\pi}\bigg)^2,
\end{split}
\end{equation}
and
\begin{equation}\label{e:g22n}
\begin{split}
&g_{22}=\bigg\|\frac{\partial{\mathbf{S}}}{\partial{z_2}}\bigg\|^2=\\
&\bigg\|-\frac{R\alpha_1}{\pi}z_1\sin{(\alpha_2
z_2)}\alpha_2,
\frac{R\alpha_1}{\pi}z_1\cos{(\alpha_2 z_2)}\alpha_2,\frac{\Delta\alpha_2}{\pi}\bigg\|^2\\
&=\bigg(\frac{R\alpha_1\alpha_2}{\pi}\bigg)^2 z_1^2
+\bigg(\frac{\Delta\alpha_2}{\pi}\bigg)^2.
\end{split}
\end{equation}
It is straight forward, using the derivatives in~(\ref{e:g11n}) and~(\ref{e:g22n}), to show that $g_{12}=0$.
One can now calculate the weak channel distortion as
\begin{equation}\label{e:cmse2}
\begin{split}
&\bar{\varepsilon}_{ch}^2 = \frac{\sigma_n^2}{3} \iint\sum_{i=1}^2 g_{ii}(\mathbf{z}) f_\mathbf{z}(\mathbf{z})\mbox{d} \mathbf{z}\\
&=\frac{\sigma_n^2}{3}\int_{-\infty}^\infty
\int_{-\infty}^\infty
\bigg[\bigg(\frac{R\alpha_1}{\pi}\bigg)^2+\bigg(\frac{R\alpha_1\alpha_2}{\pi}\bigg)^2
z_1^2
+\cdots\\
&\bigg(\frac{\Delta\alpha_2}{\pi}\bigg)^2\bigg]f_{z_1,z_2}(z_1,z_2)\mbox{d}z_1\mbox{d}z_z.
\end{split}
\end{equation}
This leads to
\begin{equation}\label{e:cmse}
\bar{\varepsilon}_{ch}^2=\frac{\sigma_n^2}{3\pi^2}((\Delta\alpha_2)^2+(R\alpha_1)^2(1+\alpha_2^2\sigma_{z_1}^2)).
\end{equation}
From~(\ref{e:cmse}) one can observe that the channel distortion increases linearly with the power on channel one ($\sigma_{z_1}^2$).
This can also be seen from the metric component $g_{22}$. I.e., the length of the velocity vector along the $z_2$ direction is dependent on $z_1$. Trying to compensate for the dependence in one direction, i.e., using arc-length along that coordinate curve, will affect the other coordinate curve and vice versa, as will be demonstrated at the end of this section.

\textbf{Approximation distortion}: The approximation distortion is not feasible to find
analytically, particularly at low SNR. The distance between two folds of $\mathbf{S}$ is determined by $\Delta$, which determines the size of the approximation distortion. The approximation distortion can then be approximated by nonlinear curvefitting: a model that coincide well when $\sigma_x=1$ is
\begin{equation}
\begin{split}
&\bar{\varepsilon}_a^2 \approx
\beta \Delta^3+\gamma \Delta^2+\vartheta \Delta=\\
&-0.0036\Delta^3+0.024\Delta^2+0.0056\Delta,\ \Delta\in[0,3], \ \sigma_x=1,
\end{split}
\end{equation}

\textbf{Optimization of the Helicoid system}: Assume $\sigma_x=1$. To find the optimal performance, one need to optimize over $R_1$,
$\Delta$, $R$ and $\alpha_2$, given a channel power constraint. The Lagrangian of this problem is
\begin{equation}
\begin{split}
&\mathcal{L}(R,\Delta,\alpha_1,\alpha_2,\lambda)=\\
&\bar{\varepsilon}_a^2(\Delta)+\bar{\varepsilon}_{ch}^2(R,\Delta,\alpha_1,\alpha_2)-\lambda
c_t(R,\Delta,\alpha_1,\alpha_2),
\end{split}
\end{equation}
where the constraint is given by
\begin{equation}
c_t(R,\Delta,\alpha_1,\alpha_2)=P_{max}-P(R,\Delta,\alpha_1,\alpha_2)\geq
0.
\end{equation}
 $P_{max}$ is the maximum allowed power per channel and
$P(R,\Delta,\alpha_1,\alpha_2)$ is given
by~(\ref{e:ch_pov_helicoid}).
The first  order necessary conditions for a minimum is given by the
the Karush-Kuhn-Tucker (KKT) equations~\cite[p.328]{nocedal/wright}.
One of the criterion that has to be satisfied to be in a KKT point
is
\begin{equation}\label{e:grad_lan}
\nabla_{R\Delta\alpha_1\alpha_2}\mathcal{L}(R,\Delta,\alpha_1,\alpha_2,\lambda)=0
\end{equation}
Solving~(\ref{e:grad_lan})  analytically is impossible since it contains an equation of high order with no algebraic solution.
Therefore numerical optimization has to be used.

\begin{figure}
    \begin{center}
             \includegraphics[width=1.0\columnwidth]{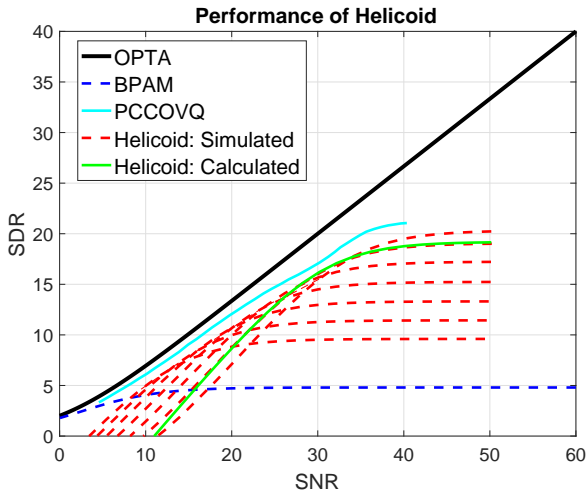}
    \end{center}
    \caption{Performance of the $3$:$2$ Helicoid based system compared to OPTA, PCCOVQ and BPAM.}\label{fig:pe3_2}
\end{figure}

Fig.~\ref{fig:pe3_2} shows the comparison between the theoretical model given above and its
simulation, as well as OPTA, PCCOVQ and BPAM.
One can observe that the performance of the theoretical
model and the simulation coincide quite well at high SNR (as expected, since higher order terms related to curvature becomes negligible). The PCCOVQ system\footnote{The reason why the PCCOVQ system declines with increasing SNR is that
only $4096$ points (centroids) are used during optimization, which is too small a number for representing a 2D space at high SNR.} clearly outperforms the Helicoid.

The Helicoid is a Ruled Surface as well as a minimal surface, but is not a developable surface since it is a PNS with a helix indicatrix $\mathbf{y}(\ell)$ (which is not a plane indicatrix).  As such, its a potentially good $3$:$2$ mapping, as seen at low to medium SNR.  However, due to the fact that the Helicoid is a ruled surface it is not isometric to the plane, and $g_{22}$ is both dependent and increases with $z_2$ as seen from~(\ref{e:g22n}). Therefore the Helicoid diverges away from OPTA as SNR gets high. Can one compensate for this effect using arc length parametrization?

\begin{figure}[h!]
    \begin{center}
            \includegraphics[width=1.0\columnwidth]{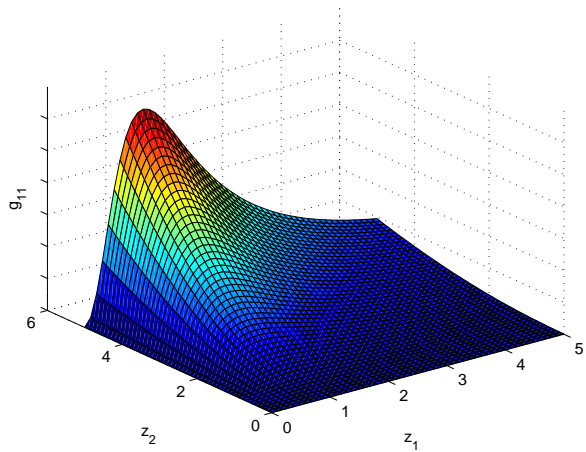}
   \end{center}
    \caption{The first diagonal component of the metric tensor $g_{11}(z_1,z_2)$ for the Helicoid after a
    arc length transformation of $z_2$.}\label{fig:transfomed_metric}
\end{figure}

\textbf{Arc length parametrization:}
Assume now that $z_1$ is mapped through a function $\varphi(z_1)$ before mapping onto the Helicoid, leading to arc length parametrization along the $z_2$ coordinate, effectively removing the dependence of $g_{22}$ w.r.t. $z_1$. Then $g_{22}$ becomes (with $\alpha_1=\alpha_2=1$)
\begin{equation}
g_{22}=\frac{\Delta^2+R^2 z_1^2}{\pi^2 \varphi^2(z_1)}.
\end{equation}
Choosing $\varphi(z_1)=\sqrt{\Delta^2+R^2 z_1^2}/\pi$ then
$g_{22}=1$. But now $g_{11}$ will become (with $R=\Delta=1$)
\begin{equation}\label{e:mod_g11}
g_{11}=\frac{z_1^4+(z_1 z_2)^2+2 z_1^2+1}{(1+z_1^2)^2 \pi^2}.
\end{equation}
Fig.~\ref{fig:transfomed_metric} depict~(\ref{e:mod_g11}).
This is even worse than $g_{22}\sim z_1^2$, since now, if the amplitudes on channel two is relatively large, the noise on channel
one will be scaled the most for the most probable channel symbols on channel one, giving a large average distortion after decoding.

This implies that the Helicoid is not a good choice of mapping. However, this interdependence is a problem for many surfaces. One exception is DS.

\subsubsection{Right Cylinder with Archimedes Spiral Directrix (RCASD)}\label{ssec:RCASD_Comp}
\begin{figure}[h]
    \begin{center}
           \includegraphics[width=1.0\columnwidth]{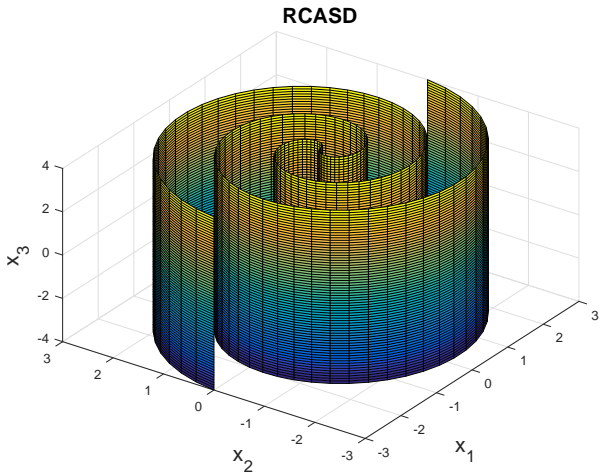}
    \end{center}
    \caption{The geometrical structure of the RCASD.}\label{fig:RCASD_Surface}
\end{figure}

The parametric equation for RCASD is~\cite[p. 51]{Surf_Encyclopedia}
\begin{equation}\label{e:S_RCASD_ParEq}
\mathbf{S}(\mathbf{x})=\big[a\varphi(z_1)\cos(\varphi(z_1)), a\varphi(z_1)\sin(\varphi(z_1)),\alpha_2 z_2 \big],
\end{equation}
with $\alpha_2$ some amplification factor and $a=\Delta/\pi$. $\Delta$ is the distance between the spiral arms. Fig.~\ref{fig:RCASD_Surface} depicts a surface plot of the RCASD with the (LoC) coordinate grid included

The tensor components of the first two fundamental forms are
\begin{equation}\label{e:RCASD_FFF_coef_Opt}
\begin{split}
g_{11}&=\bigg(\frac{\Delta}{\pi}\bigg)^2 \big(\varphi'(z_1)\big)^2 \big({1+\varphi^2(z_1)}\big),\\
g_{22}=\alpha_2^2, \ g_{12}&=g_{21}=0,
\end{split}
\end{equation}
and
\begin{equation}\label{e:RCASD_SFF_coef}
b_{11}=-a(\varphi'(z_1))^2\frac{2+\varphi^2(z_1)}{\sqrt{1+\varphi^2(z_1)}}, \ b_{12}=b_{21}=b_{22}=0.
\end{equation}
We do not show the calculation of this explicitly here, as a similar but more involved derivation is done in section~\ref{ssec:SnaSu} for the \emph{Snail Surface}.

The RCASD is a potentially good choice of $3$:$2$ mapping for several reasons. The RCASD is a developable surface, and also a bi-normal surface (BNS) with Archimedes spiral as directrix, which is a plane curve. From~(\ref{e:RCASD_FFF_coef_Opt}) and~(\ref{e:RCASD_SFF_coef}) one can conclude that the coordinate curves are LoC since $g_{12}=b_{12}=0$ (Theorem~\ref{th:LoC_Coordinates} ) and also geodesic coordinates since the directrix is a plane curve (Theorem~\ref{th:GeCo_2}). This implies that Eq.~(\ref{e:ChannelDist_2ndOrder_M_2}) describes the 2nd order behavior of this mapping exactly (LOC coordinates) and that the channel distortion is as low as possible (for the given surface). The RCASD therefore has many promising properties.

\textbf{Optimization of RCASD as $3$:$2$ mapping.}
To make the RCASD a scalable $3$:$2$ mapping we set $a=\Delta/\pi$. Note that with noise added to the channel signal $z_i$, the variable $\tilde{z}_i = z_i + n_i$ is mapped through~(\ref{e:S_RCASD_ParEq}).

\emph{Distortion:}
We compute the channel distortion according to~(\ref{e:mseort_mean_DimRed}), and for this we need the components of the metric tensor, which is given by~(\ref{e:RCASD_FFF_coef_Opt}). By choosing
\begin{equation}\label{e:RCASD_Mapping_func}
\varphi(\alpha_1 z_1)=\pm \sqrt{\frac{\alpha_1 z_1}{\eta\Delta}},
\end{equation}
with $\alpha_1$ some amplification factor depending on the SNR, one will as in~\cite{hekland_floor_ramstad_T_comm} obtain arc length parametrization along the spiral indicatrix in the $x_1,x_2$ plane of the RCASD, and therefore $g_{11}\approx \alpha_1^2, \ \forall z_1,z_2$. Then the channel distortion becomes
\begin{equation}\label{e:RCASD_ChannelDist}
\begin{split}
\bar{\varepsilon}_{ch}^2 &= \frac{\sigma_n^2}{3} \iint\sum_{i=1}^2 g_{ii}(\mathbf{z}) f_\mathbf{z}(\mathbf{z})\mbox{d} \mathbf{z}\\
&=  \frac{\sigma_n^2}{3} \big(\alpha_1^2 + \alpha_2^2 \big) f_\mathbf{z}(\mathbf{z})\mbox{d} \mathbf{z}\\
&= \frac{\sigma_n^2\big(\alpha_1^2 + \alpha_2^2 \big)}{3}.
\end{split}
\end{equation}
From this we see that the metric tensor is as in~(\ref{e:diagonal_indep_MT}), which was one of the criteria we sought.

To compute the approximation distortion one can observe from Fig.~\ref{fig:RCASD_Surface} that we have an approximation operation similar to archimedes spiral, i.e., a uniform S-K mapping (see~\cite{Floor_Ramstad09}, Definition 6), but now over 3 sources in $\mathbb{R}^3$ implying that Eq. (18) in~\cite{Floor_Ramstad09} will apply as a lower bound with channel dimension $N=2$ and source dimension $M=3$
\begin{equation}\label{e:RCASD_approx}
\bar{\varepsilon}_{q}^2 \geq \frac{M-N}{4M(M-N+2)}\Delta^2 = \frac{\Delta^2}{36}.
\end{equation}

\emph{Power:}
In the $x_1,x_2$-plane we have the same configuration as for the Archimedes spiral applied for $2$:$1$ compression in~\cite{hekland_floor_ramstad_T_comm}. That is, we map via an arc length parametrization from a two dimensional circular region (a disc) to a one dimensional representation. As shown in~\cite{hekland_floor_ramstad_T_comm}, for $y_1$ this results in a Laplace distribution with variance
\begin{equation}\label{e:RCASD_PowY1}
\sigma_{y_1}^2 = 2\bigg(2\eta \frac{\pi}{\Delta}\sigma_x^2\bigg)^2, \ \eta=0.16,
\end{equation}
at high SNR. $y_2$ is simply a linear mapping of $x_3$ so that we have a Gaussian distribution with variance $\sigma_{y_2}^2=\sigma_x^2$.

The channel signals has to be scaled by some factor $\alpha_i$ in order to satisfy the power constraint as the SNR changes. That is, $z_i=y_i\alpha_i$ and so $\sigma_{z_i}^2=\sigma_{y_i}^2/\alpha_i^2$. The total channel power then becomes
\begin{equation}\label{e:RCASD_PowTot}
P_t = \frac{1}{2}\bigg(\frac{\sigma_{y_1}^2}{\alpha_1^2}+\frac{\sigma_{y_2}^2}{\alpha_2^2}\bigg)^2.
\end{equation}

\emph{Optimization:}
With constraint $C_t = P_{\text{max}} - P_t(\Delta,\alpha_1,\alpha_2) \geq 0$, we get the following objective function
\begin{equation}\label{e:RCASD_3_2_ObjFunc}
\mathcal{L} (\Delta,\alpha_1,\alpha_2) = \bar{\varepsilon}_{q}^2(\Delta) + \bar{\varepsilon}_{ch}^2 (\alpha_1,\alpha_2)-\lambda C_t(\Delta,\alpha_1,\alpha_2).
\end{equation}
We use the same procedure to determine the optimal parameters as for the Helicoid in Section~\ref{ssec_Helicoid}.

\begin{figure}[h]
    \begin{center}
           \includegraphics[width=1.0\columnwidth]{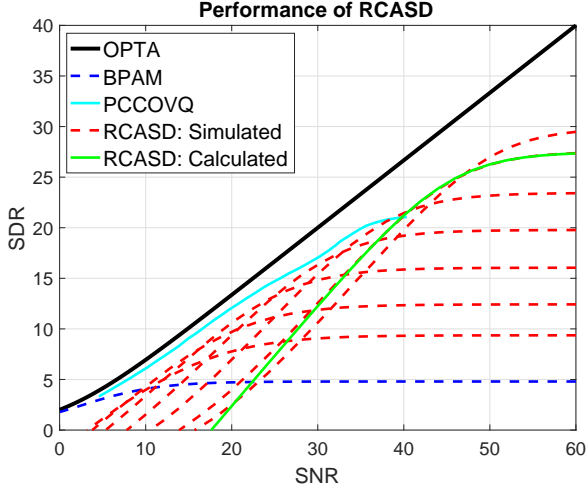}
    \end{center}
    \caption{Performance of RCASD (simulated and calculated) compared to OPTA, BPAM and PCCOVQ.}\label{fig:RCASD_Performance}
\end{figure}

The performance of the optimized RCASD is shown in Fig.~\ref{fig:RCASD_Performance}
The RCASD clearly improves with SNR, beating both the linear BPAM as well as PCCOVQ (see footnote 6) at high SNR. The RCASD is also noise robust. However, the slope at high SNR is lower than that of OPTA. We will show that this slope corresponds to that of a $2$:$1$ system:

\textbf{High SNR analysis.}
The Karush-Kuhn-Tucker (KKT) conditions~\cite{nocedal/wright} for the optimization problem in~(\ref{e:RCASD_3_2_ObjFunc}) are
\begin{equation}\label{e:KKT_RCASD}
\partial\mathcal{L}/\partial\Delta = 0, \ \partial\mathcal{L}/\partial\alpha_i = 0, \ i=1,2, \  C_t\geq 0,  \ \lambda\geq 0.
\end{equation}
From these conditions it is straight forward to show that no non-zero solution exists for inner points, and therefore  the only solution of interest lies on the boundary $C_t = 0$. Solving this equation w.r.t. $\alpha_1^2$ we get
\begin{equation}\label{e:RCASD_opt_alph1}
\alpha_1^2=\frac{\alpha_2^2 \kappa}{\Delta^2(2\alpha_2^2 P_{max} -\sigma_x^2)}, \ \kappa=2(2\eta\pi^2\sigma_x^2)^2.
\end{equation}
By inserting this in the total distortion we get the following unconstrained objective function
\begin{equation}\label{e:RCASD_unconstrained_opt}
\tilde{D}_t = \frac{\Delta^2}{36}+\frac{\sigma_n^2}{3}\bigg(\frac{\alpha_2^2\kappa}{\Delta^2(2\alpha_2^2 P_{max} -\sigma_x^2)}-\alpha_2^2\bigg).
\end{equation}
We have the system of equation $\partial \tilde{D}_t/\partial \Delta = 0$, $\partial \tilde{D}_t/\partial \alpha_2 = 0$. By solving  $\partial \tilde{D}_t/\partial \alpha_2 = 0$ w.r.t. $\alpha_2$ one get the four possible solutions
\begin{equation}\label{e:RCASD_opt_alph2}
\alpha_2=\pm \sqrt{\frac{\sigma_x}{2 P_{max} }\big(\sigma_x\pm \sqrt{\kappa}/\Delta\big)}.
\end{equation}
By inspecting this expression one can see that the first sign must be positive as $\alpha_i\geq 0$. By some further inspection one can conclude that also the sign inside the square root should be positive. By inserting~(\ref{e:RCASD_opt_alph2}) into $\partial \tilde{D}_t/\partial \Delta = 0$ we get
\begin{equation}\label{e:RCASD_opt_Delta}
\frac{\Delta^4}{18} - \frac{\sqrt{\kappa}\sigma_x^2}{3 \text{SNR}}\Delta -\frac{\kappa}{3 \text{SNR}}=0, \ \text{SNR}=P_{max}/\sigma_n^2.
\end{equation}
This equation does not have an analytical solution. However, as $\Delta$ gets small (SNR gets large)  the first and last term dominates, and so an approximate solution is
\begin{equation}\label{e:RCASD_approx_Delta}
\Delta\approx \sqrt[4]{\frac{6\kappa}{\text{SNR}}}
\end{equation}
Since at high SNR, $D_t=\bar{\varepsilon}_{q}^2=\bar{\varepsilon}_{ch}^2$, we see that SDR$\sim\text{SNR}^{-1/2}$ at high SNR, corresponding to the slope of a $2$:$1$ system. This is also exactly what one can expect from Prop.~\ref{prop:Map_Split} and Eq.~(\ref{e:OptDist_SplitComp3_2_tot}), since the RCASD is a developable surface which always can be split into a $1$:$1$ and a $2$:$1$ system (here, a line and a spiral).

\textbf{Curvature Analysis.}
As we consider LoC, the principal curvatures are easily found from the fundamental forms in~\ref{e:RCASD_FFF_coef_Opt} and~(\ref{e:RCASD_SFF_coef})
\begin{equation}\label{e:RCASD_PrincCurvat}
\kappa_1=\frac{b_{11}}{g_{11}}=-\frac{2+\varphi^2(z_1)}{a(\sqrt{1+\varphi^2(z_1)})^{3/2}}, \ \kappa_2=0.
\end{equation}
With $\varphi(z_1)$ as in~(\ref{e:RCASD_Mapping_func}) one can evaluate how curvature is affected by the free parameters $\Delta$ and $\alpha_1$. The maximal curvature as  function of both $\alpha_1$ and $\Delta$ is shown in Fig.~\ref{fig:RCASD_Curvature} in section~\ref{ssec:SnaSu} for $z_1=5$ (the result does not vary significantly with $z_1$). One can observe that the curvature of RCASD is relatively small in general. The curvature is getting smaller as $\Delta$ decreases and $\alpha_1$ increases, corresponding to the high SNR case.

As a particular example, we consider 30dB SNR. By inserting optimized parameters for 30dB SNR found by the optimization procedure below ($\Delta^\ast=0.608$, $\alpha_1^\ast=3.33$) one obtains $|\bar{\kappa}_1|<1$ averaged over the relevant range of $z_1$. By considering the distortion terms in~(\ref{e:WeakNoiseDist_2ndOrder_M_N}), with total transmission power 1, then $\sigma_n^2=0.001$, and one can see that the 1st order term is in the  order of about $0.001/0.001^2 = 1000$ over the 2nd order term. Therefore, RCASD is a mapping following Definition~\ref{def:weak_noise_exp} at high SNR.

\subsubsection{Monge Surface with Cylindrical Directrix Surface (MS-CDS)}\label{ssec:MS_CDS}
This surface has many configuration with the RCASD as a special case. The surface is described in~\cite[pp.189-197]{Surf_Encyclopedia}, and its parametric equation has components \begin{equation}\label{e:S_MSCDS_ParEq}
\begin{split}
S_1(z_1,z_2) &=  \frac{\Delta}{\pi} \cos(\varphi(z_1)) - \cdots\\ & \bigg[\frac{\Delta}{\pi} (\alpha_0 - \varphi(z_1)) - a (\alpha_2 z_2)^2\sin(\theta_0)\bigg]\sin(\varphi(z_1))\\
S_2(z_1,z_2) &=   \frac{\Delta}{\pi} \sin(\varphi(z_1)) - \cdots\\ & \bigg[\frac{\Delta}{\pi} (\alpha_0 - \varphi(z_1)) - a (\alpha_2 z_2)^2\sin(\theta_0)\bigg]\cos(\varphi(z_1))\\
S_3(z_1,z_2) &= B \alpha_2 z_2\sin(\theta_0)
\end{split}
\end{equation}
Note that we display the equation for $\theta_0=\pm \pi/2$ which will leave out some terms multiplied with $\cos(\theta_0)$. The reason is that any other angle of $\theta_0$ will lead to a non-symmetric surface across the $x_1,x_2$ plane, and therefore a sub-optimal structure w.r.t. approximation operation. We also map $z_1$ through a function $\varphi(z_1)$ for the same reason as for the RCASD (enable arc length parametrization in one of the coordinate directions). A variant of the RCASD results when $\theta_0 = \pi/2$, $a=\alpha_0=0$ and $B=1$.

\begin{figure}[h]
    \begin{center}
        \subfigure[]{
            \includegraphics[width=0.45\columnwidth]{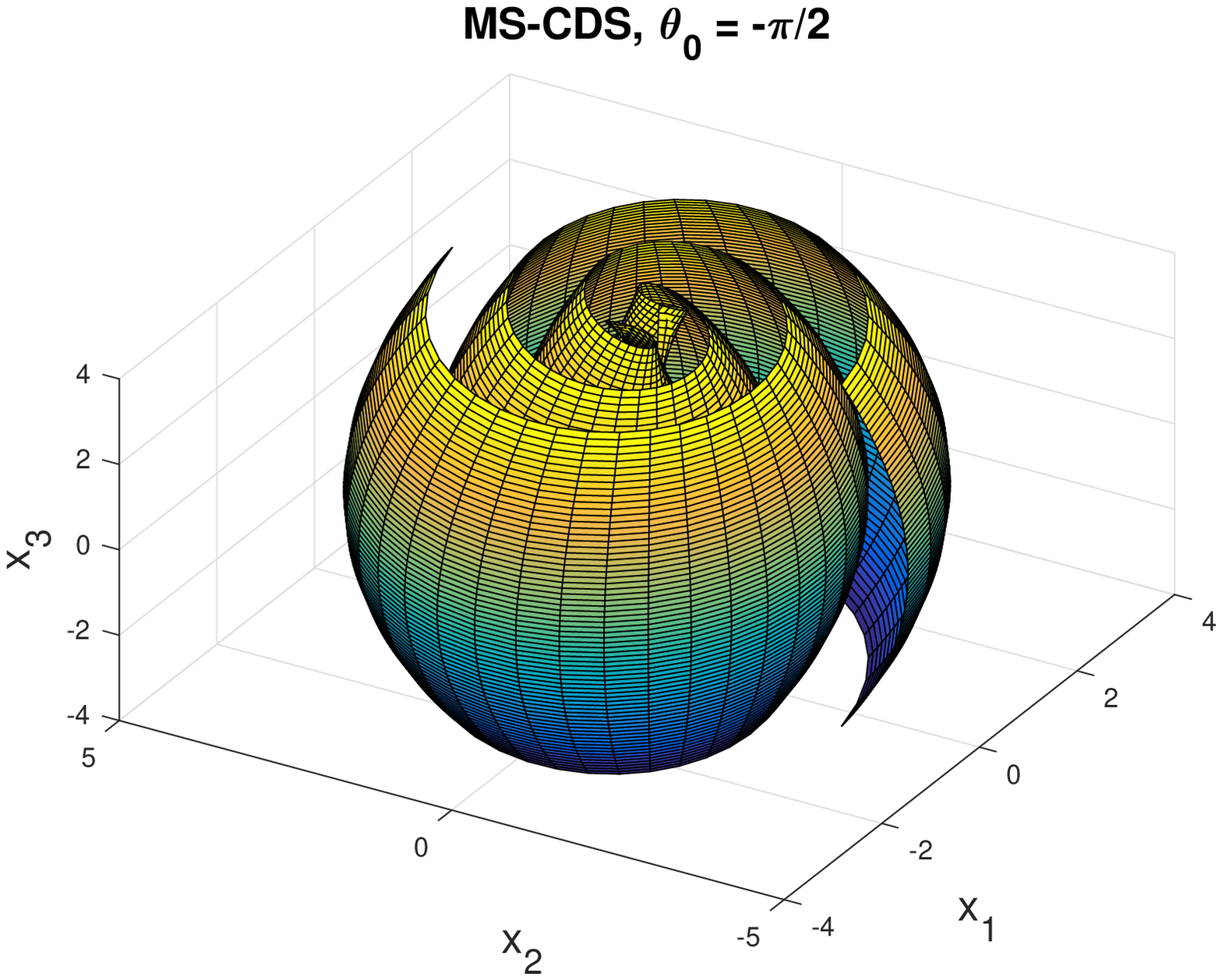}
        \label{fig:Surf_MS_CDS_th0neg}}
        \subfigure[]{
            \includegraphics[width=0.45\columnwidth]{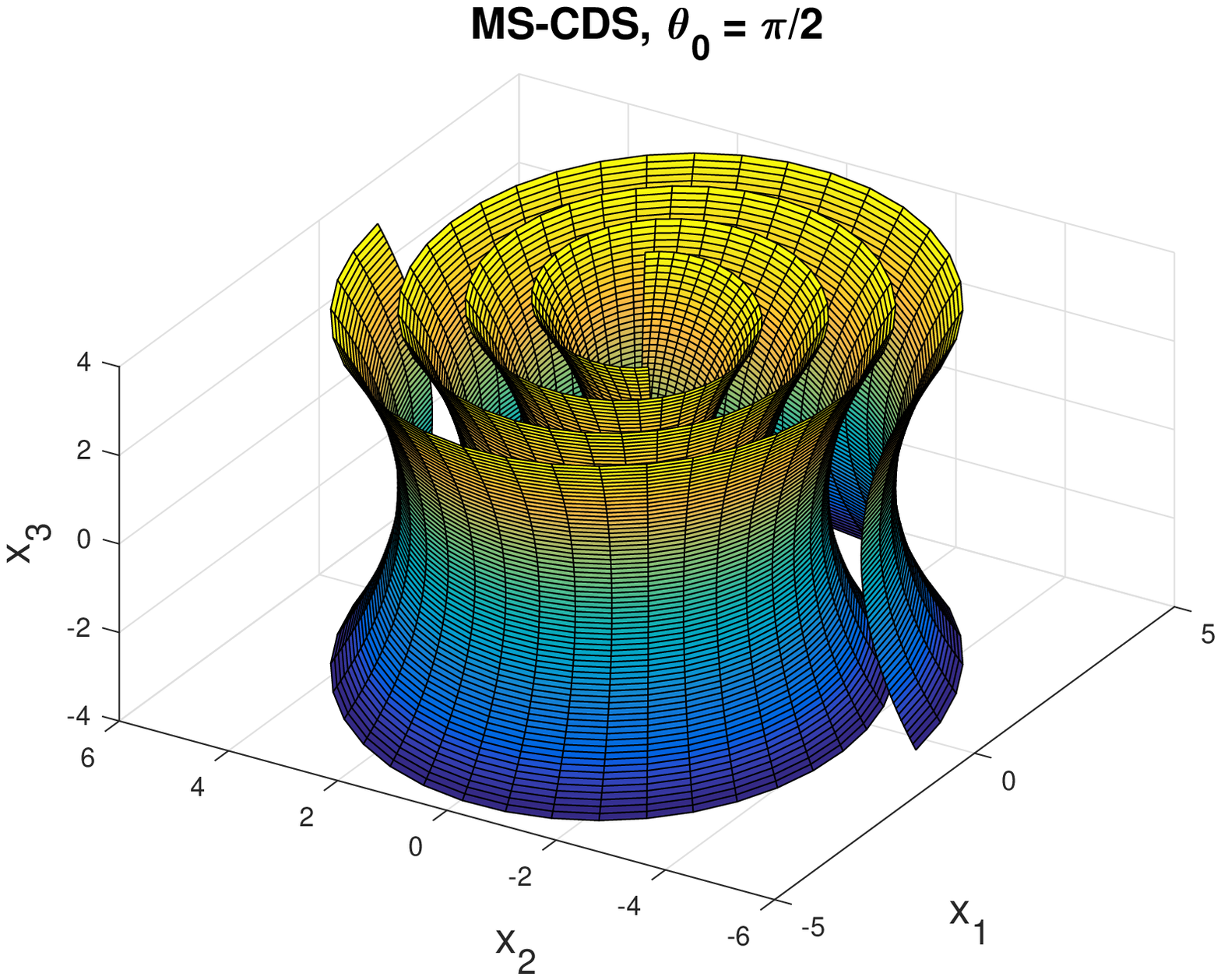}
        \label{fig:Surf_MS_CDS_th0pos}}
    \end{center}
    \caption{The MS-CDS surface for $\Delta=1$, $B=4$, $a=1$ and $\alpha_0=-0.5$ with: (a) $\theta_0=-\pi/2$. (b) $\theta_0=\pi/2$. }\label{fig:Surf_MS_CDS}
\end{figure}

Fig.~\ref{fig:Surf_MS_CDS} show the MS-CDS  for $\theta_0 = \pm \pi/2$.
The special case of MS-CDS shown in Fig.~\ref{fig:Surf_MS_CDS_th0neg} fills space in a ``spherical'' manner that should fit a 3D i.i.d Gaussian distribution quite well. We will consider the performance of this special case later.

The nonzero components of the metric tensor under LoC coordinates are ($g_{12}=0$)
\begin{equation}\label{e:MS_CDS_g11}
\begin{split}
&g_{11}(z_1,z_2)=\big(\varphi'(z_1)\big)^2 \bigg[\bigg(\frac{\Delta}{\pi}\bigg)^2\big(\alpha_0-\varphi(z_1)\big)^2+ \cdots\\
&+ a(\alpha_2 z_2)^2\sin(\theta_0)\bigg(a(\alpha_2 z_2)^2\sin(\theta_0) - 2\frac{\Delta}{\pi}\big(\alpha_0-\varphi(z_1)\big)\bigg)\bigg],
\end{split}
\end{equation}
and, since the MS-CDS is convenient to use only for $\theta_0=\pm \pi/2$,
\begin{equation}\label{e:MS_CDS_g22}
g_{22}(z_2)=B^2\alpha_2^2 + 4a^2 \alpha_2^4 z_2^2.
\end{equation}

\emph{Channel pdf's and Power:}
Since the MS-CDS is convenient to use only for $\theta_0=\pm \pi/2$, we have from~(\ref{e:S_MSCDS_ParEq}) that $x_3=\pm B \alpha_2 z_2$. Therefore $z_2\sim \mathcal{N}(0,\sigma_{z_2}^2)$ implying that the power on channel 2 becomes:
\begin{equation}\label{e:Pow_MSCDS_ch2}
P_2 =\mbox{Var}(z_2)= \sigma_{z_2}^2 = \frac{\sigma_x^2}{\alpha_2^2 B^2}.
\end{equation}

For Channel 1 we have a similar situation as for the RCASD (a spiral covering the $x_1,x_2$ plane), except that the spiral radius shrinks somewhat for larger $x_3$ values. However, this effect is generally negligible, and so the same approximation as for RCASD works at high SNR. By choosing $\varphi$ as in~(\ref{e:RCASD_Mapping_func}) we get the same power for channel 1:
\begin{equation}\label{e:Pow_MSCDS_1}
P_1 = \mbox{Var}(z_1)=\frac{2}{\alpha_1^2}\bigg(2\eta\frac{\pi}{\Delta}\sigma_x^2\bigg)^2, \ \eta=0.16.
\end{equation}

\emph{Approximation distortion:}
As for the RCASD, the MSCDS leads to a uniform structure and so the approximation distortion is given by~(\ref{e:RCASD_approx}).

\emph{Channel distortion:} From~(\ref{e:mseort_mean_DimRed}) we have
\begin{equation}\label{e:MSCDS_ChannelDist}
\begin{split}
\bar{\varepsilon}_{ch}^2 &= \frac{\sigma_n^2}{3} \iint\sum_{i=1}^2 g_{ii}(\mathbf{z}) f_\mathbf{z}(\mathbf{z})\mbox{d} \mathbf{z}\\
&= \frac{\sigma_n^2}{3} \bigg[\iint g_{11}(\mathbf{z}) f_\mathbf{z}(\mathbf{z})\mbox{d} \mathbf{z} +\int g_{22}(z_2)f_{z_2}(z_2)\mbox{d}z_2\bigg]\\
&= \frac{\sigma_n^2\big(I_1 + I_2\big)}{3},
\end{split}
\end{equation}
where
\begin{equation}\label{e:MSCDS_I_2}
\begin{split}
I_2 &=\int g_{22}(z_2)f_{z_2}(z_2)\mbox{d}z_2 \\&=\int \big(B^2\alpha_2^2 + 4 a^2 \alpha_2^4 z_2^2\big)f_{z_2}(z_2)\mbox{d}z_2\\
&= B^2\alpha_2^2 +  4 a^2 \alpha_2^4 \mbox{Var}(z_2) = \alpha_2^2\bigg(B^2 + \frac{4 a^2 \sigma_x^2}{B^2}\bigg).
\end{split}
\end{equation}
Further, we insert~(\ref{e:RCASD_Mapping_func}) into~(\ref{e:MS_CDS_g11}) and solve the relevant integral numerically.

\begin{figure}[h]
    \begin{center}
           \includegraphics[width=1.0\columnwidth]{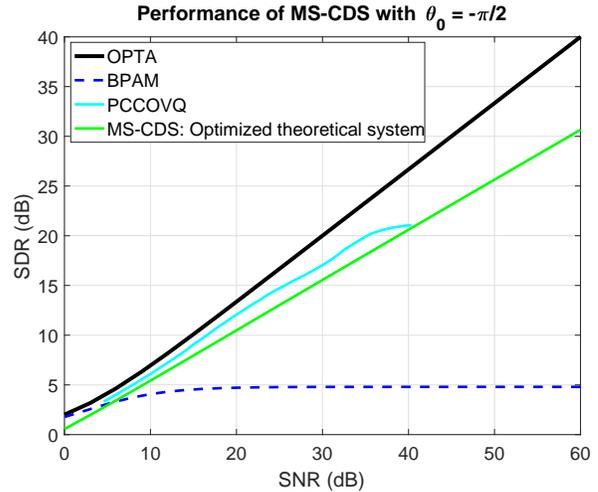}
    \end{center}
    \caption{Theoretical performance of MS-CDS when $\theta_0=-\pi/2$ compared to OPTA, BPAM and PCCOVQ.}\label{fig:MSCDS_Performance}
\end{figure}

\emph{Optimization:}
With constraint $C_t = P_{\text{max}} - P_t(\Delta,\alpha_1,\alpha_2) \geq 0$, we get the following objective function
\begin{equation}\label{e:RCASD_3_2_ObjFunc}
\mathcal{L} (\Delta,\alpha_1,\alpha_2) = \bar{\varepsilon}_{q}^2(\Delta) + \bar{\varepsilon}_{ch}^2 (\alpha_1,\alpha_2)-\lambda C_t(\Delta,\alpha_1,\alpha_2).
\end{equation}
We use the same procedure to determine the optimal parameters as for the RCASD for the special case $\theta_0=-\pi/2$. The performance is shown in Fig.~\ref{fig:MSCDS_Performance}
The MS-CDS has similar performance as RCASD, being somewhat inferior at high SNR. The slope at high SNR is again lower than that of OPTA. For the MS-CDS in general, the $1$:$1$ function mapping $x_3$ to $z_1,z_2$ has curvature $\kappa\neq 0$ , unlike the RCASD. According to~(\ref{e:ChDist_Final_2ndOrder}), the MS-CDS will then have a higher error in the ML estimate than the RCASD. As the approximation distortion is the same for the two structures, the MS-CDS will at best be on-par with the RCASD (at high SNR).

\textbf{Nonoptimality of slope of RCASD and MS-CDS}
There are two main reasons that RCASD and  MS-CDS cannot obtain the OPTA $3$:$2$ slope:

i) Both mappings can be seen as a composition of a $1$:$1$ and a $2$:$1$ mapping which is suboptimal according to Proposition~\ref{prop:Map_Split}.

ii) Any uniform S-K mapping has $\bar{\varepsilon}_{q}^2\sim \Delta^2$. According to Eq.~(\ref{e:Uniform_M_N_red_OptExp})  any uniform $3$:$2$  mapping then has to have $\bar{\varepsilon}_{ch}^2\sim 1/\Delta$ to follow the same slope as $3$:$2$ OPTA. For the RCASD we have that $\bar{\varepsilon}_{ch}^2\sim 1/\Delta^2$ according to~(\ref{e:RCASD_unconstrained_opt}), which is the wrong exponent. This is not as clear for the MS-CDS, but it is possible to verify that also for this case we have $\bar{\varepsilon}_{ch}^2\sim 1/\Delta^2$

Point ii) is easily verified for any mapping that has distortion $D_t  = \kappa_1 \Delta^2 + \kappa_2 \Delta^{-2}$ with $\kappa_1, \kappa_2$ constants. Differentiation w.r.t. $\Delta$ and solving for $\Delta$ one obtains $\Delta_{\text{opt}}=\sqrt[4]{2\kappa_1/\kappa_2}$. As the two distortion contributions $\bar{\varepsilon}_{q}^2 = \bar{\varepsilon}_{ch}^2$ at high SNR~\cite{hekland05}, we can insert $\Delta_{\text{opt}}$ in e.g. $\bar{\varepsilon}_{q}^2 $, and so SDR$\sim \sqrt{SNR}$, which is the exponent of a $2$:$1$ system.  

\subsubsection{Snail Surface}\label{ssec:SnaSu}
The \emph{Snail Surface} is a simple example of a surface that cannot be decomposed into sub-mappings. The parametric equation of the Snail Surface consists of the components~\cite[p. 280]{Surf_Encyclopedia}
\begin{equation}\label{e:SnaSU_ParEq_pos}
\begin{split}
S_1(z_1,z_2) &=  a\varphi(z_1)\sin(\varphi(z_1))\cos(\alpha_2 z_2+\phi)\\
S_2(z_1,z_2) &=  b\varphi(z_1)\cos(\varphi(z_1))\cos(\alpha_2 z_2+\phi)\\
S_3(z_1,z_2) &= -c \varphi(z_1)\sin(\alpha_2 z_2+\phi),
\end{split}
\end{equation}
which is valid for $0\leq z_1\leq k\pi$, $-\pi\leq z_2\leq\pi$. To include negative values for $z_1$ we combine the above parametric equation with
\begin{equation}\label{e:SnaSU_ParEq_neg}
\begin{split}
S_1(z_1,z_2) &=  - a\varphi(z_1)\sin(\varphi(z_1)-\psi)\cos(\alpha_2 z_2+\phi)\\
S_2(z_1,z_2) &=  - b\varphi(z_1)\cos(\varphi(z_1)-\psi)\cos(\alpha_2 z_2+\phi)\\
S_3(z_1,z_2) &=    c \sin(\alpha_2 z_2+\phi),
\end{split}
\end{equation}
valid for $-k \pi\leq z_1\leq 0$, $-\pi\leq z_2\leq\pi$, and so obtain a \emph{double} Snail Surface.

\begin{figure}[h]
    \begin{center}
            \includegraphics[width=0.8\columnwidth]{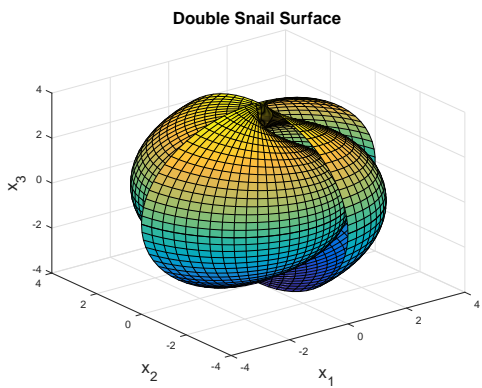}
    \end{center}
    \caption{The Double Snail Surface (SnaSu).}\label{fig:Surf_SnaSu}
\end{figure}

By choosing $\psi=\pi/2$ and $a=b=c=2\Delta/\pi$ one obtains a spherical symmetry which further results in a(close to) uniform S-K mapping (see~\cite{Floor_Ramstad09}, Definition 6), and therefore~(\ref{e:RCASD_approx}) will be a good approximation of a lower bound for the approximation distortion. Fig.~\ref{fig:Surf_SnaSu} depicts the double Snail Surface. Note the spherical symmetry, well suited for a 3D i.i.d Gaussian pdf.

Further (as will be seen later), by choosing $\phi=\pi/2$ or 0, the pdf of $z_2$ will be symmetric around the origin. The first choice leads to lower probability for the highest channel values. The choice of mapping function $\varphi(z_1)$ becomes apparent when considering the metric tensor.

For a general mapping function on $z_1$, the two non-zero metric components become
\begin{equation}\label{e:SnaSu_FFF_coef_KL}
\begin{split}
g_{11} &=(a\varphi'(z_1))^2 \big(1+ \varphi^2(z_1)\cos^2(\alpha_2 z_2 +\phi) \big),\\
g_{22} &= a^2\alpha_2^2 \varphi^2(z_1).
\end{split}
\end{equation}
By choosing $\varphi(z_1)=\alpha_1 z_1$ we get the metric tensor
\begin{equation}\label{e:SnaSu_FFF_coef_Radius}
\begin{split}
g_{11} &=a^2 \alpha_1^2 \big(1+ \alpha_1^2 z_1^2 \cos^2(\alpha_2 z_2 +\phi) \big)\\
g_{22} &= a^2\alpha_1^2\alpha_2^2 z_1^2 \\
g_{12} &= g_{21}=0.
\end{split}
\end{equation}
One can observe that $g_{ii}\sim z_1^2$, $i=1,2$, implying  that $\bar{\varepsilon}_{ch}^2$ increases with $z_1^2$. One can compensate this for both $g_{ii}$ components simultaneously by choosing $\varphi\sim \sqrt{z_1}$. As the RCASD also has $g_{11}\sim z_1^2$ when $\varphi(z_1)=\alpha_1 z_1$, and that SnaSu scales with $\Delta$ like RCASD, it makes sense to use~(\ref{e:RCASD_Mapping_func}) for here as well, the choice of $\eta$ being arbitrary.

\textbf{Optimization of Snail Surface as $3$:$2$ mapping.}

\emph{Channel Power and Density Function:}
\begin{figure}[h]
    \begin{center}
        \subfigure[]{
            \includegraphics[width=0.45\columnwidth]{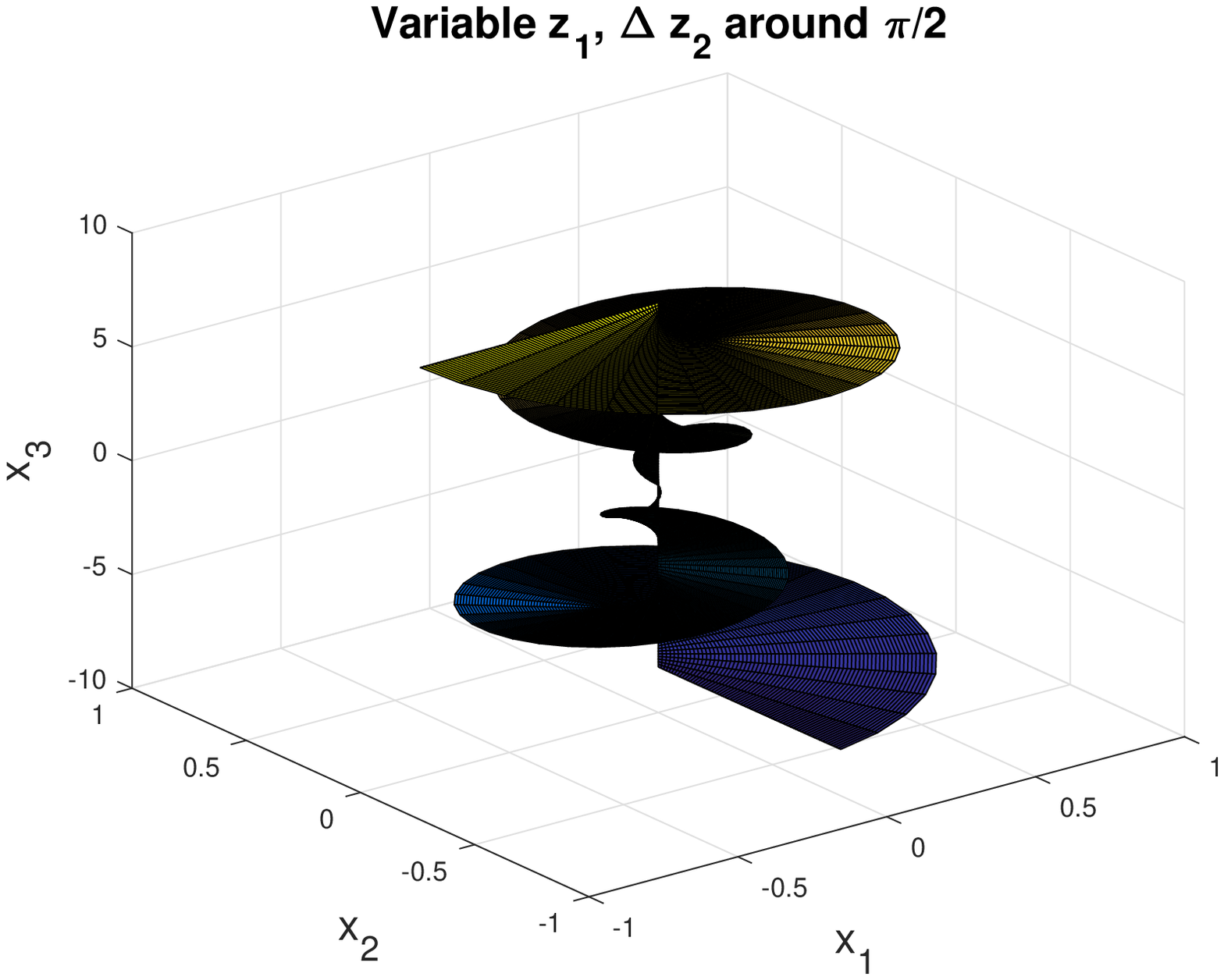}
        \label{fig:SnaSu_Var_z1_dz2}}
        \subfigure[]{
            \includegraphics[width=0.45\columnwidth]{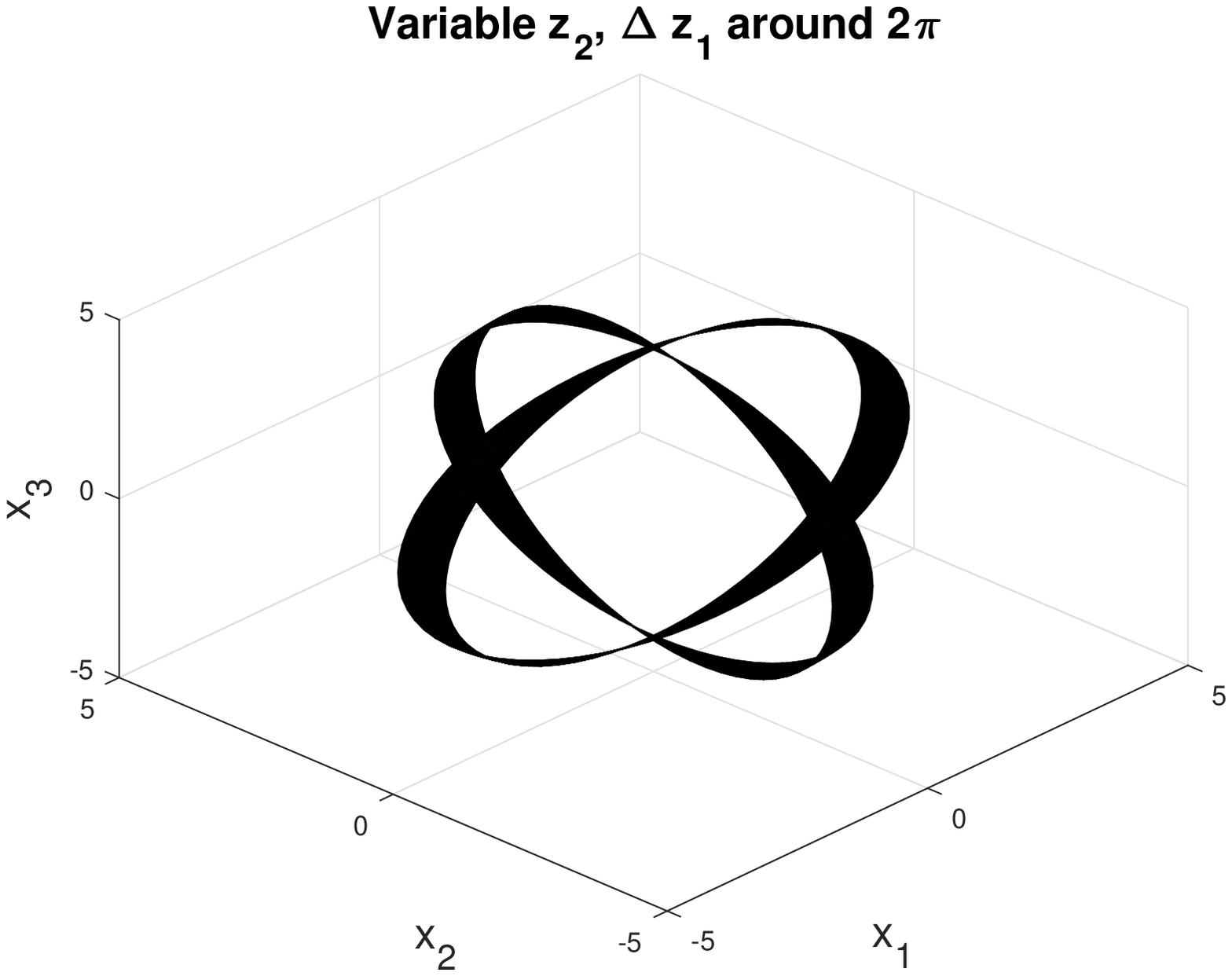}
        \label{fig:SnaSu_Var_z2_dz1}}
    \end{center}
    \caption{The geometrical behaviour for Snail Surface for the two channel signals. (a) $z_1$ variable with $z_2=\pi/2\pm\epsilon_1$. (b) $z_2$ variable with $z_1=2\pi\pm\epsilon_2$. }\label{fig:SnaSu_zi_dzj}
\end{figure}
To derive the pdf of $z_1$ consider Fig.~\ref{fig:SnaSu_Var_z1_dz2}. By letting $z_2$ vary slightly ($\pm \epsilon$) around some constant value (here $\pi/2$) with $z_1$ free we get a \emph{cork-screw} like structure. In the limit of constant $z_1$ we get a spiral with torsion $\tau\neq 0$. The rate at which the spiral \emph{rises} from the $x_1 x_2$-plane depending on the value of $z_2$. Whenever $z_2=\pm (2n+1)\pi/2, \ n\in\mathbb{N}$ the torsion $\tau$ is maximal, whereas when $z_2=\pm n\pi, \ n\in\mathbb{N}$ the spiral is a plane curve ($\tau=0$). This implies that the mapping from the snail surface to $z_1$ can be seen as a ``radius'' of a sphere that traces out all points inside this sphere depending on the value of $z_1$ and $z_2$, depending on the values of the sources $x_1, x_2,x_3$. This radius will then be $\rho=\sqrt{x_1^2 +  x_2^2 + x_3^2}$. By assuming that the snail surface is dense enough, i.e., with very small values of $\Delta$, one can approximate $\mathcal{S}$ as a mapping $h:\mathbb{R}^3 \rightarrow \mathbb{R}$ (this assumption becomes more and more accurate as the SNR gets larger since $\Delta$ becomes smaller). This mapping depends on $\varphi$. We first assume a general power function $\varphi=(\gamma z_1)^n$, $n\in \mathbb{Q}^+$.  Then we get a mapping $\mathcal{Z}: h(x_1,x_2,x_3)=\pm \gamma a^n(x_1^2+x_2^2+x_3^2)^{n/2}$,  and so $z_1=\ell(\rho)=\pm \gamma a^{-n} \rho^n = \pm \gamma a^{-n} (x_1^2+x_2^2+x_3^2)^{n/2}$.
The cumulative distribution is then given by (a straight forward generalization of the $h:\mathbb{R}^2 \rightarrow \mathbb{R}$ case in~\cite[pp. 180-181]{papoulis02})
\begin{equation}\label{e:SnaSu_CDF_z1}
\begin{split}
&F_{z_1}(z_1)=p_r\{Z_1\leq z_1\}=p_r\{(x_1,x_2,x_3)\in \mathcal{D}_{Z_1}^+\cup \mathcal{D}_{Z_1}^-\}\\
&=\iiint_{\mathcal{D}_{Z_1}^+\cup \mathcal{D}_{Z_1}^-} f_{X_1 X_2 X_3}(x_1,x_2,x_3 ) \mbox{d}x_1 \mbox{d}x_2 \mbox{d}x_3,
\end{split}
\end{equation}
where $f_{X_1 X_2 X_3}(x_1,x_2,x_3)$
 is the joint Gaussian distribution and
\begin{equation}\label{e:SnaSu_S_domain}
\begin{split}
&\mathcal{D}_{Z_1}^+ = \bigg\{(x_1,x_2,x_3)\big|(x_1^2+x_2^2+x_3^2)^\frac{n}{2}\leq \rho^n, z_1\geq 0 \bigg\},\\
& \mathcal{D}_{Z_1}^- = \bigg\{(x_1,x_2,x_3)\big|(x_1^2+x_2^2+x_3^2)^\frac{n}{2}\geq -\rho^n, z_1< 0 \bigg\}.
\end{split}
\end{equation}
The pdf is found by differentiating~(\ref{e:SnaSu_CDF_z1}) with
respect to $z_1$. To simplify, one can consider only the positive domain
in~(\ref{e:SnaSu_S_domain}) then use the fact that the channel pdf will be
symmetric about the origin with the chosen $\mathcal{S}$. Further, since the domain in~(\ref{e:SnaSu_S_domain})
is spherical, it is convenient to do the integration in {spherical coordinates}~\cite{Richter}
\begin{equation}\label{e:SnaSu_pdf_int}
f_{z_1}(z_1)=\frac{1}{2}
\frac{\mbox{d}}{\mbox{d}z_1}\int_0^{2\pi}\int_0^\pi\int_0^{a\varphi(z_1)}
f_{\rho}(\rho)\rho^{2}\sin(\theta)\mbox{d}\rho\mbox{d}\theta\mbox{d}\phi,
\end{equation}
where  $f_\rho (\rho)=\exp(-\rho^2/(2\sigma_x^2))/((2\pi)^{3/2} \sigma_x^2)$. The integrals over $\theta, \phi$ become $I_{(\theta, \phi)}=\pi$.
The integral over $\rho$ remains,
\begin{equation}\label{e:rho_int}
\begin{split}
&\frac{\mbox{d}}{\mbox{d}z_1}\int_0^{a \varphi(z_1)}f_{\rho}(\rho)\rho^{2}\mbox{d}\rho = \frac{1}{(2\pi)^{3/2}\sigma_x^3} \frac{\mbox{d}}{\mbox{d}z_1}\int_0^{a \varphi(z_1)} e^{-\frac{\rho^2}{2\sigma_x^2}}\rho^{2}\mbox{d}\rho\\
&=\frac{n a^3 \gamma^3 z_1^{3n-1}}{(2\pi)^{3/2} \sigma_x^3}e^{-\frac{a^2\varphi^2(z_1)}{2\sigma_x^2}}.
\end{split}
\end{equation}
By multiplying in $\pi$ and further using absolute value (to make the pdf symmetric around the
origin) we get
\begin{equation}\label{e:SnaSu_Z1_pdf_gen}
f_{z_1}(z_1)=\frac{n a^3 \gamma^3 |z_1|^{3n-1}}{\sqrt{2\pi}\sigma_x^3} e^{-\frac{a^2 \varphi(z_1)}{2\sigma_x^2}}.
\end{equation}
If we now assume that $\varphi(z_1)$ is as in~(\ref{e:RCASD_Mapping_func}), then $\gamma =\alpha_1/(\eta \Delta)$ and so
\begin{equation}\label{e:SnaSu_Z1_pdf}
f_{z_1}(z_1)=\frac{a^3 \alpha_1^{3/2} \sqrt{|z_1|}}{2\sqrt{2\pi}\sigma_x^3(\eta\Delta)^{3/2}} e^{-\frac{a^2 \alpha_1\varphi(z_1)|z_1|}{2\sigma_x^2\eta\Delta}}.
\end{equation}

According to~\cite[p.87]{papoulis02} the gamma distribution has the
general form $f_\gamma (x)=u(x){x^{c-1} e^{-\frac{x}{b}}}/{(\Gamma(c)b^c)}$.
From this it is easy  to realize that~(\ref{e:SnaSu_Z1_pdf}) is a \emph{double gamma distribution} with
$c=3/2$ and $b=(2\eta\Delta \sigma_x^2)/(a^2\alpha_1)$. From~\cite[p.154]{papoulis02} we have that the second moment of
of a gamma distribution is $\mbox{E}\{x^2\}=c(c+1)b^2$. Then, since~(\ref{e:SnaSu_Z1_pdf}) with
zero mean, its variance, and thereby the power of channel 1 becomes
\begin{equation}\label{e:SnaSu_Pow_z1}
P_1 = \sigma_{z_1}^2 = \frac{15(\eta\Delta\sigma_x^2)^2}{a^4\alpha_1^2}=\frac{15(\eta\pi^2\sigma_x^2)^2}{16\alpha_1^2\Delta^2}.
\end{equation}

\begin{figure}[h]
    \begin{center}
            \includegraphics[width=0.8\columnwidth]{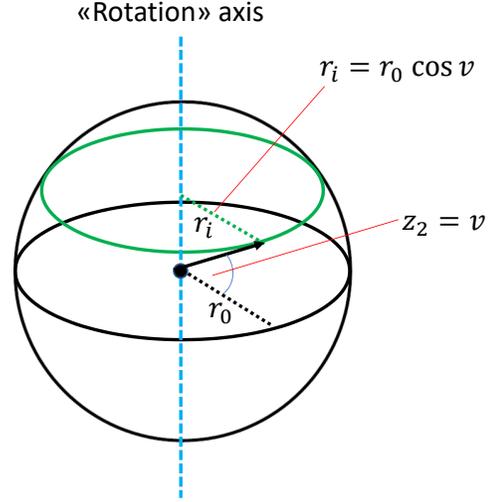}

    \end{center}
    \caption{The sphere used to compute the channel pdf of $z_2$ for the snail surface.}\label{fig:SnaSu_z2_sphere}
\end{figure}

To derive the pdf of $z_2$ consider Fig.~\ref{fig:SnaSu_Var_z2_dz1}.
 By perturbing $z_1$ with $\pm \epsilon_2$ around some constant value (here $2\pi$) with $z_2$ free, we get two M\"{o}bius strips. In the limit $\epsilon_2 \rightarrow 0$ we get a circle "rotating" about an axis whose radius increases as $2\Delta z_1/\pi$. Consider $\phi=0$. Then the rotation  axis is at $\pi/2$. The radius of the rotating circle is insignificant as $z_2\in [-\pi,\pi]$, independent of $z_1$. From the perspective of $z_2$, as the joint pdf of $\mathbf{x}$ is spherically symmetric, we have a uniform mass distribution over a virtual spherical shell of arbitrary radius, $r_0$, as depicted in Fig.~\ref{fig:SnaSu_z2_sphere}. To find the probability mass associated with different values of $z_2$, one considers the sum of all points along circles (green circle in Fig.~\ref{fig:SnaSu_z2_sphere}) resulting from intersections of this virtual sphere by planes perpendicular to the rotation axis.
The radius, $r_i$, of such a circle is  $r_i = r_0 \cos(\upsilon)$, where $\upsilon=z_2$ is the angle from the equatorial plane, $\upsilon = \pi/2 - \varrho$, and $\varrho$ the polar angle. The circumference as a function of $z_2$ is $O(z_2)=2\pi r_0 \cos(z_2)$. Since $r_0$ is arbitrary, one can set $r_0 = 1/(2\pi)$ implying that $f_{z_2}(z_2)\sim |\cos(z_2)|, \ z_2\in[-\pi,\pi]$. To avoid high probability for the largest channel amplitude values, one can set $\phi=\pi/2$ to obtain zero probability there. I.e., a \emph{sine distribution} results. To normalize, as $\int_0^\pi \sin(z_2)\mbox{d}z_2 = 2$, then
\begin{equation}\label{e:pdf_SnaSu_ch2}
f_{z_2}(z_2) = \frac{\alpha_2}{4}|\sin(\alpha_2 z_2)|.
\end{equation}
Since $f_{z_2}(z_2)$ is proportional to \emph{Gilberts sine distribution} $f_x(x)=\sin(2x)$, which according to~\cite{GilbertSine_Edwards} has variance $E\{x^2\}=(\pi^2/4-1)/2$, the power for channel 2 becomes
\begin{equation}\label{e:RCASD_PowZ2}
P_2 = \text{Var}\{z_2\} = \frac{2}{\alpha_2^2}\bigg(\frac{\pi^2}{4}-1\bigg).
\end{equation}

\emph{Weak Channel Distortion:}
From~(\ref{e:mseort_mean_DimRed}) we get
\begin{equation}\label{e:SnaSU_WeakChannelDist_pre}
\begin{split}
\bar{\varepsilon}_{ch}^2 &= \frac{\sigma_n^2}{3} \iint \big(g_{11}(\mathbf{z})+g_{22}(\mathbf{z})\big) f_\mathbf{z}(\mathbf{z})\mbox{d} \mathbf{z}\\
&= \frac{\sigma_n^2\big(I_1 + I_2\big)}{3}.
\end{split}
\end{equation}

We have, using~(\ref{e:SnaSu_FFF_coef_KL})
\begin{equation}\label{e:SnaSU_WeakChannelDist_I2}
\begin{split}
I_2 &=  \iint g_{22}(\mathbf{z}) f_\mathbf{z}(\mathbf{z})\mbox{d} \mathbf{z} = (a\alpha_2)^2\int\varphi^2(\alpha_1 z_1)f_{z_1}(z_1)\mbox{d}z_1\\
&= 2(a\alpha_2)^2 \frac{\alpha_1}{\eta\Delta}\int_{-\infty}^\infty z_1 f_{z_1}(z_1)\mbox{d}z_1 = 6 \alpha_2^2\sigma_x^2.
\end{split}
\end{equation}
The last equality comes from the fact that $z_1$ is gamma distributed and therefore the integral in~(\ref{e:SnaSU_WeakChannelDist_I2}) becomes~\cite[p.154]{papoulis02} $b c/2 = 3\eta \Delta \sigma_x^2 / (a^2 \alpha_1)$.

Further, we have (using~(\ref{e:SnaSu_FFF_coef_KL}))
\begin{equation}\label{e:SnaSU_WeakChannelDist_I1_pre}
\begin{split}
&I_1 =  \iint g_{11}(\mathbf{z}) f_\mathbf{z}(\mathbf{z})\mbox{d} \mathbf{z}= a^2\bigg(\int \varphi'(\alpha_1 z_1)^2 f_{z_1}(z_1)\mbox{d}z_1 \\
& +\int \varphi'(\alpha_1 z_1)^2 \varphi^2(\alpha_1 z_1)f_{z_1}(z_1)\mbox{d}z_1\int \cos^2(\alpha_2 z_2 +\phi)f_{z_2}(z_2)\mbox{d}z_2 \bigg)\\ &= a^2\bigg(\frac{\alpha_1}{4\eta \Delta}E\{z_1^{-1}\} + \frac{\alpha_1^2}{6(\eta \Delta)^2} \bigg)=\frac{\alpha_1\Delta}{\eta\pi^2}E\{z_1^{-1}\}+\frac{2\alpha_1^2}{3\pi^2\eta^2},
\end{split}
\end{equation}
where  we used that $\varphi'(\alpha_1 z_1)^2 = \alpha_1/(4\eta \Delta z_1)$. Through the series expansion $(1+x)^{-1} = 1-x+x^2-x^3+x^4+\cdots$, setting $x=z_1-1$, we get $E\{z_1^{-1}\}\approx E\{4-6z_1 + 4 z_1^2 -z_1^3\} = 4(1+\sigma_{z_1}^2)$ up to 3rd order, with $\sigma_{z_1}^2$ as in~(\ref{e:SnaSu_Pow_z1}).
Then
\begin{equation}\label{e:SnaSU_WeakChannelDist_I1}
I_1 \approx \frac{4\alpha_1\Delta}{\eta\pi^2}\big(1+\sigma_{z_1}^2\big)+\frac{2\alpha_1^2}{3\pi^2\eta^2}.
\end{equation}

Therefore the channel distortion becomes
\begin{equation}\label{e:SnaSU_WeakChannelDist}
\bar{\varepsilon}_{ch}^2 \approx \frac{\sigma_n^2}{3}\bigg(\frac{4\alpha_1\Delta}{\eta\pi^2}\big(1+\sigma_{z_1}^2\big)+\frac{2\alpha_1^2}{3\pi^2\eta^2} + 6\alpha_2^2\sigma_x^2\bigg).
\end{equation}

\emph{Optimization:}
With constraint $C_t = P_{\text{max}} - P_t(\Delta,\alpha_1,\alpha_2) \geq 0$, we get the following objective function
\begin{equation}\label{e:SnaSu_3_2_ObjFunc}
\mathcal{L} (\Delta,\alpha_1,\alpha_2) = \bar{\varepsilon}_{q}^2(\Delta) + \bar{\varepsilon}_{ch}^2 (\Delta,\alpha_1,\alpha_2)-\lambda C_t(\Delta,\alpha_1,\alpha_2).
\end{equation}

\begin{figure}[h]
    \begin{center}
           \includegraphics[width=1.0\columnwidth]{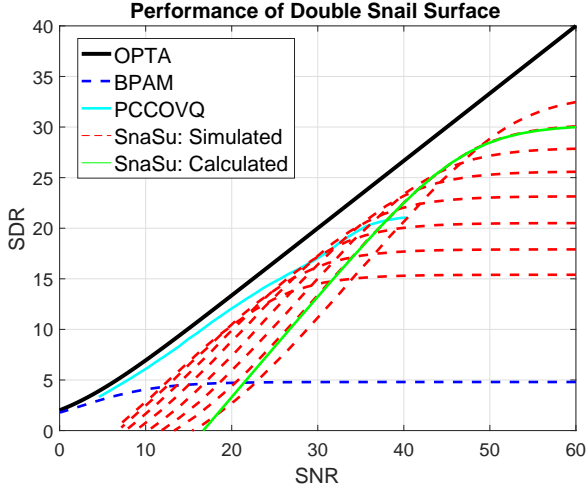}
    \end{center}
    \caption{Performance of Snail Surface (simulated and calculated) compared to OPTA, BPAM and PCCOVQ.}\label{fig:SnaSu_Performance}
\end{figure}

The performance of the optimized Snail Surface is shown in fig.~\ref{fig:SnaSu_Performance}

The Snail Surface clearly improves with SNR, beating both the linear BPAM as well as PCCOVQ (see footnote 9), and is noise robust.

\textbf{Curvature Analysis:}
By expanding~(\ref{e:SFF_SK_Red}) using the rules of determinants~\cite[p. 227]{strang} we get
\begin{equation}\label{e:SnaSu_SFF_b12_compute}
\begin{split}
&b_{ij}=b_{ji} = \frac{1}{\sqrt{g_{11} g_{22} }}\bigg(
\frac{\partial{S_1}}{\partial{z_1}}\begin{vmatrix}
\frac{\partial{S_2}}{\partial{z_2}}&\frac{\partial^2{S_2}}{\partial{z_i}\partial{z_j}}\\
\frac{\partial{S_3}}{\partial{z_2}}&
\frac{\partial^2{S_3}}{\partial{z_i}\partial{z_j}}
\end{vmatrix} -\cdots \\
&\frac{\partial{S_1}}{\partial{z_2}}
\begin{vmatrix}
\frac{\partial{S_2}}{\partial{z_1}}&\frac{\partial^2{S_2}}{\partial{z_i}\partial{z_j}}\\
\frac{\partial{S_3}}{\partial{z_1}}&
\frac{\partial^2{S_3}}{\partial{z_i}\partial{z_j}}
\end{vmatrix}+
\frac{\partial^2{S_1}}{\partial{z_i}\partial{z_j}}
\begin{vmatrix}
\frac{\partial{S_2}}{\partial{z_1}}&\frac{\partial{S_2}}{\partial{z_2}}\\
\frac{\partial{S_3}}{\partial{z_1}}&
\frac{\partial{S_3}}{\partial{z_2}}
\end{vmatrix} \bigg).
\end{split}
\end{equation}
The calculation of all derivatives and determinants above is a quite tedious process, so we do not show the detail here. With $\theta=\alpha_2 z_2+\phi$, one arrives at the following expressions
\begin{equation}\label{e:SnaSu_SFF}
\begin{split}
b_{11}&=-\frac{a(\varphi'(z_1))^2 \varphi^2(z_1)\cos^3\theta}{\sqrt{1+\varphi^2(z_1)\cos^2\theta}}, b_{22}=-\frac{a\alpha_2^2 \varphi^2(z_1)\cos\theta}{\sqrt{1+\varphi^2(z_1)\cos^2\theta}},\\ b_{12}&=\frac{a\alpha_1\alpha_2\varphi^2(z_1)\sin\theta}{\sqrt{1+\varphi^2(z_1)\cos^2\theta}}.
\end{split}
\end{equation}
From this we can conclude that we do not have LoC coordinates: According to Theorem~\ref{th:LoC_Coordinates} LoC coordinates require that $g_{12}=b_{12}=0$. From the expression for $b_{12}$ in~(\ref{e:SnaSu_SFF}) we see that $b_{12}\neq 0 $, except when $\alpha_2 z_2+\phi=0$, which is not a realistic possibility when both channel variables are used for communication. 

\begin{figure}[h]
    \begin{center}
        \subfigure[]{
            \includegraphics[width=0.45\columnwidth]{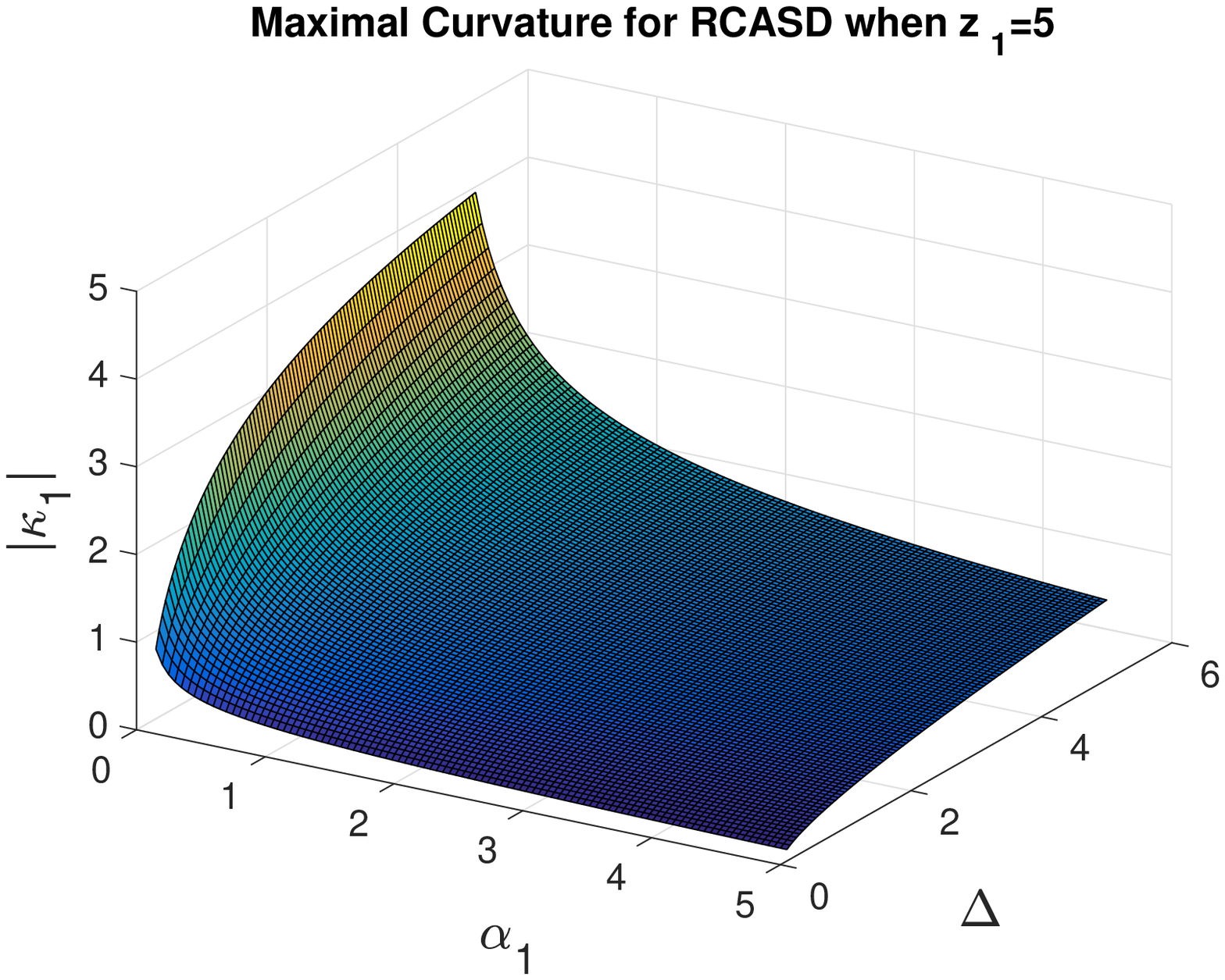}
        \label{fig:RCASD_Curvature}}
        \subfigure[]{
            \includegraphics[width=0.45\columnwidth]{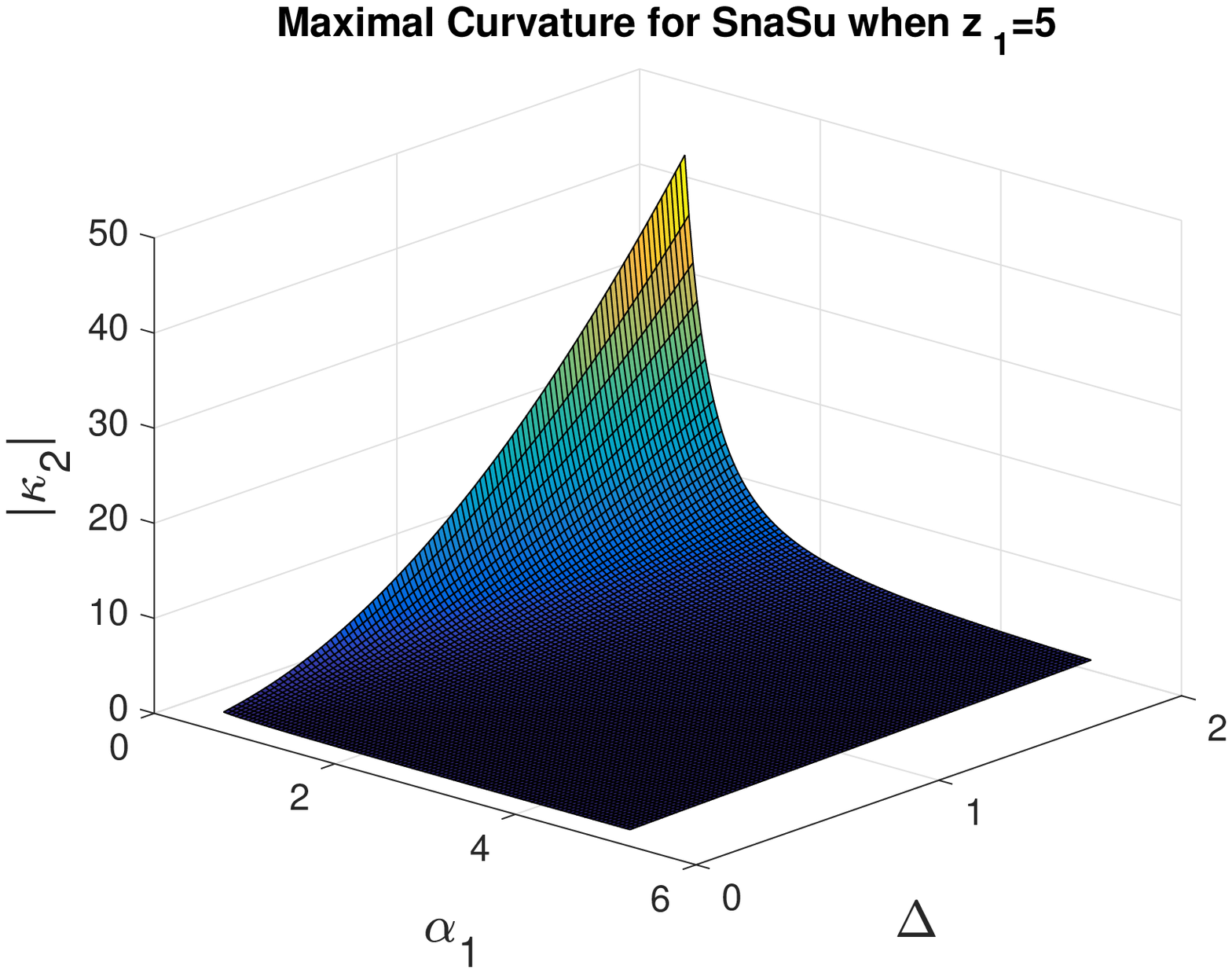}
        \label{fig:SnaSu_Curvature}}
    \end{center}
    \caption{Maximal curvature for: (a) RCASD when $z_1=5$ (b) SnaSu when $z_1=5$. All terms containing $z_2$ and $\alpha_2$ are set to their maximal value 1  }\label{fig:Curvature_3D_Plots}
\end{figure}

As the coordinates are not LoC, the principal curvatures are the roots of~(\ref{e:Princ_curvat_eq2}) 
\begin{equation}\label{e:SnaSu_PrincCurvat}
\kappa_{1/2}=\frac{1}{2}\bigg(\frac{b_{11}}{g_{11}}+\frac{b_{22}}{g_{22}} +/- \sqrt{\bigg(\frac{b_{11}}{g_{11}}+\frac{b_{22}}{g_{22}} \bigg)^2  - 4\frac{b_{11}b_{22}-b_{12}^2}{g_{11}g_{22}}}\bigg).
\end{equation}
Evaluation of $\kappa_2$ as function of $\Delta$ and $\alpha_1$ is shown in Fig.~\ref{fig:SnaSu_Curvature}.
Note that we have set all sine and cosine terms in~(\ref{e:SnaSu_FFF_coef_KL}) and~(\ref{e:SnaSu_SFF}) to one, implying that the 3D plot above shows a somewhat higher curvature than the actual value. The curvature is getting smaller as $\Delta$ decreases and $\alpha_1$ increases, corresponding to the high SNR case.  However, the curvature is quite large when $\Delta$ is large and $\alpha_1$ is small, corresponding to the low SNR case. The increase in curvature is much more dramatic than what was the case for RCASD, which may explain why the RCASD is performing better then SnaSu at lower SNR (see comparison in Fig.~\ref{fig:Performance_Map3_2All}). By studying the geometric configurations of RCASD in Fig.~\ref{fig:RCASD_Surface} and SnaSu in Fig.~\ref{fig:Surf_SnaSu}, one can expect the SnaSu to have somewhat higher curvature. 

By inserting optimized parameters for 30dB SNR found by the optimization procedure below ($\Delta^\ast=0.539$, $\alpha_1^\ast=4.76$, $\alpha_2^\ast=2.57$) one obtains maximal curvature $|\bar{\kappa}_1|<1$ averaged over the relevant range of $z_1$. By considering the distortion terms in~(\ref{e:WeakNoiseDist_2ndOrder_M_N}), with total transmission power 1, then $\sigma_n^2=0.001$, and one can see that the 1st order term is in the order of about $0.001/0.001^2 = 1000$ over the 2nd order term. Therefore, the SnaSu is also a mapping following Definition~\ref{def:weak_noise_exp} at high SNR.

\textbf{Further investigation of coordinate curves on SnaSu $3$:$2$ mapping:}
We already know that we do not have LoC coordinates. To figure out whether or not we have geodesic coordinates (GeCo) we compute the Christoffel symbols $\Gamma_{22}^1$ and  $\Gamma_{11}^2$. Using the definitions in~(\ref{e:Christoffel1}) and~(\ref{e:Christoffel2}) we get
\begin{equation}\label{e:christoffel1_dim_red}
\Gamma_{\alpha\beta\gamma}=\frac{1}{2}\bigg[\frac{\partial g_{\beta\lambda}}{\partial z^\alpha} + \frac{\partial g_{\lambda\alpha}}{\partial z^\beta}-  \frac{\partial g_{\alpha\beta}}{\partial z^\lambda} \bigg].
\end{equation}
With $g^{11}=g^{22}/g =1/g_{11}$ (since $g_{12}=0$), $g_{22} = g^{11}/g = 1/g^{22}$ and $g_{12}=-g^{12}/g = 0$, then from~(\ref{e:Christoffel1}), we get
\begin{equation}\label{e:christoffel_SnaSu1}
\Gamma_{11}^2 = -\frac{1}{2g_{22}}\frac{\partial g_{11} }{\partial z_2} = \frac{(\varphi'(z_1))^2}{\alpha_2}\cos(\alpha_2 z_2 +\phi)\sin(\alpha_2 z_2 + \phi),
\end{equation}
and
\begin{equation}\label{e:christoffel_SnaSu1}
\Gamma_{22}^1 = -\frac{1}{2g_{11}}\frac{\partial g_{22} }{\partial z_1} = \frac{\alpha_2^2 \varphi(z_1)}{\varphi'(z_1)\big(1+ \varphi(z_1) \cos^2(\alpha_2 z_2 +\phi)\big)}.
\end{equation}
By inserting $\varphi(z_1)$ from~(\ref{e:RCASD_Mapping_func}) one will see that both of these are $\neq 0$ in general ($\Gamma_{11}^2 = 0$ for $\alpha_2 z_2 + \phi = {0,\pi/2,\pi}$). Therefore, according to Theorem~\ref{th:GeCo_1}, we do not have GeCo in the SnaSu system above.

What one would like to obtain, if possible, is to find coordinates where $g_ii$ is a function of $z_i$ only and where $\bar{\varepsilon}_{ch}^2\sim \Delta^{-1}$. If this can be obtained with geodesic coordinates is unknown. This is quite a demanding problem and further research is needed to conclude.

%

\subsubsection{Comparison of $3$:$2$ Mappings}\label{ssec:Map_3_2_All}
\begin{figure}[h]
    \begin{center}
           \includegraphics[width=1.0\columnwidth]{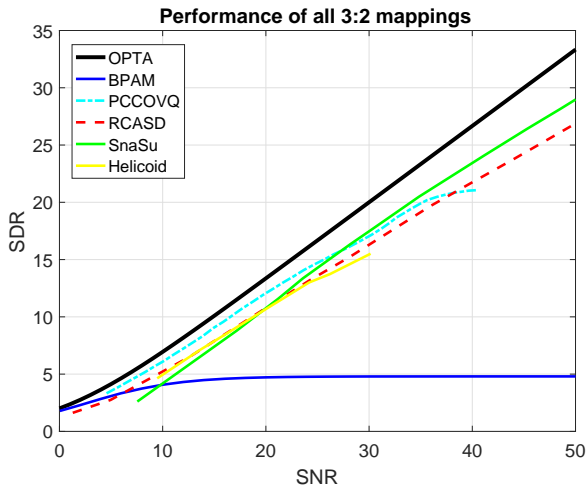}
    \end{center}
    \caption{Performance of three suggested $3$:$2$ mappigs compared to OPTA, BPAM and PCCOVQ. }\label{fig:Performance_Map3_2All}
\end{figure}

In Fig.~\ref{fig:Performance_Map3_2All} we compare all $3$:$2$ mappings proposed in this section.
We see that Helicoid and RCASD is quite well performing in the range 10-22 dB channel SNR. However, the Snail Surface is clearly superior at higher channel SNR, following the same slope as OPTA to a much higher SNR. However, it will eventually also diverge from OPTA.

\appendices
\section{Unit speed parametrization.}\label{sec:app_usp}
Assume a parametrization of a curve $\mathbf{s}(\varphi(x))$. For every
continuous parameter curve there exists a special $\varphi$ that
makes all tangent vectors along the curve unit vectors.

Let $\mathbf{s}: x\in [a,b]\subseteq \mathbb{R}\rightarrow
\mathbf{s}(x)\in\mathbb{R}^N$ be a parametrization for the curve $C$.
Let $\ell(x)$ denote the curve length function of $\mathbf{s}$ given by
\begin{equation}
\ell(x)=\int_a^x\|\mathbf{s}'(q)\|\mbox{d}q
\end{equation}
 and $\varphi$ denote its inverse.

\begin{theorem}\label{th:clpar}
Let $\mathbf{y}(\ell)$ be a curve length parametrization of $C$. Then
$\mathbf{y}(\ell)$ and $\mathbf{s}(\varphi(x))$ will have the same image, and
$\|\mathbf{y}'(\ell)\|=\| \mathbf{s}'(\varphi(\ell))\|\equiv1,\hspace{0.5cm}
\forall\ell $.
\end{theorem}
\begin{pf}
See\cite[pp. 115-116]{callahan00}.
\end{pf}

\section{The Metric Tensor}\label{sec:app_one_mt}
Consider a hyper surface $\mathcal{S}$ realized by the
parametric equation~(\ref{e:par_surf_eq}) (or~\ref{e:par_surf_eq2}).
The \emph{metric tensor} (also named \emph{Riemannian Metric})~\cite[pp.301-343]{Spivak99}  for a smooth embedding $\mathbf{S}$ in $\mathbb{R}^N$ ($M\leq N$)
is given by:
\begin{equation}
G=J^T J=
\begin{bmatrix}
g_{11}&g_{12}&\cdots&g_{1M}\\
g_{21}&g_{22}&\cdots&g_{2M}\\
\vdots&\vdots&\ddots&\vdots\\
g_{M1}&g_{M2}&\cdots&g_{MM},
\end{bmatrix}
\end{equation}
where $J$ is the Jacobian~\cite[p.47]{munkr91} of $\mathcal{S}$,
given by
\begin{equation}\label{e:jacobian}
J=
\begin{bmatrix}
\frac{\partial{s_1}}{\partial{x_1}}&
\frac{\partial{s_2}}{\partial{x_1}}
& \cdots & \frac{\partial{s_N}}{\partial{x_1}}\vspace{1mm}\\
\frac{\partial{s_1}}{\partial{x_2}}&
\frac{\partial{s_2}}{\partial{x_2}}& \cdots&
\frac{\partial{s_N}}{\partial{x_2}}\vspace{1mm}\\
\vdots & \vdots & \ddots & \vdots  \vspace{1mm}\\
\frac{\partial{s_1}}{\partial{x_M}}&
\frac{\partial{s_2}}{\partial{x_M}}& \cdots&
\frac{\partial{s_N}}{\partial{x_M}}\vspace{1mm}
\end{bmatrix}
^T
\end{equation}
The metric tensor $G$ is symmetric and positive definite~\cite[pp. 302-303]{Spivak99}. $g_{ii}$ can be interpreted  as the squared
length of the tangent vector in the direction of parameter $x_i$,
where $x_i$ is the i'th parameter in a parametric description of
$\mathcal{S}$. All ``cross terms'' $g_{ij}$, are the inner product
of the tangent vectors in the direction of $x_i$ and $x_j$.
See~\cite[pp.301-347]{Spivak99} for details.

\bibliographystyle{IEEEtran}
\bibliography{IEEEabrv,./references}


%








\end{document}